\documentclass[showpacs,showkeys,preprintnumbers,twocolumn,eqsecnum,prd]{revtex4}
\usepackage{graphicx}
\usepackage{latexsym}
\usepackage{amsmath}
\usepackage{amssymb}
\usepackage{bm}
\usepackage{hyperref}
\usepackage[utf8]{inputenc}

\allowdisplaybreaks

\begin{document}

\title{Effective action approach to higher-order relativistic tidal
  interactions in binary systems and their effective one body description}

\author{Donato Bini}
\affiliation{
Istituto per le Applicazioni del Calcolo ``M. Picone,'' CNR, I-00185 Rome,
Italy}

\author{Thibault Damour}
  \affiliation{Institut des Hautes \'{E}tudes Scientifiques, F-91440
    Bures-sur-Yvette, France}

\author{Guillaume Faye}
  \affiliation{Institut d'Astrophysique de Paris, UMR 7095 CNRS, Universit\'e
    Pierre et Marie Curie, 98 bis Boulevard Arago, 75014 Paris, France}

\date{\today}

\begin{abstract}
  The gravitational-wave signal from inspiralling neutron-star--neutron-star
  (or black-hole--neutron-star) binaries will be influenced by tidal coupling
  in the system. An important science goal in the gravitational-wave detection
  of these systems is to obtain information about the equation of state of
  neutron star matter via the measurement of the tidal polarizability
  parameters of neutron stars. To extract this piece of information will
  require accurate analytical descriptions  both of the motion and the
  radiation of tidally interacting binaries. We improve the analytical
  description of the late inspiral dynamics by computing the
  next-to-next-to-leading order relativistic correction to the tidal
  interaction energy. Our calculation is based on an effective-action approach
  to tidal interactions, and on its transcription within the
  effective-one-body formalism. We find that second-order relativistic effects
  (quadratic in the relativistic gravitational potential $u=G(m_1 +m_2)/(c^2
  r)$) significantly increase the effective tidal polarizability of neutron
  stars by a distance-dependent amplification factor of the form $1 + \alpha_1
  \, u + \alpha_2 \, u^2 + \cdots $ where, say for an equal-mass binary,
  $\alpha_1=5/4=1.25$ (as previously known) and $\alpha_2=85/14\simeq6.07143$
  (as determined here for the first time). We argue that higher-order
  relativistic effects will lead to further amplification, and we suggest a
  Pad\'e-type way of resumming them. We recommend testing our results by
  comparing resolution-extrapolated numerical simulations of
  inspiralling-binary neutron stars to their effective one body description.
\end{abstract}


\maketitle

\section{Introduction}\label{sec1}

Inspiralling binary neutron stars are among the most promising sources for the
advanced versions of the currently operating ground-based gravitational-wave
(GW) detectors LIGO/Virgo/GEO. These detectors will be maximally sensitive to
the inspiral part of the GW signal, which will be influenced by tidal
interaction between two neutron stars. An important science goal in the
detection of these systems (and of the related mixed black-hole--neutron-star
systems) is to obtain information about the equation of state of neutron-star
matter via the measurement of the tidal polarizability parameters of neutron
stars. The analytical description of tidally interacting compact-binary
systems (made of two neutron stars or one black hole and one neutron star) has
been initiated quite recently \cite{Flanagan:2007ix, Hinderer:2007mb,
  Damour:2009vw, Binnington:2009bb, Damour:2009wj, Vines:2010ca, Vines:2011ud,
  Ferrari:2011as}. In addition, these analytical descriptions have been
compared to accurate numerical simulations \cite{Damour:2009wj,
  Baiotti:2010xh, Baiotti:2011am, Lackey:2011vz}, and have been used to
estimate the sensitivity of GW signals to the tidal polarizability parameters
\cite{Lackey:2011vz, Read:2009yp, Hinderer:2009ca, Pannarale:2011pk,
  DamourNagarVillain2012}.

\smallskip

Here, we shall focus on one aspect of the analytical description of tidally
interacting {\it relativistic} binary systems, namely the role of the
higher-order post-Newtonian (PN) corrections in the tidal interaction energy,
as described, in particular, within the effective one body (EOB) formalism
\cite{Buonanno:1998gg, Buonanno:2000ef, Damour:2000we, Damour:2001tu}. Indeed,
the analysis of Ref.~\cite{Damour:2009wj}, which compared the prediction of
the EOB formalism for the binding energy of tidally interacting neutron stars
to (nonconformally flat) numerical simulations of quasi-equilibrium circular
sequences of binary neutron stars \cite{Uryu:2005vv, Uryu:2009ye}, suggested
the importance of higher-order PN corrections to tidal effects, beyond the
first post-Newtonian (1PN) level, and their tendency to {\it significantly
  increase} the ``effective tidal polarizability'' of neutron stars.

\smallskip

In the EOB formalism, the gravitational binding of a binary system is
essentially described by a certain ``radial potential'' $A(r)$. In the tidal
generalization of the EOB formalism proposed in Ref.~\cite{Damour:2009wj}, the
EOB radial potential $A(r)$ is written as the sum of three contributions,
\begin{equation}
\label{eq1.1}
A(r) = A^{\rm BBH} (r) + A_A^{\rm tidal} (r) + A_B^{\rm tidal} (r) \, ,
\end{equation}
where $A^{\rm BBH} (r)$ is the radial potential describing the dynamics of
binary black holes, and where $A_A^{\rm tidal} (r)$ and $A_B^{\rm tidal} (r)$
are the additional radial potentials associated, respectively, with the tidal
deformations of body $A$ and body $B$. [For binary neutron-star systems both
$A_A^{\rm tidal}$ and $A_B^{\rm tidal}$ are present, while for mixed
neutron-star--black-hole systems only one term, corresponding to the neutron
star, is present; see following]. Here, we consider a binary system of
(gravitational) masses $m_A$ and $m_B$, and denote
\begin{equation}
\label{eq1.2}
M \equiv m_A + m_B \, , \quad \nu \equiv \frac{m_A \, m_B}{(m_A + m_B)^2} \, .
\end{equation}
[A labelling of the two bodies by the letters $A$ and $B$ will be used in this
Introduction for writing general formulas. We shall later use the alternative
labelling $A=1$, $B=2$ when explicitly dealing with the metric generated by
the two bodies.]
The binary black-hole (or point mass) potential $A^{\rm BBH} (r)$ is known up
to the third post-Newtonian (3PN) level \cite{Damour:2000we}, namely
\begin{equation}
\label{eq1.3}
A_{3{\rm PN}}^{\rm BBH} (r) = 1-2 \, u + 2 \, \nu \, u^3 + a_4 \, \nu \, u^4 \, ,
\end{equation}
where $a_4 = 94/3 - (41/32) \, \pi^2 \simeq 18.68790269$, and
\begin{equation}
\label{eq1.4}
u \equiv \frac{GM}{c^2 \, r} \, ,
\end{equation}
with $c$ being the speed of light in vacuum and $G$ the Newtonian constant of
gravitation.

It was recently found \cite{Damour:2009kr, Buonanno:2009qa} that an excellent
description of the dynamics of binary black-hole systems is obtained by
augmenting the 3PN expansion Eq.~(\ref{eq1.3}) with additional fourth post-Newtonian (4PN) and 
fifth post-Newtonian (5PN)
terms, and by Pad\'e resumming the corresponding 5PN Taylor expansion.

\smallskip

The tidal contributions $A_{A,B}^{\rm tidal} (r)$ can be decomposed according
to multipolar order $\ell$, and type, as
\begin{align}
\label{eq1.5}
A_A^{\rm tidal} (r) &= \sum_{\ell \geq 2} \Bigl\{ A_{A \, {\rm electric}}^{(\ell) \, {\rm LO}} (r) 
\, \widehat A_{A \, {\rm electric}}^{(\ell)} (r) \nonumber\\
&+ A_{A \, {\rm magnetic}}^{(\ell) \, {\rm LO}} (r) \, \widehat A_{A \, {\rm magnetic}}^{(\ell)} (r) 
+ \ldots \Bigl\} \, .
\end{align}

Here, the label ``electric'' refers to the gravito-electric tidal polarization
induced in body $A$ by the tidal field generated by its companion, while the
label ``magnetic'' refers to a corresponding gravito-magnetic tidal
polarization. On the other hand, the label LO refers to the leading-order
approximation (in powers of $u$) of each (electric or magnetic) multipolar
radial potential. For instance, the gravito-electric contribution at
multipolar order $\ell$ is equal to \cite{Damour:2009wj}
\begin{equation}
\label{eq1.6}
A_{A \, {\rm electric}}^{(\ell) \, {\rm LO}} (r) = - \kappa_A^{(\ell)} \, u^{2\ell + 2}
\end{equation}
where
\begin{equation}
\label{eq1.7}
\kappa_A^{(\ell)} = 2 \, k_A^{(\ell)} \, \frac{m_B}{m_A} 
\left( \frac{R_A \, c^2}{G(m_A + m_B)} \right)^{2\ell + 1} \, .
\end{equation}
Here, $R_A$ denotes the radius of body $A$, and $k_A^{(\ell)}$ denotes a
dimensionless ``tidal Love number''. [Note that $k_A^{(\ell)}$ was denoted
$k_{\ell}^A$ in our previous work. Here we shall always put the multipolar
index $\ell$ within parentheses to avoid ambiguity with our later use of the
labelling $A,B=1,2$ for the two bodies.] The corresponding leading-order
radial potential of the gravito-magnetic type is proportional to $u^{2\ell +
  3}$ (instead of $u^{2\ell + 2}$), and to $j_A^{(\ell)} R_A^{2\ell + 1}$,
where $j_A^{(\ell)}$ denotes a dimensionless ``magnetic tidal Love number''.
It was found \cite{Damour:2009vw, Binnington:2009bb} that both types of Love
numbers have a strong dependence upon the {\it compactness} ${\mathcal C}_A
\equiv G \, m_A / (c^2 R_A)$ of the tidally deformed body, and that both
$k_A^{(\ell)}$ and $j_A^{(\ell)}$ contain a factor $1-2 \, {\mathcal C}_A$, so
that they would formally vanish in the limit where body $A$ becomes as compact
as a black hole (i.e. ${\mathcal C}_A \to {\mathcal C}_{\rm BH} =
\frac{1}{2}$). This is consistent with the decomposition Eq. (\ref{eq1.1}), where
the binary black-hole radial potential $A^{\rm BBH} (r)$ is the only remaining
contribution when one formally takes the limit where both ${\mathcal C}_A$ and
${\mathcal C}_B$ tend to the black-hole value ${\mathcal C}_{\rm BH} = 1/2$.
Finally, the supplementary factors $\widehat A_{A \, {\rm electric}}^{(\ell)}
(r)$ and $\widehat A_{A \, {\rm magnetic}}^{(\ell)} (r)$ denote the
distance-dependent {\it amplification factors} of the leading-order tidal
interaction by higher-order PN effects. They have the general form
\begin{equation}
\label{eq1.8}
\widehat A_{A \, {\rm electric}}^{(\ell)} (r) = 1 + \alpha_{1 \, {\rm electric}}^{A(\ell)} u 
+ \alpha_{2 \, {\rm electric}}^{A(\ell)} u^2 +
\ldots \, ,
\end{equation}
\begin{equation}
\label{eq1.9}
\widehat A_{A \, {\rm magnetic}}^{(\ell)} (r) = 1 + \alpha_{1 \, {\rm magnetic}}^{A(\ell)} u 
+ \ldots \, ,
\end{equation}
where $u$ is defined by Eq.~(\ref{eq1.4}).

\smallskip

The main aim of the present investigation will be to compute the electric-type
amplification factors $\widehat A_{A \, {\rm electric}}^{(\ell)}$, for 
$\ell = 2$ (quadrupolar tide) and $\ell = 3$ (octupolar tide), at the {\it
  second order} in $u$, i.e. to compute both $\alpha_{1 \, {\rm electric}}^{A(\ell)}$
and $\alpha_{2 \, {\rm electric}}^{A(\ell)}$. We shall also compute the
magnetic-type amplification factor $\widehat A_{A \, {\rm magnetic}}^{(\ell)}$, for 
$\ell = 2$, at the first order in $u$.

\smallskip

The analytical value of the first-order electric amplification coefficient
$\alpha_{1 \, {\rm electric}}^{A(\ell)}$ was computed some time ago for $\ell
= 2$ (see Ref.~[29] in \cite{Damour:2009wj}) and was reported in Eq.~(38) of
\cite{Damour:2009wj}, namely
\begin{equation}
\label{eq1.10}
\alpha_{1 \, {\rm electric}}^{A(\ell=2)}= \frac{5}{2} \, X_A \, ,
\end{equation}
where $X_A \equiv m_A / (m_A + m_B)$ is the mass fraction of body $A$. The
analytical result (\ref{eq1.10}) has been recently confirmed
\cite{Vines:2010ca}. On the other hand, several comparisons of the analytical
description of tidal effects with the results of numerical simulations have
indicated that the amplification factor $\widehat A_{A \, {\rm
    electric}}^{(\ell = 2)} (r)$ is larger that its 1PN value $1 + \alpha_{1
  \, {\rm electric}}^{A(\ell = 2)} \, u$, and have suggested that the
higher-order coefficients $\alpha_{2 \, {\rm electric}}^{A(\ell)} , \ldots$
take large, positive values. More precisely, the analysis of
Ref.~\cite{Damour:2009wj} suggested (when taking into account the value
(\ref{eq1.10}) for $\alpha_1$) a value of order $\alpha_{2 \, {\rm
    electric}}^{A(\ell=2)} \sim + \, 40$ (for the equal-mass case) from a
comparison with the numerical results of Refs.~\cite{Uryu:2005vv, Uryu:2009ye}
on quasi-equilibrium adiabatic sequences of binary neutron stars. Recently, a
comparison with dynamical simulations of inspiralling binary neutron stars
confirmed the need for such a large value of $\alpha_{2 \, {\rm electric}}^A$
\cite{Baiotti:2010xh, Baiotti:2011am}. [Note that, while the comparison to the
highest resolution numerical data suggests the need of even larger values of
$\alpha_{2 \, {\rm electric}}^{A(\ell=2)}$, of order $+ \, 100$, the
comparison to approximate resolution-extrapolated data  call only for
$\alpha_2$ values of order $+ \, 40$. See Fig.~13 in \cite{Baiotti:2011am}.]

\section{Effective action approach to tidal effects}\label{sec2}
\setcounter{equation}{0}

\subsection{Finite-size effects and nonminimal worldline couplings}

It was shown long ago \cite{Damour:1982wm}, using the technique of matched
asymptotic expansions, that the motion and radiation of $N$ (non-spinning)
compact objects can be described, up to the 5PN
approximation, by an effective action of the type
\begin{equation}
\label{eq2.1}
S_0 = \int \frac{d^D x}{c} \, \frac{c^4}{16 \, \pi \, G} \, \sqrt g \, R(g) +
S_{\rm point \, mass} \, ,
\end{equation}
where $R(g)$ represents the scalar curvature associated with the metric
$g_{\mu\nu}$, with determinant $-g$, and where
\begin{equation}
\label{eq2.2}
S_{\rm point \, mass} = - \sum_A \int m_A \, c^2 \, d \tau_A
\end{equation}
is the leading order skeletonized description of the compact objects, as point
masses. Here $d\tau_A$ denotes the proper time along the worldline
$y_A^{\mu}(\tau_A)$ of $A$, namely $d\tau_A \equiv c^{-1} (-g_{\mu\nu} (y_A)
\, dy_A^{\mu} \, dy_A^{\nu})^{1/2}$. To give meaning to the notion of point
mass sources in General Relativity one needs to use a covariant regularization
method. The most convenient one is {\it dimensional regularization}, i.e.
analytic continuation in the value of the spacetime dimension
$D=4+\varepsilon$, with $\varepsilon \in {\mathbb C}$ being continued to zero
only at the end of the calculation. The consistency and efficiency of this
method has been shown in the calculations of the motion
\cite{Damour:2001bu, Blanchet:2003gy} and radiation \cite{Blanchet:2004ek} of
binary black holes at the  3PN approximation.

\smallskip

It was also pointed out in Ref. \cite{Damour:1982wm} that finite-size effects
(linked to tidal effects, and the fact that neutron stars have, contrary to
black holes, non-zero Love numbers $k_A^{(\ell)}$) enter at the 5PN level. In
effective field theory, finite-size effects are treated by augmenting the
point-mass action of Eq. (\ref{eq2.2}) by nonminimal worldline couplings involving
higher-order derivatives of the field
\cite{Damour:1995kt, Damour:1998jk, Goldberger:2004jt}. In a gravitational
context this means considering worldline couplings involving the 4-velocity
$u_A^{\mu} \equiv dy_A^{\mu} / d\tau_A$ (satisfying $g_{\mu\nu} \, u_A^{\mu}
\, u_A^{\nu} = -c^2$) together with the Riemann tensor $R_{\alpha\beta\mu\nu}$
and its covariant derivatives. To classify the possible worldline scalars that
can be constructed one can appeal to the relativistic theory of tidal
expansions \cite{Zhang86, Damour:1990pi, Damour:1991yw}. In the notation of
Refs.~\cite{Damour:1990pi, Damour:1991yw} the tidal expansion of the ``external
metric'' felt by body $A$ can be entirely expressed in terms of two types of
external tidal gradients evaluated along the central worldline of this body:
the gravito-electric $G_L^A (\tau_A) \equiv G_{a_1 \ldots a_{\ell}}^A
(\tau_A)$ and gravito-magnetic $H_L^A (\tau_A) \equiv H_{a_1 \ldots
  a_{\ell}}^A (\tau_A)$ symmetric trace-free (spatial) tensors, together with
their time-derivatives. [The spatial indices $a_i = 1,2,3$ refer to a local
frame $X_A^0 \equiv c \, \tau_A$, $X_A^a$ attached to body $A$.] This implies
that the most general nonminimal worldline action has the form
\begin{eqnarray}
\label{eq2.3}
S_{\rm nonminimal} &= &\sum_A \ \sum_{\ell \geq 2} \ \biggl\{ \frac{1}{2} \, 
\frac{1}{\ell!} \, \mu_A^{(\ell)} \int d\tau_A (G_L^A (\tau_A))^2 \nonumber \\
&+ &\frac{1}{2} \, \frac{\ell}{\ell + 1} \, \frac{1}{\ell!} \, \frac{1}{c^2} 
\, \sigma_A^{(\ell)} \int d\tau_A (H_L^A (\tau_A))^2 \nonumber \\
&+ &\frac{1}{2} \, \frac{1}{\ell!} \, \frac{1}{c^2} \, \mu'^{(\ell)}_A 
\int d\tau_A (\dot G_L^A (\tau_A))^2 \nonumber \\
&+ &\frac{1}{2} \, \frac{\ell}{\ell + 1} \, \frac{1}{\ell!} \, \frac{1}{c^4} 
\, \sigma'^{(\ell)}_A \int d\tau_A (\dot H_L^A (\tau_A))^2 \nonumber \\
&+ &\ldots \biggl \} \, ,
\end{eqnarray}
where $\dot G_L^A (\tau_A) \equiv d \, G_L^A / d\tau_A$, and where the
ellipsis refer either to higher proper-time derivatives of $G_L^A$ and
$H_L^A$, or to higher-than-quadratic invariant monomials made from $G_L^A$,
$H_L^A$ and their proper-time derivatives. For instance, the leading-order
non-quadratic term would be
\begin{equation}
\label{eq2.4}
\int d\tau_A \ G_{ab}^A \ G_{bc}^A \ G_{ca}^A \, .
\end{equation}
Note that the allowed monomials in $G_L$, $H_L$ and their time derivatives are
restricted by symmetry constraints. When considering a non-spinning neutron
star (which is symmetric under time and space reflections) one should only
allow monomials invariant under time and space reversals. For instance $G_{ab}
\, \dot G_{ab}$ and $G_{ab} \, H_{ab}$ are not allowed.

\subsection{Tidal coefficients}

The electric-type tidal moments $G_L^A$ are normalized in a Newtonian way,
i.e. such that, in lowest PN order, they reduce to the usual Newtonian tidal
gradients: $G_L^A = [\partial_L \, U (X^a)]_{X^a = 0} + O \left( \frac{1}{c^2}
\right)$, where $U(X)$ is the Newtonian potential and
$\partial_L\equiv \partial_{i_1} \partial_{i_2} \ldots \partial_{i_\ell}$
represents multiple ordinary space derivatives. The magnetic-type ones $H_L^A$
are defined (in lowest PN order) as repeated gradients of the gravitomagnetic
field $c^3 g_{0a}$. With these normalizations the coefficients
$\mu_A^{(\ell)}$ and $\sigma_A^{(\ell)}$ in the nonminimal action in Eq. 
(\ref{eq2.3}) both have dimensions $[{\rm length}]^{2\ell + 1}/G$. They are
related to the dimensionless Love numbers $k_A^{(\ell)}$ and $j_A^{(\ell)}$,
and to the radius of body $A$, via \cite{Damour:2009vw}
\begin{equation}
\label{eq2.5}
G \mu_A^{(\ell)} = \frac{1}{(2\ell - 1)!!} \, 2 \, k_A^{(\ell)} \, R_A^{2\ell + 1} \, ,
\end{equation}
\begin{equation}
\label{eq2.6}
G \sigma_A^{(\ell)} = \frac{\ell - 1}{4 (\ell + 2)} \, 
\frac{1}{(2\ell - 1)!!} \, j_A^{(\ell)} \, R_A^{2\ell + 1} \, .
\end{equation}
Note that the coefficients associated with the first time derivatives of
$G_L^A$ and $H_L^A$ have dimensions $G \mu'^{(\ell)}_A \sim [{\rm
  length}]^{2\ell + 3} \sim G \sigma'^{(\ell)}_A$. The nonminimal action in Eq. 
(\ref{eq2.3}) has a double ordering in powers of $R_A$ and in powers of
$1/c^2$. The lowest-order terms in the $R_A$ expansion are proportional to
$R_A^5$ and correspond to the electric and magnetic quadrupolar tides, as
measured by $G_{ab}^A$ and $H_{ab}^A$, respectively.

\subsection{Tidal tensors}

We have written the most general nonminimal action Eq. (\ref{eq2.3}) in terms of
the irreducible symmetric trace-free spatial tensors [with respect to the
local space associated with the worldline $y_A^{\mu} (\tau_A)$] describing the
tidal expansion of the ``external metric'' felt by body $A$, as defined in Ref. 
\cite{Damour:1990pi}. These tidal tensors played a useful role in simplifying
the (1PN-accurate) relativistic theory of tidal effects. In our present
investigation, it will be convenient to express them in terms of the Riemann
tensor and its covariant derivatives. Eq.~(3.40) in Ref. \cite{Damour:1990pi} shows
that (in the case where one can neglect corrections proportional to the
covariant acceleration of the worldline) the first two electric spatial tidal
tensors, $G_{ab}$ and $G_{abc}$, are simply equal (modulo a sign) to the
non-vanishing spatial components (in the local frame) of the following
spacetime tensors (evaluated along the considered worldline)
\begin{equation}
\label{eq2.7}
G_{\alpha\beta} \equiv - \, R_{\alpha\mu\beta\nu} \, u^{\mu} \, u^{\nu} \, ,
\end{equation}
\begin{equation}
\label{eq2.8}
G_{\alpha\beta\gamma} \equiv - \, {\rm Sym}_{\alpha\beta\gamma}
(\nabla_{\alpha}^{\perp} \, R_{\beta\mu\gamma\nu}) \, u^{\mu} \, u^{\nu} \, .
\end{equation}

Here the notation $G_{\alpha\beta}$ for (minus) the electric part of the
curvature tensor should not be confused with the Einstein tensor, ${\rm
  Sym}_{\alpha\beta\gamma}$ denotes a symmetrization (with weight one) over
the indices $\alpha \, \beta \, \gamma$, while $\nabla_{\alpha}^{\perp} \equiv
P (u)^{\mu}_{\, \alpha} \, \nabla_{\mu}$ denotes the projection of the
spacetime gradient $\nabla_{\mu}$ orthogonally to $u^{\mu}$ ($P(u)^{\mu}_{\,
  \nu} \equiv \delta_{\nu}^{\mu} + c^{-2} \, u^{\mu} \, u_{\nu}$). [Note that
in the Newtonian limit $u^{\mu} \simeq c \, \delta_0^{\mu}$ so that the
Newtonian limit of $G_{\alpha\beta}$ is $- \, c^2 \, R_{\alpha0\beta0}$, where
the factor $c^2$ cancels the $O(1/c^2)$ order of the curvature tensor.] By
contrast, the presence of the extra term $- \, 3 \, c^{-2} \, E^*_{\langle a}
\, E^*_{b\rangle}$ on the right-hand side of Eq.~(3.40) in Ref. 
\cite{Damour:1990pi} shows that the $\ell = 4$ electric spatial tidal tensor
$G_{abcd} = \partial_{\langle abc} \, E^*_{d\rangle}$ would differ from the
symmetrized spatial projection of $(\nabla_{\alpha} \nabla_{\beta} \,
R_{\gamma\mu\delta\nu}) \, u^{\mu} \, u^{\nu}$ by a term proportional to
$G_{\langle\alpha\gamma} \, G_{\beta\delta \rangle}$. (Here, the angular
brackets denote a (spatial) symmetric trace-free projection.) In addition, the
electric time derivatives, such as $\dot G_{ab}$ can be replaced by
corresponding spacetime tensors such as $u^{\mu} \, \nabla_{\mu} \,
G_{\alpha\beta}$. Similarly to Eqs.~(\ref{eq2.7}), (\ref{eq2.8}), one finds
that the $\ell = 2$ and $\ell = 3$ magnetic tidal tensors (as defined in
Refs.~\cite{Damour:1990pi, Damour:1991yw}) are equal to the nonvanishing
local-frame spatial components of the spacetime tensors
\begin{equation}
\label{eq2.9}
H_{\alpha\beta} \equiv + \, 2 \, c \, R^*_{\alpha\mu\beta\nu} \, u^{\mu} \, u^{\nu} \, ,
\end{equation}
\begin{equation}
\label{eq2.10}
H_{\alpha\beta\gamma} \equiv + \, 2 \, c \, {\rm Sym}_{\alpha\beta\gamma}
(\nabla_{\alpha}^{\perp} \, R^*_{\beta\mu\gamma\nu}) \, u^{\mu} \, u^{\nu} \, ,
\end{equation}
where $R^*_{\mu\nu\alpha\beta} \equiv \frac{1}{2} \,
\epsilon_{\mu\nu\rho\sigma} \, R^{\rho\sigma} \, _{\alpha\beta}$ is the dual
of the curvature tensor, $\epsilon_{\mu\nu\rho\sigma}$ denoting here the
Levi-Civita tensor (with $\epsilon_{0123} = + \sqrt{g}$). Note the factor $+2$
entering the link between the magnetic tidal tensors $H_{\alpha\beta},
\ldots$ (normalized as in Refs.~\cite{Damour:1990pi, Damour:1991yw}) and the
dual of the curvature tensor, which contrasts with the factor $-1$ entering
the corresponding electric tidal-tensor links, Eqs.~(\ref{eq2.7}),
(\ref{eq2.8}). (The definition of ${\cal B}_{\alpha\beta}^A$ in the text below
Eq. (5) of Ref. \cite{Damour:2009wj} should have included such a factor 2 in its
right-hand side. On the other hand, the corresponding magnetic-quadrupole
tidal action, Eq. (13) there, was computed with $H_{ab}$ and was correctly
normalized.) Let us also note that the expressions in Eqs. 
(\ref{eq2.7})--(\ref{eq2.10}) assume that the Ricci tensor vanishes (e.g. to
ensure the tracelessness of $G_{\alpha\beta}$). One could have, alternatively,
defined $G_{\alpha\beta}$ etc. by using the Weyl tensor
$C_{\alpha\mu\beta\nu}$ instead of $R_{\alpha\mu\beta\nu}$. However, as
discussed in Ref. \cite{Damour:1998jk}, the terms in an effective action which are
proportional to the (unperturbed) equations of motion (such as Ricci terms)
can be eliminated (modulo contact terms) by suitable field redefinitions.

\subsection{Covariant description of tidal interactions}

Finally, the covariant form of the effective action describing tidal
interactions reads
\begin{equation}
\label{eq2.11}
S_{\rm tot} = S_0 + S_{\rm point \, mass} + S_{\rm nonminimal}
\end{equation}
where $S_0$ and $S_{\rm point \, mass}$ are given by Eqs.~(\ref{eq2.1}),
(\ref{eq2.2}), and where the covariant form of the nonminimal worldline
couplings starts as
\begin{eqnarray}
\label{eq2.12}
S_{\rm nonminimal} &= &\sum_{A} \biggl\{ \frac{1}{4} \, \mu_A^{(2)} \int
d\tau_A \, G_{\alpha\beta}^A \, G_A^{\alpha\beta}\nonumber\\
&+ & \frac{1}{6 \, c^2} \, \sigma_A^{(2)} \int d\tau_A \, H_{\alpha\beta}^A \,
H_A^{\alpha\beta} \nonumber \\
&+ & \frac{1}{12} \, \mu_A^{(3)} \int d\tau_A \, G_{\alpha\beta\gamma}^A \,
G_A^{\alpha\beta\gamma} \nonumber \\
&+ & \frac{1}{4 \, c^2} \, \mu'^{(2)}_A \int d\tau_A (u_A^{\mu} \nabla_{\mu} 
G_{\alpha\beta}^A) (u_A^{\nu} \nabla_{\nu} G_A^{\alpha\beta}) \nonumber \\
&+ & \ldots \biggl\} \, ,
\end{eqnarray}
where $G_A^{\alpha\beta} \equiv g^{\alpha\mu} \, g^{\beta\nu} \,
G_{\mu\nu}^A$, etc. [evaluated along the $A$ worldline].

\smallskip

In principle, one can then derive the influence of tidal interaction on the
motion and radiation of binary systems by solving the equations of motion
following from the action of Eqs. (\ref{eq2.11}), (\ref{eq2.12}). More precisely, this
action implies both a dynamics for the worldlines where the geodesic equation
is modified by tidal forces [coming from $\delta S_{\rm nonminimal} / \delta
\, y_A^{\mu} (\tau_A)$], and modified Einstein equations for the gravitational
field of the type
\begin{equation}
\label{eq2.13}
R_{\mu\nu} - \frac{1}{2} \, R \, g_{\mu\nu} = \frac{8 \pi \, G}{c^4} \,
\left\{ T_{\mu\nu}^{\rm point \, mass} + T_{\mu\nu}^{\rm nonminimal} \right\}
\, ,
\end{equation}
where the new tidal sources $T_{\rm nonminimal}^{\mu\nu} (x) = (2c/\sqrt g) \,
\delta S^{\rm nonminimal} / \delta \, g_{\mu\nu} (x)$ are, essentially, sums
of derivatives of worldline Dirac-distributions:
$$
T_{\rm nonminimal}(x) \sim \sum_A \sum_{\ell} \partial^{\ell} \, \delta (x-y_A) \, .
$$

\subsection{A simplifying, general property of {\it reduced} actions}

The task of solving the coupled dynamics of the worldlines and of the
gravitational field, both being modified by tidal effects, at the second post-Newtonian (2PN) level,
i.e. at the next-to-next-to-leading order in tidal effects, and then of
computing the looked for higher-order terms in the amplification factors of Eqs. 
(\ref{eq1.8}), (\ref{eq1.9}) is quite non-trivial. Happily, one can
drastically simplify the needed work by using a general property of {\it
  reduced actions}. Indeed, we are interested here in knowing the influence of
tidal effects on the {\it reduced} dynamics of a compact binary, that is,  the
dynamics of the two worldlines $y_A^{\mu} (\tau)$, $y_B^{\mu} (\tau)$,
obtained after having ``integrated out'' the gravitational field (i.e., after
having explicitly solved $g_{\mu\nu} (x)$ as a functional of the two
worldlines). When considering, as we do here, the {\it conservative} dynamics
of the system (without radiation reaction), it can be obtained from a {\it
  reduced action}, which is traditionally called the ``Fokker action''. See
Ref. \cite{Damour:1995kt} and references therein for a detailed discussion (using a
diagrammatic approach) of Fokker actions (at the 2PN level, and with the
inclusion of scalar couplings in addition to the pure Einsteinian tensor
couplings). If we denote the fields mediating the interaction between the
worldlines $y = \{ y_A , y_B \}$ as $\varphi$ (in our case $\varphi =
g_{\mu\nu}$), the reduced worldline action $S_{\rm red} [y]$ (a functional of
the worldlines $y$) that corresponds to the complete action $S [\varphi , y]$
describing the coupled dynamics of $y$ and $\varphi$ is formally defined as:
\begin{equation}
\label{eq2.14}
S_{\rm red} \, [y] \equiv S \, [\varphi_{\rm sol} \, [y], \, y] \, ,
\end{equation}
where $\varphi_{\rm sol} \, [y]$ is the functional of $y$ obtained by solving
the $\varphi$-field equation,
\begin{equation}
\label{eq2.15}
\delta \, S \, [\varphi , y] / \delta \varphi = 0 \, ,
\end{equation}
considered as an equation for $\varphi$, with given source-worldlines. (This
must be done with time-symmetric boundary conditions and, in the case of
$g_{\mu\nu}$, the addition of a suitable gauge-fixing term; see
Ref. \cite{Damour:1995kt} for details.)

\smallskip

Having recalled the concept of reduced (or Fokker) action, let us now consider
the case where the complete action is of the form
\begin{equation}
\label{eq2.16}
S \, [\varphi , y] = S^{(0)} [\varphi , y] + \epsilon \, S^{(1)} [\varphi , y] \, ,
\end{equation}
where $\epsilon$ denotes a ``small parameter''. In our case, $\epsilon$ can be
either a formal parameter associated with all the nonminimal tidal terms in
$S_{\rm nonminimal}$, Eq.~(\ref{eq2.12}), or, more concretely, any of the
tidal parameters entering Eq. (\ref{eq2.12}): $\mu_A^{(\ell = 2)}$, $\mu_B^{(\ell
  = 2)}$, etc. As said previously, when turning on $\epsilon$, the equations of
motion, and therefore the solutions of both $\varphi$ and $y$ get perturbed by
terms of order $\epsilon \, $: $\varphi = \varphi^{(0)} + \epsilon \,
\varphi^{(1)} + \ldots$, $y = y^{(0)} + \epsilon \, y^{(1)} + \ldots$, but a
simplification occurs when considering the reduced action Eq. (\ref{eq2.14}).
Indeed, it is true that the field equation (\ref{eq2.15}) for $\varphi$ gets
modified into
\begin{equation}
\label{eq2.17}
0 = \frac{\delta \, S \, [\varphi , y]}{\delta \, \varphi} = \frac{\delta \,
  S^{(0)} [\varphi , y]}{\delta \, \varphi} + \epsilon \, \frac{\delta \,
  S^{(1)} [\varphi , y]}{\delta \, \varphi} \, ,
\end{equation}
so that its solution $\varphi_{\rm sol} \, [y]$ gets perturbed:
\begin{equation}
\label{eq2.18}
\varphi_{\rm sol} \, [y] = \varphi_{\rm sol}^{(0)} \, [y] + \epsilon \,
\varphi_{\rm sol}^{(1)} \, [y] + O (\epsilon^2) \, .
\end{equation}
However, when inserting the perturbed solution of Eq. (\ref{eq2.18}) into the
complete, perturbed action of Eq. (\ref{eq2.16}), one finds
\begin{eqnarray}
\label{eq2.19}
S_{\rm red} \, [y] &= &S \, [\varphi_{\rm sol}^{(0)} \, [y] + \epsilon \,
\varphi_{\rm sol}^{(1)} \, [y] + O(\epsilon^2) , y] \nonumber \\
&= & S \, [\varphi_{\rm sol}^{(0)} \, [y] , y] + \epsilon \, 
\varphi_{\rm sol}^{(1)} \, [y] \, \frac{\delta \, S}{\delta \, \varphi} \, 
[\varphi_{\rm sol}^{(0)} \, [y] , y] + O(\epsilon^2) \nonumber \\
&= & S \, [\varphi_{\rm sol}^{(0)} \, [y] , y] \nonumber\\
&& + \epsilon \, \varphi_{\rm sol}^{(1)} \, [y] \, 
\frac{\delta \, S^{(0)}}{\delta \, \varphi} \, [\varphi_{\rm sol}^{(0)} \, 
[y] , y] + O(\epsilon^2) \nonumber \\
&= &S \, [\varphi_{\rm sol}^{(0)} \, [y] , y] + O(\epsilon^2) \, ,
\end{eqnarray}
because, by definition, $\varphi_{\rm sol}^{(0)}$ is a solution of $\delta \,
S^{(0)} / \delta \, \varphi = 0$. 
Note that, in Eq. (\ref{eq2.19}), while the functional $S$ is the {\it complete, perturbed} action, 
the functional argument is the {\it unperturbed} solution. Decomposing the functional $S$ into 
its unperturbed plus perturbed parts [see Eq. (\ref{eq2.16})] then leads to the
final result:
\begin{eqnarray}
\label{eq2.20}
S_{\rm red} \, [y] &= &S^{(0)} [\varphi_{\rm sol}^{(0)} \, [y],y] 
+ \epsilon \, S^{(1)} [\varphi_{\rm sol}^{(0)} \, [y] , y] 
+ O(\epsilon^2) \nonumber \\
&= &S_{\rm red}^{(0)} \, [y] + \epsilon \, S^{(1)} [\varphi_{\rm sol}^{(0)} \,
[y] , y] + O(\epsilon^2) \, .
\end{eqnarray}
In words: the order $O(\epsilon)$ perturbation
$$
\epsilon \, S_{\rm red}^{(1)} \, [y] \equiv S_{\rm red} \, [y] - S_{\rm red}^{(0)} \, [y]
$$
of the {\it reduced} action is correctly obtained, modulo terms of order
$O(\epsilon^2)$, by replacing in the $O(\epsilon)$ perturbation
$$
\epsilon \, S^{(1)} \, [\varphi , y]
$$
of the {\it complete} (unreduced) action the field $\varphi$ by its {\it
  unperturbed} solution $\varphi_{\rm sol}^{(0)} \, [y]$.

\smallskip

In our case, the ordering parameter $\epsilon$ is either the collection
$\mu_A^{(2)} , \mu_B^{(2)} , \mu_A^{(3)} , \mu_B^{(3)}, \ldots , \sigma_A^{(2)} \, c^{-2}
,\ldots ,$ $\mu'^{(2)}_A \, c^{-2} , \ldots$, or the corresponding sequence of
powers of $R_A$ and $R_B \, $: $R_A^5 , R_B^5 , R_A^7 , R_B^7 , \ldots$ The
terms quadratic in $\epsilon$ would therefore involve at least {\it ten}
powers of the radii (and would mix with higher-than-quadratic worldline
contributions akin to (\ref{eq2.4})). Neglecting such terms, we conclude that
the higher-PN corrections to the tidal effects are correctly obtained by
replacing in Eq.~(\ref{eq2.12}), considered as a functional of $g_{\mu\nu}
(x)$ and $y_A^{\mu} (\tau_A)$, the metric $g_{\mu\nu} (x)$ by the {\it
  point-mass metric} obtained by solving Einstein's equations with point-mass
sources. [This was the method used by one of us (T.D.) to compute the 1PN
coefficient of Eq. (\ref{eq1.10}) from the calculation by Damour, Soffel and Xu of
the 1PN-accurate value of $G_{ab}$ \cite{Damour:1992qi, Damour:1993zn}.]

\section{The 2PN point-mass metric and its regularization}\label{sec3}

\subsection{Form of the 2PN point-mass metric}

\setcounter{equation}{0}

The result of the last Section allows one to compute the tidal corrections to
the reduced action for two tidally interacting bodies $A,B$ with the same
accuracy at which one knows the metric generated by two (structureless) point
masses $m_A , y_A^\mu ; m_B , y_B^\mu$. The metric generated by two point
masses has been the topic of many works over many years. It has been known (in
various forms and gauges) at the 2PN approximation for a long time
\cite{OOKH73, DamourMG3, Schaefer:1986rd}. Here, we shall use the convenient,
explicit harmonic-gauge form of Ref.~\cite{Blanchet:1998vx}, with respect to
the (harmonic) coordinates $x^{\mu} = (x^0 \equiv c t , x^i)$, i.e. the metric
\begin{equation}
\label{eq3.1}
ds^2 = g_{00} (dx^0)^2 + 2 \, g_{0i} \, dx^0 dx^i + g_{ij} \, dx^i dx^j \, ,
\end{equation}
where, at 2PN, the metric components are written as
\begin{align}
\label{eq3.2}
g_{00}&= -1+2 \, \epsilon^2 \, V - 2 \, \epsilon^4 \, V^2 \nonumber\\
&+ 8 \, \epsilon^6 \left(\hat X+\delta^{ij} \, V_i V_j 
+ \frac16 \, V^3\right) + O(8) \, , \nonumber \\
g_{0i}&=- \, 4 \, \epsilon^3 \, V_i - 8 \, \epsilon^5 \hat R_i+O(7)\nonumber
\, , \\
g_{ij}&=\delta_{ij}\left(1+ 2 \, \epsilon^2 \, V + 2 \, 
\epsilon^4 \, V^2\right) + 4 \, \epsilon^4 \, \hat W_{ij} + O(6) \, .
\end{align}
Here, as below, we sometimes use the alternative notation $\epsilon \equiv
1/c$ for the small PN parameter. We used also the shorthand notation $O(n)
\equiv O(\epsilon^n) \equiv O(c^{-n})$.

\smallskip

The various 2PN brick potentials $V , V_i , \hat W_{ij} , \hat R_i$ and $\hat
X$ are the (time-symmetric) solutions of
\begin{align}
\label{eq3.3}
\Box \, V &=  - \, 4 \, \pi \, G \sigma \, , \nonumber \\
\Box \, V_i &= - \, 4 \, \pi \, G \sigma_i \, , \nonumber \\
\Box \, \hat W_{ij} &= - \, 4 \, \pi \, G (\sigma_{ij} - 
\delta_{ij} \, \sigma_{kk}) - \partial_i V \, \partial_j V \, , \nonumber \\
\Box \, \hat R_i &= - \, 4 \, \pi \, G (V \sigma_i - V_i \sigma) 
- \, 2 \, \partial_k V \, \partial_i V_k - \frac32 \, \partial_t V 
\, \partial_i V \, , \nonumber \\
\Box \, \hat X &= - \, 4 \, \pi \, G \, V \sigma_{ii} + 2 \, V_i 
\, \partial_t \, \partial_i V + V \, \partial_t^2 \, V + \frac32 
\, (\partial_t V)^2 \nonumber \\
&- \, 2 \, \partial_i V_j \, \partial_j V_i + \hat W_{ij} \, \partial_{ij} V \, ,
\end{align}
where $\partial_t$ denotes a time derivative (while we remind that
$\partial_i$, for instance, denotes a spatial one), and where the
compact-supported source terms are \cite{Blanchet:1989ki}
\begin{equation}
\label{eq3.4}
\sigma \equiv \frac{T^{00} + T^{ii}}{c^2} \, , \quad \sigma_i \equiv
\frac{T^{0i}}{c} \, , \quad \sigma_{ij} \equiv T^{ij} \, ,
\end{equation}
with $T^{\mu\nu}$ being the stress-energy tensor of two point masses:
\begin{equation}
\label{eq3.5}
T^{\mu\nu} = \mu_1 (t) \, v_1^{\mu} (t) \, v_1^{\nu} (t) \, \delta ({\bm x} -
{\bm y}_1 (t)) + 1 \leftrightarrow 2 \, ,
\end{equation}
where
\begin{equation}
\label{eq3.6}
\mu_1 (t) = m_1 \left[ g^{-1/2} (g_{\mu\nu} \, v_1^{\mu} \, v_1^{\nu}/c^2)^{-1/2}
\right]_1 \, .
\end{equation}
Here, $v_1^{\mu} = \frac{dy_1^{\mu}}{dt} = (c , v_1^i)$ and the index $1$ on
the bracket in Eq.~(\ref{eq3.6}) refers to a regularized limit where the field
point $x^i$ tends towards the (point-mass) source point $y_1^i$. Note that, in
this section, we shall generally label the two particles as $(m_1 , y_1^i)$,
$(m_2 , y_2^i)$, instead of $(m_A , y_A^i)$, $(m_B , y_B^i)$ as above. The
notation $1 \leftrightarrow 2$ means adding the terms obtained by exchanging
the particle labels $1$ and $2$.

\smallskip

The explicit forms of the 2PN-accurate brick potentials $V$, $V_i$, $\hat W_{ij}$, $\hat R_i$, $\hat X$ 
were given in Ref. \cite{Blanchet:1998vx}. Their time-symmetric parts are recalled in Appendix A. 
These brick potentials are expressed as explicit functions of  ${\bm r}_1 \equiv {\bm x} - {\bm y}_1$, $r_1 \equiv \vert {\bm r}_1
\vert$, ${\bm n}_1 \equiv {\bm r}_1 / r_1$, ${\bm r}_2 \equiv {\bm x} - {\bm
  y}_2$, etc., ${\bm y}_{12} \equiv {\bm y}_1 - {\bm y}_2$, $r_{12} \equiv
\vert {\bm y}_{12} \vert$, ${\bm n}_{12} \equiv {\bm y}_{12} / r_{12}$, ${\bm
  v}_{12} \equiv {\bm v}_1 - {\bm v}_2$, $(n_{12} \, v_1) \equiv {\bm n}_{12}
\cdot {\bm v}_1$. Note the appearance  of the auxiliary quantity $S$, which denotes the perimeter of the triangle
defined by ${\bm x}$, ${\bm y}_1$ and ${\bm y}_2$, viz
\begin{equation}
\label{eq3.12}
S \equiv r_1 + r_2 + r_{12} \, .
\end{equation}

In all the PN expressions, the spacetime points $x^{\mu} , y_1^{\mu} ,
y_2^{\mu}$ (and the velocities $v_A^{\mu}$) are taken at the same instant $t$,
i.e. $x^0 = y_1^0 = y_2^0 = ct$.

\subsection{Regularization of the 2PN metric and of the 2PN tidal actions}

Let us now discuss in more detail the crucial operation (already implicit in
Sec.~\ref{sec2} above) of regularization of all the needed field
quantities, such as $g_{\mu\nu} (x)$, $g(x)$, $R_{\mu\alpha\nu\beta}(x) ,
\ldots$, when they are to be evaluated on a worldline: $x^{\mu} \to
y_A^{\mu}$. As mentionned at the beginning of Sec.~\ref{sec2}, all the
quantities $[G_{\mu\nu} (x)]_1 , \ldots , [R_{\mu\alpha\nu\beta} (x)]_1$ are
defined by dimensional continuation. It was shown long ago \cite{D80,
  Damour:1982wm} that, at 2PN, dimensional regularization is equivalent to the
Riesz' analytic regularization, and is a technical shortcut for computing the
physical answer obtained by the matching of asymptotic expansions. In
addition, because of the restricted type of singular terms that appear at 2PN
[see Eqs.~(25), (30) and (33) in Ref. \cite{Damour:1982wm}], the
analytic-continuation regularization turns out to be equivalent to Hadamard
regularization (used, at 2PN, in Refs. \cite{Bel:1981be, Schaefer:1986rd,
  Blanchet:1998vx}); see below. Here, it will be technically convenient to use
Hadamard regularization (which is defined in $D=4$) because the explicit form of Eqs. 
(\ref{eq3.7})--(\ref{eq3.11}) of the 2PN metric that we shall use applies only
in the physical dimension $D=4$ and has lost the information about its
dimensionally continued kin in $D = 4+\varepsilon$.

\smallskip

Let us summarize here the (Hadamard-type) definition of the regular part of
any field quantity $\varphi (x)$ (which might be a brick potential, $V(x) ,
V_i (x) , \ldots$, a component of the metric $g_{\mu\nu} (x)$, or a specific
contribution to a tidal moment, $G_{\alpha\beta} , \ldots$). We consider the
behavior of $\varphi (x)$ near particle $1$, i.e. when $r_1 = \vert {\bm x} -
{\bm y}_1 \vert \to 0$. To ease the notation, we shall provisionally put the
origin of the (harmonic) coordinate system at ${\bm y}_1$ (at some instant
$t$), i.e. we shall assume that ${\bm y}_1 = 0$, so that $r_1 = \vert {\bm x}
\vert \equiv r$ and ${\bm n}_1 = {\bm r}_1 / r_1 = {\bm x} / r \equiv {\bm
  n}$. We consider the expansion of $\varphi ({\bm x})$ in (positive and
negative) integer powers $k$ of $r_1 = r$, and in spherical harmonics of the
direction ${\bm n}_1 = {\bm n}$, say (for $k \in {\mathbb Z}$, $\ell \in
{\mathbb N}$, $N \in {\mathbb N}$)
\begin{equation}
\label{eq3.13}
\varphi ({\bm x}) = \sum_{k \geq -N} \, \sum_{\ell \geq 0} \, r^k \, \hat n^L 
\, f_L^k \, ,
\end{equation}
where $\hat n^L \equiv \hat n^{a_1 \ldots a_{\ell}}$ denotes the symmetric
trace-free projection of the tensor $n^L \equiv n^{a_1} \ldots n^{a_{\ell}}$.
[The angular function $f_L^k \, \hat n^L$ is equivalent to a sum of
$\underset{m=-\ell}{\overset{+\ell}{\sum}} \, c_m \, Y_{\ell m}$.] We
(uniquely) decompose the field $\varphi ({\bm x})$ in a {\it regular} part
$(R)$ and a {\it singular} one $(S)$,
\begin{equation}
\label{eq3.14}
\varphi ({\bm x}) = R \, [\varphi ({\bm x})] + S \, [\varphi ({\bm x})] \, ,
\end{equation}
by defining ($n \in {\mathbb N}$)
\begin{equation}
\label{eq3.15}
R \, [\varphi ({\bm x})] \equiv \sum_{\ell \geq 0} \, \sum_{n \geq 0} \, 
r^{\ell + 2n} \, \hat n^L f_L^{\ell + 2n} \, ,
\end{equation}
\begin{equation}
\label{eq3.16}
S \, [\varphi ({\bm x})] \equiv \sum_{k \ne \ell + 2n} \, r^k \, \hat n^L f_L^k \, .
\end{equation}
Note that $R \, [\varphi ({\bm x})]$ can be rewritten as a sum of infinitely
differentiable terms of the type $\hat x^L ({\bm x}^2)^n$. By contrast $S \,
[\varphi ({\bm x})]$ is such that it (if $N$, in Eq.~(\ref{eq3.13}), is
strictly positive), or, one of its (repeated) spatial derivatives, tends
towards infinity as $r \to 0$. Note also that the $R+S$ decomposition commutes
with linear combinations (with constant coefficients), as well as with spatial
derivatives, in the sense that $R \, [a \, \varphi ({\bm x}) + b \, \psi ({\bm
  x})] = a \, R \, [\varphi ({\bm x})] + b \, R \, [\psi ({\bm x})]$, $S \, [a
\, \varphi ({\bm x}) + b \, \psi ({\bm x})] = a \, S \, [\varphi ({\bm x})] +
b \, S \, [\psi ({\bm x})]$, $R \, [\partial_i \, \varphi ({\bm x})] = \partial_i \,
R \, [\varphi ({\bm x})]$ and $S \, [\partial_i \, \varphi ({\bm x})]
= \partial_i \, S \, [\varphi ({\bm x})]$. By contrast, the $R+S$
decomposition (as defined above, in the Hadamard way) does not commute with
nonlinear operations (e.g. $R \, [\varphi \, \psi] \ne R \, [\varphi] \, R \,
[\psi]$), nor even with multiplication by a smooth $(C^{\infty})$ function
$f({\bm x})$ (e.g. $R \, [f \varphi] \ne f \, R \, [\varphi]$). This is a
well-known inconsistency of the Hadamard regularization, which created many
ambiguities when it was used at the 3PN level \cite{Jaranowski:1997ky,
  Blanchet:2000nv}. One might worry that our present calculation (which aims
at regularizing nonlinear quantities quadratic in $R_{\mu\alpha\nu\beta}
\sim \partial^2 g + g^{-1} \, \partial g \, \partial g$) might be intrinsically
ambiguous already at the 2PN level. Actually, this turns out not to be the
case because of the special structure of the 2PN metric which is at work in
the Riesz-analytic-continuation derivation of the 2PN dynamics in Ref. 
\cite{Damour:1982wm}. This structure guarantees, in particular, that the Riemann
tensor (or its derivatives) is regularized unambiguously.

\subsection{On the special structure of the 2PN metric guaranteeing its unambiguous regularization}

Let us first recall why the Riesz-analytic-continuation
method, or, equivalently (when considering the regularization of the metric
and its derivatives), the dimensional-continuation method, {\it is} consistent
under nonlinear operations. The dimensional-continuation analog of Eqs. 
(\ref{eq3.14})--(\ref{eq3.16}) consists of distinguishing, within $\varphi ({
  \bm x})$, the terms that (in dimension $4+\varepsilon$) contain powers of
$r$ of the type $r^{k-n\varepsilon}$, with $n=1,2,3,\ldots$ [which define the
$\varepsilon$-singular part of $\varphi ({\bm x})$], and the terms that are
(formally) $C^{\infty}$ in $4+\varepsilon$ dimensions [which define the
$\varepsilon$-regular part of $\varphi ({\bm x})$]. It is then easily seen in
dimensional continuation (simply by considering the continuation to large,
negative values of the real part of $\varepsilon$) that the
$\varepsilon$-singular terms give vanishing contributions when evaluated at $r
\to 0$, and that they do so consistently in nonlinear terms such as
$\partial\varphi \, \partial\psi$. Let us now indicate why the special
structure of the 2PN metric ensures that the decomposition into
$\varepsilon$-singular parts and $\varepsilon$-regular parts of the various
brick potentials $V(x) , V_i (x) , \ldots$ coincides with their above-defined
decomposition into Hadamard-singular ($S \, [V(x)], S \, [V_i (x)] , \ldots$)
and Hadamard-regular parts ($R \, [V(x)], R \, [V_i (x)] , \ldots$) in the
four-dimensional case. This is trivially seen to be the case for most of the
2PN contributions to the brick potentials (because one easily sees how those
contributions smoothly evolve when analytically continuing the dimension).
However, the most nonlinear contributions to the 2PN metric, namely the terms,
say ${\hat X}^{(VVV)}$, in ${\hat X}$ that are generated by the cubically
nonlinear terms contained in the last source term, $\hat W_{ij}^{(VV)}
\, \partial_{ij} V $, on the right hand-side of the last Eq.~(\ref{eq3.3})
(where $\hat W_{ij}^{(VV)}$ is the part of $\hat W_{ij}$ generated by $
- \partial_i V \, \partial_j V$) are more delicate to discuss. Actually, among
the contribution ${\hat X}^{(VVV)}$, only the terms proportional either to
$m_1^2 m_2$ or to $m_1 m_2^2$, i.e., the terms whose cubically nonlinear source
$\sim \partial^2 V \Delta^{-1} \partial V \partial V$ involve two $V$
potentials generated by one worldline and one $V$ potential generated by the
other worldline, such as ${\hat X}^{(V_1V_1V_2)} \propto m_1^2 m_2$, pose a
somewhat delicate problem. More precisely, it is easily seen that the only
dangerous part in ${\hat X}^{(V_1V_1V_2)}$, considered near the first
worldline, is of the form $f({\bm x})/r_1^{(2 + 2 \varepsilon)}$ in dimension
$4 + \varepsilon$, where $f({\bm x})$ denotes a smooth function. [Here, we add
back the particle label indicating whether the expansions Eqs. (\ref{eq3.15}),
(\ref{eq3.16}) refer to the first $(A=1)$, or the second $(A=2)$ particle. The
appropriate label should be added both on $r$ and $n$ in Eqs.~(\ref{eq3.15}),
(\ref{eq3.16}): $r^k \, \hat n^L \to r_A^k \, \hat n_A^L$.] The problem
is that the power of $1/r_1$ in this $\varepsilon$-singular term becomes an
{\it even integer} when $\varepsilon \to 0$. When inserting the Taylor
expansion of $f({\bm x})$, say $f({\bm x}) \sim \sum r_1^{\ell + 2n} \, \hat
n_1^L f_L^{\ell + 2n}$, some of the terms in the $\varepsilon$-singular
contribution $f({\bm x})/r_1^{(2 + 2 \varepsilon)}$ might be of the form $
r_1^{\ell + 2n'- 2 \varepsilon} \, \hat n_1^L$, with $n'=n-1 \geq0$, and might
then contribute to the Hadamard-regular part of ${\hat X}^{(V_1V_1V_2)}$ in
the limit $\varepsilon \to 0$. This would mean that the Hadamard-regular part
of ${\hat X}^{(V_1V_1V_2)}$ would not coincide with its $\varepsilon$-regular
part. We already know from Refs.~\cite{Schaefer:1986rd, Blanchet:1998vx}),
which used Hadamard regularization to derive the 2PN-accurate dynamics and
found the same result (modulo gauge effects) as the analytic-continuation
derivation of Ref.~\cite{Damour:1982wm}, that this is not the case for the
regularized values of ${\hat X}^{(V_1V_1V_2)}$ and of its first derivatives on
the first worldline. [Indeed, these quantities enter the computation of the
equations of motion.] On the other hand, the computations that we shall do
here involve higher spatial derivatives of ${\hat X}$, and it is important to
check that we can safely use Hadamard regularization to evaluate them. This
can be proven by using the techniques explained in Ref. \cite{Damour:1982wm}, based
on iteratively considering the singular terms in $\hat W_{ij}^{(VV)}$ and
${\hat X}^{(VVV)}$ generated by the singular local behaviour (near the first
worldline) of their respective source terms. One finds then that the smooth
function $f({\bm x})$ entering the dangerous terms $f({\bm x})/r_1^{(2 + 2
  \varepsilon)}$ in ${\hat X}^{(V_1V_1V_2)}$ is of the {\it special} form
$f({\bm x}) \sim \sum c_{\ell} G_L r_1^{\ell } \, n_1^L$ in dimension
$4+\varepsilon$, with $\ell \geq 1$, where $G_L \equiv \partial_L V_2$ denotes
the $\ell$-th tidal gradient (considered near the first worldline) of the $V$
potential generated by the second worldline. When working (as we do) at the
2PN accuracy, we can take $V$ at Newtonian order, and the gradients $G_L
\simeq [\partial_L (Gm_2/r_2^{(1+\varepsilon)})]_1$ are then {\it traceless}:
$G_L = G_{a_1 a_2 \cdots a_\ell}= G_{\langle a_1 a_2 \cdots a_\ell \rangle}$.
As a consequence, it is immediately seen that, in the limit $\varepsilon \to
0$, the potentially dangerous term $f({\bm x})/r_1^{(2 + 2 \varepsilon)}$ in
${\hat X}^{(V_1V_1V_2)}$ {\it does not} give any contribution to the
Hadamard-regular part of ${\hat X}$. This means that we can compute the
$\varepsilon$-regularized reduced tidal action in (\ref{eq2.12}) by replacing,
from the start, the brick potentials $V(x) , V_i (x) , \ldots$, by their
Hadamard-regularized counterparts, $R \, [V(x)], R \, [V_i (x)] , \ldots$

\bigskip

Summarizing: The $A$-worldline part of the tidal action Eq. (\ref{eq2.12}) can be
obtained by computing all its elements ($d\tau_A = c^{-1} (-g_{\mu\nu} (y_A)
\, dy_A^{\mu} \, dy_A^{\nu})^{1/2} , G_{\alpha\beta}^A , \ldots$) within the
$A$-regular metric $g_{\mu\nu}^{A-{\rm reg}} (x)$ obtained by replacing each
2PN brick potential $V(x) , V_i (x) , \ldots$ by its $A$-Hadamard-regular part
$R_A [V(x)] , R_A [V_i (x)] , \ldots$

As a check on our results (and on the many complicated algebraic operations
needed to derive them) we have also re-computed the electric-quadrupole tidal Lagrangian,
$L_{\mu_A^{(2)}}=\frac{1}{4} \,
\left(d\tau_A/dt\right) \, G_{\alpha\beta}^A \, G_A^{\alpha\beta}$ 
by effecting the Hadamard 
regularization
in a different way. 
Our alternative
computation was done by  separately Hadamard-regularizing
  each factor entering the Lagrangian,
$L_{\mu_A^{(2)}}$, when it is expressed in terms of $d\tau_A/dt$, the 
contravariant metric,
the covariant Riemann
tensor, and the contravariant 4-velocity. 
More precisely, we first 
calculated $G_{\alpha\beta}(y_A)$
as $-R_A[R_{\alpha\mu\beta\nu}]\, R_A[u^\mu] R_A[u^\nu]$, then we computed
$[G_{ab}^2](y_A) \equiv G_{\alpha\beta}(y_A) G_{\mu\nu}(y_A)
R_A[g^{\alpha\mu}] R_A[g^{\beta\nu}]$, which we inserted into the expression
of $L_{\mu_A^{(2)}}$ just written. The remaining factor,
$(d\tau_A/dt)/4$, was taken to be $R_A[d\tau/dt]/4$. Note in passing that,
while one can a priori prove that the alternative regularization of  
$G_{\alpha\beta}(y_A)$ (and
subsequently $[G_{ab}^2](y_A) \equiv G_{\alpha\beta}(y_A) G_{\mu\nu}(y_A)
R_A[g^{\alpha\mu}] R_A[g^{\beta\nu}]$) just explained, must coincide 
with the
one explained above, namely $(G_{\alpha\beta} \, G^{\alpha\beta})[R_A 
(V) , R_A (V_i) , \ldots]$
(because both of them agree with the Riesz-analytic-regularization 
and/or dimensional-regularization)
  a different result would have been obtained if one had postponed
the Hadamard regularization of the squared tidal quadrupole to the last 
moment,
i.e. if one had computed $R_A[
G_{\alpha\beta} \, G^{\alpha\beta}]$. [Such a difference occurs because of
the appearance of a dangerous nonlinear mixing of Hadamard-regular and
Hadamard-singular parts in $\partial_{ij} V \partial_{ij} {\hat
   X}^{(V_1V_1V_2)}$ (with the special structure of the delicate terms in
${\hat X}^{(V_1V_1V_2)}$ given above). 
This shows again the
consistency problems of the Hadamard regularization, when it is used
beyond the types of calculations where it is equivalent to the
Riesz analytic regularization (or to dimensional regularization).]

\subsection{Explicit rules for computing the regular parts of the 2PN brick potentials}

Let us now give some indications on the computation of the regular parts of
the various brick potentials $V(x) , V_i (x) , \ldots$ 

\subsubsection{Regularizing $V$ and $V_i$}

The situation is very
simple for the ``linear potentials'' $V$ and $V_i$, which satisfy linear
equations with delta-function sources [see Eqs.~(\ref{eq3.3})]. Near, say, the
particle $A=1$, the $A$-regular parts of $V$ and $V_i$ are the terms in
Eqs.~(\ref{eq3.7}), (\ref{eq3.8}) which are generated by the source terms
$\propto \delta ({\bm x} - {\bm y}_2)$ of the second particle. It is indeed
easily seen [from the definition in Eq. (\ref{eq3.15})] that the $1$-regular part of
all the terms explicitly written in Eq.~(\ref{eq3.7}) vanishes, while all the
non-explicitly written terms obtained by the $1 \leftrightarrow 2$ exchange
are regular near the particle $1$. The same is true for $V_i$,
Eq.~(\ref{eq3.8}). A simple rule for obtaining these results is to note that,
from the definition in Eq. (\ref{eq3.16}), any term of the form
\begin{equation}
\label{eq3.17}
r_1^{2k+1} \, f(x) \, , \quad k \in {\mathbb Z} \, ,
\end{equation}
where $f( x)$ is a smooth function of $x^\mu$ (near ${\bm x} = {\bm y}_1$
at fixed instant $t$), and where the power of $r_1$ is {\it odd}, is purely
singular.

The situation is more complicated for the higher-order potentials $\hat
W_{ij}$ and $\hat R_i$, whose sources contain both compact terms $\propto
\delta ({\bm x} - {\bm y}_A)$, and quadratically nonlinear non-compact ones
$\propto \partial V \, \partial V$, and still more complicated for the $\hat
X$ potential whose source even depends on the previous $\hat W_{ij}$
potential.

\subsubsection{Regularizing $\hat W_{ij}$}

The potential $\hat W_{ij}$ can be decomposed in powers of the masses. It
contains terms proportional to $m_1 , m_2 , m_1^2 , m_2^2$ and $m_1 \, m_2$.
It is easily seen that while the terms proportional to $m_1$ and $m_1^2$ are
$1$-singular, the terms proportional to $m_2$ and $m_2^2$ are $1$-regular. It
is more delicate to decompose the mixed terms $\propto m_1 \, m_2$ into
$1$-regular $(R_1)$ and $1$-singular $(S_1)$ parts. More precisely the $m_1 \,
m_2$ part of $\hat W_{ij}$ has the form
\begin{equation}
\label{eq3.18}
\hat W_{ij}^{[m_1m_2]} = \hat W_{ij(0)}^{[m_1m_2]} + \tilde{\hat W}_{ij(0)}^{[m_1m_2]}
\end{equation}
where
\begin{align}
\label{eq3.19}
{\hat W}_{ij}{}_{(0)}^{[m_1m_2]}&= \frac{1}{r_{12}S}\delta^{ij}
+\left\{ \frac{1}{S^2} \left(n_1^{(i} n_2^{j)} 
+ 2n_1^{(i} n_{12}^{j)} \right)\right.\nonumber\\
&- \left.
n_{12}^i n_{12}^j \left(\frac{1}{S^2}+\frac{1}{r_{12} S} \right) \right\}\nonumber\\
&\equiv  \frac{1}{r_{12} S}P(n_{12})^{ij} \nonumber \\
&+ \frac{1}{S^2} \left(n_1^{(i} n_2^{j)} 
+ 2n_1^{(i} n_{12}^{j)} -n_{12}^i n_{12}^j\right) \, ,
\end{align}
\begin{align}
\label{eq3.20}
\tilde {\hat W}_{ij}{}_{(0)}^{[m_1m_2]}&= \frac{1}{r_{12}S}\delta^{ij}
+\left\{ \frac{1}{S^2} \left(n_2^{(i} n_1^{j)} 
- 2n_2^{(i} n_{12}^{j)} \right) \right.\nonumber\\
&-\left.
n_{12}^i n_{12}^j \left(\frac{1}{S^2}+\frac{1}{r_{12} S} \right) \right\}\nonumber\\
&\equiv  \frac{1}{r_{12}S} P(n_{12})^{ij} \nonumber \\
&+ \frac{1}{S^2} \left(n_2^{(i} n_1^{j)} 
- 2n_2^{(i} n_{12}^{j)}-n_{12}^i n_{12}^j  \right) \, , 
\end{align}
and where $P(n_{12})^{ij} \equiv \delta^{ij} - n_{12}^i \, n_{12}^j$ denotes
the projector orthogonal to the unit vector ${\bm n}_{12}$. [The decomposition in Eq. 
(\ref{eq3.18}) simply corresponds to the decomposition of Eq.~(\ref{eq3.10})
into an explicitly written term and its $1 \leftrightarrow 2$ counterpart.]
Here we see that there appear (modulo $x$-independent factors, such as
$r_{12}^{-1}$, $n_{12}^i$, $P(n_{12})^{ij} , \ldots$) terms of the type
\begin{equation}
\label{eq3.21}
\frac1S \, , \quad \frac{1}{S^2} \, , \quad \frac{n_1^i}{S^2} \, , 
\quad \frac{n_2^i}{S^2} \, , \quad \frac{n_1^i \, n_2^j}{S^2} \, ,
\end{equation}
where we recall that $S \equiv r_1 + r_2 + r_{12}$. Near particle $1$, $n_2^i$
is a smooth function, while $n_1^i = r_1^i / r_1$ is the ratio of a smooth
function $(r_1^i = x^i - y_1^i)$ by $r_1$. In other words, the five terms
listed in Eq.~(\ref{eq3.21}) are of {\it three} different types:
\begin{equation}
\label{eq3.22}
\frac1S \, , \quad \frac{f(x)}{S^2}  \qquad \mbox{and} \qquad  
\frac{f(x)}{r_1 \, S^2} \, , 
\end{equation}
where $f(x)$ denotes a generic smooth function near particle $1$. [As we
always consider the neighborhood of particle $1$, we do not add an index to
$f(x)$ to recall that it is $1$-regular, but might be singular near particle
$2$.] Because $S = r_1 + r_2 + r_{12}$ is a function of ``mixed character''
(partly regular and partly singular), it is not immediate to decompose the
functions in Eq. (\ref{eq3.22}) into $1$-regular and $1$-singular parts. [This mixed
character of $S$ is deeply linked with the fact that it enters the 2PN metric
because of the basic fact that a solution of $\Delta g = r_1^{-1} \, r_2^{-1}$
is $g = \ln \, S$.] A simple (though somewhat brute-force) way of extracting
the regular parts of the functions in Eq. (\ref{eq3.22}) consists of decomposing $S$
into
\begin{equation}
\label{eq3.23}
S \equiv S_0 + r_1 = S_0 \left( 1 + \frac{r_1}{S_0} \right) \, ,
\end{equation}
with
\begin{equation}
\label{eq3.24}
S_0 \equiv r_2 + r_{12} \, ,
\end{equation}
(note that $S_0$ is a smooth function near particle $1$), and then expanding
$S^{-n}$ in powers of $r_1 / S_0$. Namely
\begin{equation}
\label{eq3.25}
\frac1S = \frac{1}{S_0} \left( 1 - \frac{r_1}{S_0} + \frac{r_1^2}{S_0^2} 
- \frac{r_1^3}{S_0^3} + \ldots \right) \, ,
\end{equation}
\begin{equation}
\label{eq3.26}
\frac{1}{S^2} = \frac{1}{S_0^2} \left( 1 - 2 \, \frac{r_1}{S_0} 
+ 3 \, \frac{r_1^2}{S_0^2} - 4 \, \frac{r_1^3}{S_0^3} + \ldots \right) \, ,
\end{equation}
and more generally 
\begin{align}
\label{eq3.27}
\frac{1}{S^n} &= \frac{1}{S_0^n}\left(1-n\frac{r_1}{S_0}
+\frac{(n+1)n}{2}\left(\frac{r_1}{S_0}\right)^2\right.\nonumber\\
& \qquad \quad - \left. \frac{(n+2)(n+1)n}{3!}
\left(\frac{r_1}{S_0}\right)^3 \right. \nonumber \\
&  \qquad \quad 
+ \left.\frac{(n+3)(n+2)(n+1)n}{4!}\left(\frac{r_1}{S_0}\right)^4 
+\ldots\right)\,,\,\nonumber\\
&  n=1,2,\ldots
\end{align}
Using these expansions, together with the rule that terms of the form in Eq. 
(\ref{eq3.17}) are purely singular, it is easy to derive the following results
for the $1$-regular parts of functions of the type in Eq. (\ref{eq3.22}), and, more
generally, of the types $f(x)/S$, $f(x)/S^2$, $f(x)/(r_1 \, S)$ and $f(x) /
(r_1 \, S^2)$:
\begin{align}
\label{eq3.28}
\left(\frac{f(x)}{S}\right)_R&= \frac{f(x)}{S_0}
\left(1+\left(\frac{r_1}{S_0}\right)^2+\left(\frac{r_1}{S_0}\right)^4
\right.\nonumber\\
& \qquad \quad~ + \left. \left(\frac{r_1}{S_0}\right)^6+\ldots\right)\nonumber\\
&\equiv  f(x) \left(\frac1S\right)_R \, ,
\end{align}
\begin{align}
\label{eq3.29}
\left(\frac{f(x)}{S^2}\right)_R&= \frac{f(x)}{S_0^2}
\left(1+3\left(\frac{r_1}{S_0}\right)^2
+5\left(\frac{r_1}{S_0}\right)^4\right.\nonumber\\
& \qquad \quad ~+\left.
7\left(\frac{r_1}{S_0}\right)^6+\ldots\right)\nonumber\\
&\equiv  f(x) \left(\frac1{S^2}\right)_R \, ,
\end{align}
\begin{align}
\label{eq3.30}
\left(\frac{f(x)}{r_1S} \right)_R &= -\frac{f(x)}{S_0^2}
\left(1+\left(\frac{r_1}{S_0}\right)^2+\left(\frac{r_1}{S_0}\right)^4 
+\ldots \right) \nonumber\\
&\equiv  f(x) \left(\frac1{r_1S}\right)_R \, ,
\end{align}
\begin{align}
\label{eq3.31}
\left(\frac{f(x)}{r_1S^2}\right)_R   &= -\frac{f(x)}{S_0^3}
\left(2+4\left(\frac{r_1}{S_0}\right)^2+6\left(\frac{r_1}{S_0}\right)^4
\right.\nonumber\\
& \qquad \qquad~ +\left.
8\left(\frac{r_1}{S_0}\right)^6+\ldots  \right)\,\nonumber\\
&\equiv  f(x) \left(\frac1{r_1 S^2}\right)_R \, .
\end{align}
Here, we use a lower $R$ subscript $(\varphi (x))_R$ to denote the $1$-regular
part of a function $\varphi (x)$ (above denoted as $R \, [\varphi (x)]$). (We
omit decorating $R$ with a label $1$, but one should remember that we are always
talking about the $1$-regular part of $\varphi (x)$.)

\smallskip

Note that, as indicated, all the terms above have the simple property that the
regular-projection operator $R$ {\it commutes} with the multiplication by a
smooth function, e.g. $R \, [f(x) \, S^{-1}] = f(x) \, R \, [S^{-1}]$. Beware
that this property is true only for the special singular terms considered
here. We shall later see that more-complicated singular terms (entering the
$\hat X$ potential) do not satisfy this simple commutativity property.

\smallskip

Note that the number of terms one needs to retain in the above expansions
depends on the quantity one wants to evaluate on the first worldline. For
instance, when evaluating $G_{\alpha\beta}^1$, which involves the curvature
tensor, and therefore two spatial derivatives of the metric (and, in
particular, of $R \, [\hat W_{ij}]$), we need to include enough terms to ensure
that $R \, [\hat W_{ij}]$ is $C^2$ near ${\bm x} = {\bm y}_1$. Actually, we
shall push our calculations up to the level of $G_{\alpha\beta\gamma}^1$,
which depends on the first covariant derivative of the curvature tensor, and
we shall therefore need all the brick potentials to be at least $C^3$ near
${\bm x} = {\bm y}_1$.

\smallskip

The application of the above results yields the following explicit expressions
for the $1$-regular part of the two separate $O(m_1 m_2)$ delicate
contributions to $\hat W_{ij}$ [defined in
Eqs.~(\ref{eq3.18})--(\ref{eq3.20})]:
\begin{eqnarray}
\label{eq3.32}
[{\hat W}_{ij}{}_{(0)}^{[m_1m_2]}]_R&=& \frac{P(n_{12})^{ij}}{r_{12}}
\frac{1}{S_0}\left(1+\frac{r_1^2}{S_0^2}+\frac{r_1^4}{S_0^4}
+\ldots\right)\nonumber\\{}
&+&
\frac{{\mathbf r}_1^{(i} n_2^{j)}}{S_0^2} \left(-\frac{2}{S_0}-4\frac{r_1^2}{S_0^3}
-6\frac{r_1^4}{S_0^5}+\ldots  \right)\nonumber\\
&+&2\frac{{\mathbf r}_1^{(i} n_{12}^{j)}}{S_0^2} \left(-\frac{2}{S_0}
-4\frac{r_1^2}{S_0^3}-6\frac{r_1^4}{S_0^5}+\ldots  \right)\nonumber\\
&-&n_{12}^i n_{12}^j\frac{1}{S_0^2}\left(1+3\frac{r_1^2}{S_0^2}
+5\frac{r_1^4}{S_0^4} +\ldots\right) \, , \nonumber\\
\end{eqnarray}
\begin{eqnarray}
\label{eq3.33}
[\tilde {\hat W}_{ij}{}_{(0)}^{[m_1m_2]}]_R&=& \frac{P(n_{12})^{ij}}{r_{12}}
\frac{1}{S_0} \left(1+\frac{r_1^2}{S_0^2}+\frac{r_1^4}{S_0^4}
+\ldots \right)\nonumber\\
&+& \frac{n_2^{(i} {\mathbf r}_1^{j)}}{S_0^2} \left(   -\frac{2}{S_0}
-4\frac{r_1^2}{S_0^3}-6\frac{r_1^4}{S_0^5}+\ldots  \right)\nonumber\\
&-& 2n_2^{(i} n_{12}^{j)}\frac{1}{S_0^2}\left(1+3\frac{r_1^2}{S_0^2}
+5\frac{r_1^4}{S_0^4} +\ldots\right) \nonumber\\
&-&n_{12}^i n_{12}^j\frac{1}{S_0^2}\left(1+3\frac{r_1^2}{S_0^2}
+5\frac{r_1^4}{S_0^4} +\ldots\right) \, . \nonumber\\
\end{eqnarray}

\subsubsection{Regularizing $\hat R_{i}$}

As the potential $\hat R_i$ has a source of the same type as $\hat W_{ij}$
(namely $\delta ({\bm x} - {\bm y}_A)$ terms plus a non-compact term quadratic
in the $V$ potentials), the calculation of its regular part can be done in
exactly the same way as $\hat W_{ij}$. $\hat R_i$ contains terms $\propto
m_1^2 , m_2^2$ and $m_1 \, m_2$. The $O(m_1^2)$ piece is purely singular, the
$O(m_2^2)$ one is purely regular, while the $O(m_1 \, m_2)$ one is a mix of
regular and singular terms. As above, we can decompose the $m_1 \, m_2$ part
of $\hat R_i$ in two pieces, say
\begin{equation}
\label{eq3.34}
\hat R_i^{[m_1m_2]} = \hat R_{i(0)}^{[m_1 m_2]} + \tilde{\hat R}_{i(0)}^{[m_1 m_2]} \, ,
\end{equation}
where
\begin{align}
\label{eq3.35}
{\hat R}_{i}{}_{(0)}^{[m_1m_2]}&=  n_{12}^i \left\{ -\frac{(n_{12}v_1)}{2S} 
\left(\frac{1}{S}+\frac{1}{r_{12}} \right) \right.\nonumber\\
& \qquad \quad -\left.
\frac{2 (n_2v_1)}{S^2}+\frac{3(n_2v_2)}{2S^2} \right\}\nonumber\\
&+n_1^i\frac{1}{S^2} \left( 2 (n_{12}v_1) - \frac{3(n_{12}v_2)}{2}
\right.\nonumber\\
& \qquad \quad ~+ \left. 2 (n_2v_1) - \frac{3(n_2v_2)}{2} \right)\nonumber\\
&+v_1^i \left( \frac{1}{r_1 r_{12}} + \frac{1}{2 r_{12}S} \right)
-v_2^i\frac{1}{r_1 r_{12}} \, ,
\end{align}
\begin{align}
\label{eq3.36}
\tilde {\hat R}_{i}{}_{(0)}^{[m_1m_2]}&=  -n_{12}^i 
\left\{ \frac{(n_{12}v_2)}{2S} \left(\frac{1}{S}+
\frac{1}{r_{12}} \right)\right.\nonumber\\
& \qquad \quad - \left. 
\frac{2 (n_1v_2)}{S^2}+\frac{3(n_1v_1)}{2S^2} \right\}\nonumber\\
&+n_2^i\frac{1}{S^2} \left( -2 (n_{12}v_2) 
+ \frac{3(n_{12}v_1)}{2}\right.\nonumber\\
& \qquad \quad~ + \left. 2 (n_1v_2) - \frac{3(n_1v_1)}{2} \right)\nonumber\\
&+v_2^i \left( \frac{1}{r_2 r_{12}} + \frac{1}{2 r_{12}S} \right)
-v_1^i\frac{1}{r_2 r_{12}}  \,.
\end{align}

Applying the above results then yields the following expressions for the 
$1$-regular parts of these quantities:
\begin{align}
\label{eq3.37}
[{\hat R}_{i}{}_{(0)}^{[m_1m_2]}]_R&=  n_{12}^i \left\{
  \left[-\frac{(n_{12}v_1)}{2} 
- 2 (n_2v_1) \right. \right. \nonumber\\ & \left. \qquad 
+\frac{3(n_2v_2)}{2}\right] 
\left( \frac{1}{S^2} \right)_R
\left.  -\frac{(n_{12}v_1)}{2r_{12}}\left(\frac{1}{S}\right)_R
\right\}\nonumber\\ 
&+\left( 2 (n_{12}v_1) - \frac{3(n_{12}v_2)}{2}+2 (n_2v_1) \right.\nonumber\\
& \left. -\frac{3(n_2v_2)}{2} \right){\mathbf r}_1^i
\left(\frac{1}{r_1S^2}\right)_R \nonumber\\
&+ \frac{v_1^i}{2 r_{12}} \left(\frac{1}{S}\right)_R \, ,
\end{align}
\begin{align}
\label{eq3.38}
[\tilde {\hat R}_{i}{}_{(0)}^{[m_1m_2]}]_R&=  -n_{12}^i 
\left\{ \frac{(n_{12}v_2)}{2} \left(\left(\frac{1}{S^2}\right)_R \right.
\right. \nonumber \\ & \left. \qquad \quad
+\frac{1}{r_{12}} \left(\frac{1}{S}\right)_R\right) 
-2 ({\mathbf r}_1v_2)\left(\frac{1}{r_1S^2}\right)_R   \nonumber \\
&\left. \qquad \quad +\frac{3}{2}({\mathbf r}_1v_1)
\left(\frac{1}{r_1S^2}\right)_R \right\}\nonumber\\
&+n_2^i \left[ \left(-2 (n_{12}v_2) + 
\frac{3(n_{12}v_1)}{2}\right)\left(\frac{1}{S^2}\right)_R \right.
\nonumber\\
& \left. \qquad
+\left(2 ({\mathbf r}_1v_2) - \frac{3({\mathbf r}_1v_1)}{2}\right)
\left(\frac{1}{r_1S^2}\right)_R  \right]\nonumber\\
&+v_2^i \left( \frac{1}{r_2 r_{12}} + \frac{1}{2 r_{12}}
\left(\frac{1}{S}\right)_R \right)-v_1^i\frac{1}{r_2 r_{12}}  \, .
\end{align}
One should substitute the expansions in Eqs. (\ref{eq3.28})--(\ref{eq3.31}) into these
results to get their explicit forms.

\subsubsection{Regularizing $\hat X$}

Finally, we come to the most complicated 2PN brick potential, namely $\hat X$.
It contains contributions proportional to $m_1^2 , m_1 \, m_2 , m_2^2$; $m_1^3
, m_1^2 \, m_2 , m_1 \, m_2^2$ and $m_2^3$ (see Eq.~(\ref{eq3.11})). The terms
in $m_1^2 , m_2^2 , m_1^3 , m_2^3$ are easily dealt with (they are either
purely singular or purely regular). Many, but not all, of the $m_1 \, m_2$
terms can be dealt with in the same way as the $m_1 \, m_2$ terms in $\hat
W_{ij}$ and $\hat R_i$. If we again decompose $\hat X^{[m_1 m_2]}$ in two
pieces
\begin{equation}
\label{eq3.39}
\hat X^{[m_1 m_2]} = \hat X_{(0)}^{[m_1 m_2]} + \tilde{\hat X}_{(0)}^{[m_1 m_2]} \, ,
\end{equation}
we have the following results for their regular parts:
\begin{widetext}
\begin{eqnarray}
\label{eq3.40}
[{\hat X}_{(0)}^{[m_1 m_2]}]_R&=& 
v_1^2 \left[ \left(\frac{1}{r_1S}\right)_R + \frac{1}{r_{12}}
\left(\frac{1}{S}\right)_R \right]
+
v_2^2 \left[ \left(\frac{1}{r_1S}\right)_R + \frac{1}{r_{12}}
\left(\frac{1}{S}\right)_R \right]
-(v_1v_2) \left(
2\left(\frac{1}{r_1S}\right)_R+
\frac{3}{2 r_{12}} \left(\frac1S\right)_R\right)
\nonumber\\  &&
-(n_{12}v_1)^2 \left(\left(\frac{1}{S^2}\right)_R+\frac{1}{r_{12}}
\left(\frac1S\right)_R\right)  
-(n_{12}v_2)^2 \left(\left(\frac{1}{S^2}\right)_R+\frac{1}{r_{12}}
\left(\frac1S\right)_R\right)
\nonumber\\ &&
+\frac{3(n_{12}v_1)(n_{12}v_2)}{2} \left(\left(\frac{1}{S^2}\right)_R
+\frac{1}{r_{12}}\left(\frac1S\right)_R\right) 
+2(n_{12}v_1)({\mathbf r}_1v_1)\left(\frac{1}{r_1S^2}\right)_R
\nonumber \\ 
& & 
-5(n_{12}v_2)({\mathbf r}_1v_1)\left(\frac{1}{r_1S^2}\right)_R 
-({\mathbf r}_1v_1)^2 \left(\frac{1}{r_1^2S^2} + \frac{1}{r_1^3 S} \right)_R
+2(n_{12}v_2)({\mathbf r}_1v_2)\left( \frac{1}{r_1S^2} \right)_R 
\nonumber \\ & & 
+2({\mathbf r}_1v_1)({\mathbf r}_1v_2) \left(\frac{1}{r_1^2S^2} 
+ \frac{1}{r_1^3S} \right)_R
-({\mathbf r}_1v_2)^2 \left(\frac{1}{r_1^2S^2} + \frac{1}{r_1^3S}\right)_R
-2(n_{12}v_2)(n_2v_1)\left(\frac{1}{S^2}\right)_R \nonumber \\ 
& &  +2({\mathbf r}_1v_2)(n_2v_1)\left(\frac{1}{r_1S^2}\right)_R
-\frac{3}{2}({\mathbf r}_1v_1)(n_2v_2)\left(\frac{1}{ r_1S^2}\right)_R \, ,
\end{eqnarray}
\begin{eqnarray}
\label{eq3.41}
[\tilde {\hat X}_{(0)}^{[m_1 m_2]}]_R&=& 
v_2^2 \left(\frac{1}{r_2 r_{12}} 
+ \frac{1}{r_2 }\left(\frac{1}{S}\right)_R 
+ \frac{1}{r_{12} }\left(\frac{1}{S}\right)_R \right)
+
v_1^2 \left(-\frac{1}{r_2 r_{12}}+\frac{1}{r_2 }\left(\frac{1}{S}\right)_R 
+ \frac{1}{r_{12} }\left(\frac{1}{S}\right)_R \right)\nonumber\\
&& 
-(v_1v_2) \left(
\frac{2}{r_2}+
\frac{3}{2 r_{12}} \right)\left(\frac{1}{S}\right)_R
-(n_{12}v_2)^2 \left(\left(\frac{1}{S^2}\right)_R
+\frac{1}{r_{12}}\left(\frac1S\right)_R\right) 
\nonumber \\ 
& & 
-(n_{12}v_1)^2 \left(\left(\frac{1}{S^2}\right)_R
+\frac{1}{r_{12}}\left(\frac1S\right)_R\right)
+\frac{3(n_{12}v_2)(n_{12}v_1)}{2} \left(\left(\frac{1}{S^2}\right)_R
+\frac{1}{r_{12}}\left(\frac1S\right)_R\right)\nonumber\\
&&
-2(n_{12}v_2)(n_2v_2)\left(\frac{1}{S^2}\right)_R
+5(n_{12}v_1)(n_2v_2)\left(\frac{1}{S^2}\right)_R\nonumber\\
&&
-(n_2v_2)^2 \left(\left(\frac{1}{S^2}\right)_R 
+ \frac{1}{r_2}\left(\frac1S\right)_R \right)
-2(n_{12}v_1)(n_2v_1)\left(\frac{1}{S^2} \right)_R\nonumber \\ 
& & 
+2(n_2v_2)(n_2v_1) \left(\left(\frac{1}{S^2}\right)_R 
+ \frac{1}{r_2}\left(\frac1S\right)_R \right)
- (n_2v_1)^2  \left(\left(\frac{1}{S^2}\right)_R 
+ \frac{1}{r_2}\left(\frac1S\right)_R\right)\nonumber\\
&&
+2(n_{12}v_1)({\mathbf r}_1v_2)\left(\frac{1}{r_1S^2}\right)_R 
+2(n_2v_1)({\mathbf r}_1v_2)\left(\frac{1}{r_1S^2}\right)_R 
-\frac32 (n_2v_2)({\mathbf r}_1v_1)\left(\frac{1}{r_1S^2}\right)_R\,.
\end{eqnarray}

\end{widetext}

One should substitute the expansions in Eqs. (\ref{eq3.28})--(\ref{eq3.31}) into the
corresponding terms in Eqs.~(\ref{eq3.40}), (\ref{eq3.41}). However, these
equations involve new types of terms, not discussed above. These new terms are
of the form
\begin{equation}
\label{eq3.42}
f(x) \left( \frac{1}{r_1^2 \, S^2} + \frac{1}{r_1^3 \, S} \right) \, .
\end{equation} 
The fact that we have a specific combination of $r_1^{-2} \, S^{-2}$ and
$r_1^{-3} \, S^{-1}$ simplifies things. Indeed, using the expansions in Eqs. 
(\ref{eq3.25}), (\ref{eq3.26}) above we have
\begin{equation}
\label{eq3.43}
\frac{f(x)}{r_1^2 \, S^2} = \frac{f(x)}{S_0^4} 
\left[ \frac{S_0^2}{r_1^2} - 2 \, \frac{S_0}{r_1} + 3 
- 4 \, \frac{r_1}{S_0} + 5 \, \frac{r_1^2}{S_0^2} + \ldots \right] \, ,
\end{equation} 
\begin{equation}
\label{eq3.44}
\frac{f(x)}{r_1^3 \, S} = \frac{f(x)}{S_0^4} \left[ \frac{S_0^3}{r_1^3} 
- \frac{S_0^2}{r_1^2} + \frac{S_0}{r_1} -1 + \frac{r_1}{S_0} 
- \frac{r_1^2}{S_0^2} + \ldots \right] \, .
\end{equation} 
When summing these two equations we see that the terms $\propto 1/r_1^2$
cancel. We shall deal later with these terms, which turn out to be delicate to
handle but, anyway, in the sum of Eqs. (\ref{eq3.43}) and (\ref{eq3.44}), they cancel
out. The remaining terms  contain either an odd power of $r_1$ [and are therefore
{\it purely} singular, Eq.~(\ref{eq3.17})] or a {\it positive}, {\it even}
power of $r_1$ (which makes them purely regular). As a consequence, the
regular part of the combination of Eq. (\ref{eq3.42}) reads
\begin{align}
\label{eq3.45}
\left[f(x)\left(\frac{1}{r_1^2S^2}+\frac{1}{r_1^3S}\right)\right]_R 
&= \frac{f(x)}{S_0^4}\left( 2+4\left(\frac{r_1}{S_0}\right)^2
\right. \nonumber\\
& \left. +6\left(\frac{r_1}{S_0}\right)^4+8\left(\frac{r_1}{S_0}\right)^6
+\ldots\right)\nonumber\\
&\equiv  f(x) \left(\frac{1}{r_1^2S^2}+\frac{1}{r_1^3S}\right)_R \, .
\end{align}
Note that, thanks to the cancellation of the $1/r_1^2$ terms, we have again a
property of commutativity $R \, [f(x) \, \varphi (x)] = f(x) \, R \, [\varphi
(x)]$, for the special type of terms $\varphi (x)$ entering
Eq.~(\ref{eq3.42}).

\smallskip

Concerning the $m_1 \, m_2^2$ contribution to $\hat X$, it is the sum of
\begin{equation}
\label{eq3.46}
{\hat X}_{(0)}^{[m_1 m_2^2]} =  -\frac{1}{2 r_{12}^3} 
+ \frac{r_2}{2 r_1 r_{12}^3} -\frac{1}{2 r_1 r_{12}^2}
\end{equation}
and
\begin{eqnarray}
\label{eq3.47}
\tilde {\hat X}_{(0)}^{[m_1m_2^2]}&=& \frac{1}{2 r_2^3} + \frac{1}{16 r_1^3} + 
\frac{1}{16 r_2^2 r_1} - \frac{r_1^2}{2 r_2^2 r_{12}^3} \nonumber\\
&& \left.
+ \frac{r_1^3}{2 r_2^3 r_{12}^3} - \frac{r_2^2}{32 r_1^3 r_{12}^2} 
- \frac{3}{16 r_1 r_{12}^2} + \frac{15 r_1}{32 r_2^2 r_{12}^2} \right. \nonumber \\ 
&&  - \frac{r_1^2}{2 r_2^3 r_{12}^2} - \frac{r_1}{2 r_2^3 r_{12}} 
- \frac{r_{12}^2}{32 r_2^2 r_1^3} \,.
\end{eqnarray}
Using the rule of Eq. (\ref{eq3.17}), we easily see that each term clearly is  either
purely regular or purely singular. Computing the regular part of $\hat X^{[m_1
  m_2]}$ is then easy.

\smallskip

The most delicate contribution to $\hat X$ is its $O(m_1^2 \, m_2)$ one, which
can again be written as the sum of
\begin{align}
\label{eq3.48}
{\hat X}_{(0)}^{[m_1^2m_2]}=& \frac{1}{2 r_1^3} 
+ \frac{1}{16 r_2^3} + \frac{1}{16 r_1^2 r_2} 
- \frac{r_2^2}{2 r_1^2 r_{12}^3}\nonumber\\
& \left. + \frac{r_2^3}{2 r_1^3 r_{12}^3} 
- \frac{r_1^2}{32 r_2^3 r_{12}^2} - \frac{3}{16 r_2 r_{12}^2} + 
\frac{15 r_2}{32 r_1^2 r_{12}^2} \right. \nonumber \\ 
&  - \frac{r_2^2}{2 r_1^3 r_{12}^2} 
- \frac{r_2}{2 r_1^3 r_{12}} - \frac{r_{12}^2}{32 r_1^2 r_2^3}
\end{align}
and
\begin{equation}
\label{eq3.49}
\tilde {\hat X}_{(0)}^{[m_1^2m_2]}= -\frac{1}{2 r_{12}^3} 
+ \frac{r_1}{2 r_2 r_{12}^3} 
-\frac{1}{2 r_2 r_{12}^2} \, .
\end{equation}
Actually the part $\tilde {\hat X}_{(0)}^{[m_1^2m_2]}$ is easy to discuss: Its
regular part is obtained simply by discarding the term: $r_1 / (2 \, r_2 \,
r_{12}^3)$. Similarly, most of the terms in $\hat X_{(0)}^{[m_1^2 m_2]}$ are
easy to treat, being either purely regular or purely singular because of
Eq.~(\ref{eq3.17}). However, the third, fourth, eighth and last terms in the
right hand side of Eq.~(\ref{eq3.48}) are somewhat tricky. [These terms
correspond to the ``dangerous terms'' in ${\hat X}^{(V_1V_1V_2)}$ that were
discussed in Sec. III when making the link between the
$\varepsilon$-regularization and the Hadamard one.] The third term is
\begin{equation}
\label{eq3.50}
Q \equiv \frac{1}{16 \, r_1^2 \, r_2} \, ,
\end{equation}
while the sum of the fourth, eighth and last terms reads
\begin{equation}
\label{eq3.51}
P \equiv - \frac{r_2^2}{2 \, r_1^2 \, r_{12}^3} + \frac{15 \, r_2}{32 \, 
r_1^2 \, r_{12}^2} - \frac{r_{12}^2}{32 \, r_1^2 \, r_2^3} \, .
\end{equation}
Both $Q$ and $P$ are of the form $f(x) / r_1^2$ (but we shall see that $Q$ is
special compared with $P$). The computation of the regular part of $f(x) /
r_1^2$ is a bit subtle. It can, however, be done by brute force, namely by
replacing the smooth function $f(x)$ by its Taylor expansion around ${\bm
  y}_1$:
\begin{eqnarray}
\label{eq3.52}
f(x) &=& f({\bm y}_1) + r_1^i \, \partial_i \, f({\bm y}_1) 
+ \frac{1}{2!} \, r_1^i \, r_1^j \, \partial_{ij} \, f({\bm y}_1) \nonumber\\
&+& \frac{1}{3!} \, r_1^i \, r_1^j \, \, r_1^k \, 
\partial_{ijk} \, f({\bm y}_1) + \ldots
\end{eqnarray}
When replacing $r_1^i \to r_1 \, n_1^i$ and dividing by $r_1^2$, one sees that
the regular part of $f(x)/r_1^2$ will only come from the terms $r_1^L \equiv
r_1^{i_1 i_2 \ldots i_{\ell}}$ with $\ell = 2,4,6,\ldots$ Moreover, by
decomposing $r_1^L = r_1^{\ell} \, n_1^L$ in irreducible tensorial parts, as
in
\begin{equation}
\label{eq3.53}
r_1^{ij} = r_1^2 \, n_1^{ij} = r_1^2 \left[n_1^{\langle ij \rangle} 
+ \frac13 \, \delta^{ij} \right] \, ,
\end{equation}
where $n_1^{\langle ij \rangle} \equiv \hat n_1^{ij} \equiv n_1^{ij} - \frac13
\, \delta^{ij}$ denotes the symmetric trace-free projection of $n_1^{ij}
\equiv n_1^i \, n_1^j$, we see [in view of the definition in Eq. (\ref{eq3.15})] that
only the pieces containing at least one Kroneker $\delta$ in the decomposition
of $n_1^L$ will contribute to the regular part. For instance, in the case
$\ell = 2$, only the $\delta^{ij}$ in Eq. (\ref{eq3.52}) will contribute to the
regular part of $f(x)/r_1^2$. More generally, we have that $R[r_1^L/r_1^2] =
(r_1^L - \hat{r}_1^L)/r_1^2$.

\smallskip

Applying this method yields the following result (here written with the
simplified notation used around Eq. (\ref{eq3.13})) for the regular part of
$f(x) / r_1^2$:
\begin{eqnarray}
\label{eq3.54}
\left(\frac{f(x)}{r^2} \right)_R&=& \frac16 \Delta f(0)+\frac{1}{10}x^i 
\partial_i \Delta f(0)+\frac{1}{28}\hat x^{ij}\partial_{ij}\Delta f(0) \nonumber\\
&& +\frac{1}{120}r^2 \Delta^2 f(0)+ \frac{1}{108} \hat x^{ijk}\partial_{ijk}
\Delta f(0) \nonumber\\
&& +\frac{1}{280}r^2 x^i \partial_i \Delta^2 f(0)+O(x^4)\,.
\end{eqnarray}

As one sees in Eq.~(\ref{eq3.54}) (and as can be proven to all orders), all
the terms on the right-hand side of Eq. (\ref{eq3.54}) are derivatives of the
Laplacian of $f(x)$ (taken at ${\bm x}={\bm y}_1$). As a consequence, in the
particular case where $\Delta f(x) = 0$, the regular part of $f(x)/r_1^2$ is
exactly zero. This is the case for the term $Q$ in $\hat X^{[m_1^2 m_2]}$,
Eq.~(\ref{eq3.50}). [Let us point out in passing that the discussion in
Sec. IIIC  of the link between the $\varepsilon$-regularization and the
Hadamard one essentially consisted in remarking that all the ``dangerous''
terms in $\hat X^{[m_1^2 m_2]}$ had this innocuous structure $f(x)/r_1^{(2 +
  2\varepsilon)}$ with $\Delta f(x) = 0$.] Therefore, we have simply
\begin{equation}
\label{eq3.55}
Q_R = 0 \, .
\end{equation}

On the other hand, this is not the case for the term $P$, Eq.~(\ref{eq3.51}).
The evaluation of the regular part of $P$ needs to appeal to the result
in Eq. (\ref{eq3.54}) and yields (modulo terms of order $O({\bm r}_1^4)$ that will
not be needed in our calculations)
\begin{eqnarray}
\label{eq3.56}
P_R &=&-\frac{1}{2 r_{12}^3} \left(\frac{r_2^2}{r_1^2 }\right)_R
+\frac{15}{32r_{12}^2}
\left(\frac{ r_2}{r_1^2 }\right)_R 
- \frac{r_{12}^2}{32 }\left(\frac{1}{r_1^2 r_2^3}\right)_R\nonumber\\
&=& -\frac{3}{8r_{12}^3}+\frac{1}{r_{12}^5}\left[ \frac{3}{224}r_1^2 
-\frac{15}{112}({\mathbf r}_1 n_{12})^2 \right]\nonumber\\
& +& \frac{5}{12 r_{12}^6}({\mathbf r}_1 n_{12})
\left[ ({\mathbf r}_1  n_{12})^2 -\frac{3}{8}r_1^2 \right] \, .
\end{eqnarray}

\bigskip

Summarizing: We have explicitly displayed all the rules needed to compute
(near particle 1) the {\it regular} parts of the various brick potentials
$V,V_i,\hat W_{ij} , \hat R_i , \hat X$ entering the 2PN metric. By replacing
$V \to V_R , \ldots , \hat X \to \hat X_R$, in Eq.~(\ref{eq3.2}), we define a
{\it regularized} version of the 2PN metric generated by two point masses,
$g_{\mu\nu}^R(x) \equiv g_{\mu\nu} [V_R (x) , \ldots , \hat X_R (x)]$, which
is smooth near particle $1$.

\section{Computation of the invariants entering the tidal action}\label{sec4}
\setcounter{equation}{0}

As we explained above, when neglecting terms quadratic in the tidal parameters
$\mu^{(\ell)}$, etc., the tidal part of the two-body action is simply obtained
by evaluating the $S_{\rm nonminimal}$, Eq.~(\ref{eq2.12}), as a function of
the worldlines, by replacing the metric $g_{\mu\nu} (x)$ entering the
right-hand side of Eq. (\ref{eq2.12}) by the (regular part of the) point-mass
metric $g_{\mu\nu}^{\rm point \, mass} (x,y_1,y_2,m_1,m_2)$. This reduced
action is a sum over the various tidal parameters, $\mu_A^{(\ell)}$,
$\sigma_A^{(\ell)}$, $\mu'^{(\ell)}_A , \ldots$. We can therefore compute
separately the part of the reduced action associated with each of them. This
is what we shall do in this section for the actions associated with the
parameters $\mu_{A=1}^{(\ell = 2)}$, $\mu_{A=1}^{(\ell = 3)}$,
$\sigma_{A=1}^{(\ell = 2)}$ and $\mu'^{(\ell = 2)}_{A=1}$. [We shall only
explicitly write down the results for $A=1$ but, evidently, they also yield the
results for $A=2$ by exchanging $1 \leftrightarrow 2$.]

\smallskip

First, let us note that each action, say, associated with the parameter $\mu_1$
related to the first worldline, is of the form
\begin{equation}
\label{eq4.1}
\mu_1 \int dt \, L_{\mu_1} ({\bm y}_1 , {\bm y}_2 , {\bm v}_1 , {\bm v}_2)
\end{equation}
where the Lagrangian $L_{\mu_1}$ is the product of a geometrical invariant by
$d\tau_1 / dt$. For instance
\begin{equation}
\label{eq4.2}
L_{\mu_1^{(2)}} =\frac14 \, \frac{d\tau_1}{dt} \left[G_{\alpha\beta} \, 
G^{\alpha\beta} \right]_1 \equiv \frac14 \, 
\frac{d\tau_1}{dt} \left[ G_{ab}^2 \right]_1 \, .
\end{equation}

We shall separately evaluate each geometrical invariant, $G_{ab}^2 , G_{abc}^2
, \ldots$, before multiplying it by the (regularized) proper-time redshift
factor $d\tau_1 /dt$ (``Einstein time dilation''). Note also that we
systematically work with the {\it order-reduced} 2PN metric, i.e. the 2PN
metric in which the higher time derivatives of ${\bm y}_1$ and ${\bm y}_2$
have been expressed in terms of positions and velocities only, $({\bm y}_1 ,
{\bm y}_2 , {\bm v}_1 , {\bm v}_2)$, by iterative use of the (harmonic-gauge)
equations of motion. As was discussed long ago, such an order reduction
of the action is allowed, when it is understood that it corresponds to a
certain additional change of coordinate gauge \cite{Schaefer84,
  DamourSchaefer85, Damour:1990jh}. As we shall ultimately be interested in
computing gauge-invariant quantities associated with the EOB reformulation of
the dynamics, we do not need to keep track of this coordinate change.

\subsection{Explicit 2PN-accurate tidal actions for general orbits}

Let us start by discussing the simplest (and physically most important)
geometric invariant, namely the one associated with the electric-type
quadrupolar tide, say
\begin{eqnarray}
\label{eq4.3}
J_{2e} &\equiv& [G_{ab} \, G_{ab}]_1 = [G_{\alpha\beta} \, G^{\alpha\beta}]_1 \nonumber\\
&=& [R_{\alpha\mu\beta\nu} \, R_{\gamma\kappa\delta\lambda} \,
g^{\alpha\gamma} \, 
g^{\beta\delta} \, u^{\mu} \, u^{\nu} \, u^{\kappa} \, u^{\lambda}]_1 \, ,
\end{eqnarray}
where $u_1^{\mu} \equiv dy_1^{\mu} / d\tau_1 = (c,{\bm v}_1) \, dt / d\tau_1$, and where the subscript $2e$
on $J_{2e}$ refers to \lq\lq $\ell =2$ electric."
Using two independently written codes (one based on the Maple system, and the
other one based on the Mathematica software supplemented by the package xAct
\cite{JMM02}) we have computed the right-hand side of Eq. (\ref{eq4.3}) within the
(regularized) 2PN metric. (Actually, as explained above, the Mathematica code alternatively regularized, \`a la
Hadamard, the value of $G_{\alpha\beta}$ computed with the full
(non-regularized) 2PN metric.)

\smallskip

As the PN expansion of the quadrupolar tidal tensor Eq. (\ref{eq2.7}) starts as 
\begin{eqnarray*}
G_{ab} &=& - \, c^2 \, R_{a0b0} + \ldots \nonumber\\
&=& + \, \frac12 \, c^2 (\partial_{ab} \, g_{00} - \partial_{a0} \, g_{b0}
- \partial_{b0} \, g_{a0} + \partial_{00} \, g_{ab}) \nonumber\\ & & + \ldots \, ,
\end{eqnarray*}
one sees that the 2PN-accurate metric [i.e., the knowledge of $g_{00}$ up to
$O(1/c^6)$ terms included, of $g_{0a}$ up to $O(1/c^5)$, and of $g_{ab}$ up to
$O(1/c^4)$] is exactly what is needed to be able to compute $G_{ab}$ to the
2PN (fractional) accuracy, i.e., $G_{ab} = + \, \partial_{ab} V + c^{-2}
(\ldots) + c^{-4} (\ldots)$. The same is true for the higher {\it electric}
tidal moments $G_{abc} , \ldots$. However, one can easily see that one loses a
PN order when evaluating either the {\it magnetic} tidal moments $H_{ab}
,H_{abc} , \ldots$ or the time-differentiated electric one $\dot
G_{ab},\ldots$. The result we obtained, for general orbits, is
\begin{eqnarray}
\label{eq4.4}
J_{2e} &=&  \frac{6 G^2 m_2^2}{r_{12}^6} \bigg\{1+\frac{1}{c^2}
  \Big(-3 (n_{12} v_{12})^2 - 3 (n_{12} v_2)^2 +3 v_{12}^2 
\nonumber\\
&& \qquad \qquad \qquad \quad
- \frac{G}{r_{12}}
  (5 m_1 + 6 m_2) \Big) \nonumber \\ 
&+&\frac{1}{c^4}
  \bigg[3 (n_{12} v_{12})^4+ 
  12 (n_{12} v_2)^2 (n_{12} v_{12})^2 + 6 (n_{12} v_2)^4 \nonumber\\
&& \quad -9 v_{12}^2 (n_{12} v_{12})^2 -6
  (n_{12} v_{12})^2 (v_2 v_{12})   \nonumber \\ 
&& \quad -6 (n_{12} v_2) (n_{12} v_{12}) (v_2 v_{12}) -3 v_2^2 
  (n_{12} v_{12})^2 \nonumber\\
&& \quad -9 v_{12}^2 (n_{12} v_2)^2 -3 v_2^2 (n_{12} v_2)^2 + 6 v_{12}^4
  \nonumber \\ 
&& \quad +6 v_{12}^2 (v_2
  v_{12}) + 3 (v_2 v_{12})^2+3 v_2^2 v_{12}^2  \nonumber\\
&& \quad + 
  \frac{G m_1}{r_{12}} \Big(-\frac{109}{4}
  (n_{12} v_{12})^2 + \frac{41}{2} (n_{12} v_2)^2 +\frac{21}{4} v_{12}^2\Big)
  \nonumber \\ 
&& \quad + \frac{G m_2}{r_{12}} \Big(6 (n_{12}
  v_{12})^2 + 21 (n_{12} v_2)^2 -6 v_{12}^2\Big) \nonumber\\
&& \quad +\frac{G^2}{r_{12}^2}
  \Big(\frac{365 m_1^2}{28}+\frac{125 m_1
    m_2}{2}+21 m_2^2\Big) \bigg] \bigg\} \, .
\end{eqnarray}
Similarly, we computed the further geometrical invariants  \lq\lq $\ell =3$ electric"
\begin{equation}
\label{eq4.16}
J_{3e} \equiv \left[ G_{abc}^2 \right]_1 
= \left[ G_{\alpha\beta\gamma} \, G^{\alpha\beta\gamma} \right]_1 \, ,
\end{equation}
and \lq\lq $\ell =2$ magnetic"
\begin{eqnarray}
\label{eq4.15}
J_{2m} &\equiv& \frac14 \, \left[H_{ab}^2\right]_1 
= \frac14 \, \left[H_{\alpha\beta} \, H^{\alpha\beta}\right]_1 \nonumber\\
&=& c^2 \left[R_{\alpha\mu\beta\nu}^* \, R_{\gamma\kappa\delta\lambda}^* 
\, g^{\alpha\gamma} \, g^{\beta\delta} \, u^{\mu} \, u^{\nu} \, u^{\kappa} \, 
u^{\lambda}\right]_1 \, .
\end{eqnarray}
Note the factor $\frac14$, introduced in the definition of $J_{2m}$ to have 
$J_{2m}= (c R_{\alpha\mu\beta\nu}^* u^\mu u^\nu)^2$, analogously to 
$J_{2e}= (R_{\alpha\mu\beta\nu} u^\mu u^\nu)^2$. Let us also note in passing that, in
evaluating $J_{3e}$, i.e., the square of the electric octupole
$G_{\alpha\beta\gamma}$, Eq.~(\ref{eq2.8}), it is important to use the
orthogonally projected covariant derivative $\nabla_{\alpha}^{\perp}$. If,
instead of $(G_{\alpha\beta\gamma})^2$, one evaluates
$(C_{\alpha\beta\gamma})^2$ where $C_{\alpha\beta\gamma} 
= {\rm Sym}_{\alpha\beta\gamma} \, \nabla_{\alpha} \, R_{\beta\mu\gamma\nu} \,
u^{\mu} \, u^{\nu} $, one finds a result which differs from
$(G_{\alpha\beta\gamma})^2$ by a term proportional to $J_{\dot 2 e} = (\dot G_{ab})^2$
[see Eq.~(\ref{eq6.51})].

The results for these invariants (along general orbits) are 
\begin{eqnarray}
\label{eq4.5}
J_{3e}  &=&
 \frac{90 G^2 m_2^2}{r_{12}^8} \bigg\{1
  + \frac{1}{c^2} \bigg[-2 (n_{12} v_{12})^2 -4 (n_{12} v_2)^2 +3 v_{12}^2 \nonumber\\
&& \qquad \qquad \qquad \quad  -
  \frac{G (4 m_1+10 m_2)}{r_{12}} \bigg] \nonumber \\ 
&& +
  \frac{1}{c^4} \bigg[10 (n_{12} v_2)^2 (n_{12} v_{12})^2 +10 (n_{12} v_2)^4
\nonumber\\
&& \qquad
  -\frac{14}{3} v_{12}^2 (n_{12} v_{12})^2 -4 (n_{12} v_{12})^2 (v_2 v_{12})
  \nonumber \\ 
&&  \qquad  -12 v_{12}^2 (n_{12} v_2)^2 -4 (n_{12}
  v_2) (n_{12} v_{12}) (v_2 v_{12}) \nonumber\\
&& \qquad -2 v_2^2 (n_{12} v_{12})^2-4 v_2^2 (n_{12}
  v_2)^2  \nonumber \\ 
&& \qquad
  +\frac{17}{3} v_{12}^4 + 6 v_{12}^2 (v_2 v_{12}) +3 (v_2 v_{12})^2 +
  3 v_2^2 v_{12}^2 \nonumber \\ 
&& \qquad + \frac{G m_1}{r_{12}} \Big(-32 (n_{12}
  v_{12})^2+ 2 (n_{12} v_2) (n_{12} v_{12})
\nonumber\\
&& \qquad \qquad \quad +22 (n_{12} v_2)^2 +\frac{16}{3}
  v_{12}^2\Big) \nonumber \\ 
&& \qquad + \frac{G m_2}{r_{12}} \Big(12 (n_{12} v_{12})^2+ 45
  (n_{12} v_2)^2-18 v_{12}^2\Big) \nonumber\\
&& \qquad +\frac{G^2}{r_{12}^2} \Big(9
  m_1^2+\frac{259 m_1 m_2}{3}+54 m_2^2 \Big) \bigg] \bigg\} \, ,
\end{eqnarray}
and
\begin{eqnarray}
\label{eq4.6}
J_{2m} &=& \frac{18 G^2 m_2^2}{r_{12}^6} \bigg\{- (n_{12}
  v_{12})^2 + v_{12}^2 \nonumber \\ 
&  +&
  \frac{1}{c^2}\bigg[ (n_{12} v_{12})^4+ 4 (n_{12} v_2)^2 (n_{12} v_{12})^2 \nonumber\\
&& \quad -3
  v_{12}^2 (n_{12} v_{12})^2 
-2 (n_{12} v_{12})^2 (v_2 v_{12})- v_2^2 (n_{12}
  v_{12})^2 \nonumber \\ 
&& \quad -2 (n_{12} v_2) (n_{12}
  v_{12}) (v_2 v_{12}) -3 v_{12}^2 (n_{12} v_2)^2 \nonumber\\
&& \quad +2 v_{12}^4 +2 v_{12}^2 (v_2
  v_{12}) + (v_2 v_{12})^2+ v_2^2 v_{12}^2 \nonumber \\ 
&& \quad + \frac{2G}{r_{12}}\Big(\frac{m_1}{3}+ m_2 \Big) 
\Big( (n_{12} v_{12})^2 - v_{12}^2\Big) \bigg] \bigg\} \, .
\end{eqnarray}

The result Eq. (\ref{eq4.4}), after multiplication by the redshift factor
\begin{equation}
\label{eq4.9}
\frac{d\tau_1}{dt} = \left( - \, g_{00} - 2 \, g_{0i} \, \frac{v_1^i}{c} 
- g_{ij} \, \frac{v_1^i \, v_1^j}{c^2} \right)^{1/2} \, ,
\end{equation}
which evaluates to (we use again the notation $\epsilon \equiv 1/c$, and 
henceforth often set Newton's constant to one)
\begin{align}
\label{eq4.10}
\frac{d\tau_1}{dt} & =  1  - \epsilon^2 \left( \frac12 \,  v_1^2 + 
\frac{m_2}{r_{12}} \right) \nonumber \\ 
& + \epsilon^4 \left[ - 
\frac{1}{8} v_1^4 + \frac{m_2}{r_{12}}\left(\frac{1}{2} (n_{12}v_2)^2-
\frac{3}{2} v_1^2  
\right.\right.\nonumber\\
& \left. \quad~
+ 4 (v_1 v_2) -2 v_2^2\bigg) + 
\frac{m_2}{2r_{12}^2}(3 m_1 + m_2) \right] \, ,
\end{align}
provides the $O (\mu_1^{(\ell = 2)})$ piece (``gravitoelectric tidal
quadrupole'') of the reduced two-body action at the 2PN approximation level;
i.e., including tidal correction terms that are $(v/c)^4$ smaller than the
leading order tidal Lagrangian which is simply given by $J_{2e}^{(0)} = 6 \,
m_2^2 / r_{12}^6$. Similarly, multiplying the results of Eqs. (\ref{eq4.5}) and
(\ref{eq4.6}) by the redshift factor in Eq. (\ref{eq4.10}) provides the reduced tidal
actions associated with $J_{3e} \equiv [G_{abc}^2]_1$ and $J_{2m} \equiv
\frac14 [H_{ab}^2]_1$, at the 2PN approximation for the
electric-octupole term $J_{3e}$, and at the 1PN approximation for the
magnetic-quadrupole term $J_{2m}$.

\bigskip

In view of their complexity, the results of Eqs. (\ref{eq4.4}), (\ref{eq4.5}),
(\ref{eq4.6}), which provide the action for general orbits, are not very
useful as they are. In what follows, we shall extract the physically most
useful information they contain by: (i) focusing our attention on {\it
  circular orbits} and (ii) reformulating our results in terms of the EOB
description of binary systems. Note in passing that though circular orbits are
only special solutions of binary dynamics, they are the ones of prime physical
importance in many situations, most notably radiation-reaction-driven
inspiralling binary systems.

\subsection{Tidal actions along circular orbits}

In the following, we shall therefore restrict our attention to circular
motions. [However, we shall show below how this restricted result can
crucially inform the EOB description of tidally interacting binary systems.]
We shall also focus on the {\it relative} dynamics in the center of mass
frame. As we see in Eqs.~(\ref{eq4.4}), (\ref{eq4.5}), (\ref{eq4.6}),
(\ref{eq4.10}), the various Lagrangians depend only on the relative position
${\bm y}_{12} = {\bm y}_1 - {\bm y}_2$ and start depending on (individual)
velocities only at 1PN (for general orbits), and even at 2PN for the
invariants themselves (in the case of circular orbits). This implies that we
shall not really need to use to its full 2PN accuracy the relation between
center-of-mass variables ${\bm y}_1^{\rm CM} , {\bm y}_2^{\rm CM} , {\bm
  v}_1^{\rm CM} , {\bm v}_2^{\rm CM}$, and relative ones ${\bm y}_{12} , {\bm
  v}_{12}$, namely (in the circular case)
\begin{eqnarray}
\label{eq4.11}
y_1^i &= &\left[ X_2 + 3  \left( \frac{M}{r_{12} \, c^2} \right)^2 \nu 
\, X_{12} \right] y_{12}^i \, , \nonumber \\
y_2^i &= &\left[ - \, X_1 + 3  \left( \frac{M}{r_{12} \, c^2} \right)^2 \nu 
\, X_{12} \right] y_{12}^i \, ,
\end{eqnarray}
and the corresponding velocity relations obtained by time-differentiating
them, using the fact that ${\bm y}_{12} = r_{12} \, {\bm n}_{12}$ where
$r_{12}$ is constant and ${\bm n}_{12}$ rotates with an angular velocity given
by
\begin{eqnarray}
\label{eq4.12}
\Omega^2 &=& \frac{M}{r_{12}^3} \left[ 1 + \epsilon^2 (\nu - 3) 
\, \frac{M}{r_{12}} \right.\nonumber\\
&&\left.+ \epsilon^4 \left( 6+\frac{41}{4} \, \nu + \nu^2 \right) 
\left( \frac{M}{r_{12}} \right)^2 \right] \, .
\end{eqnarray}

Here and below we use the notation
\begin{equation}
\label{eq4.13}
X_1 \equiv \frac{m_1}{M}\, , \quad X_2 \equiv \frac{m_2}{M} \, , \quad 
\nu \equiv X_1 \, X_2 \, , \quad X_{12} \equiv X_1 - X_2 \, ,
\end{equation}
(recall that $M \equiv m_1 + m_2$ so that $X_1 + X_2 = 1$). In our
calculations, the $\epsilon^4 = 1/c^4$ contributions in
Eqs.~(\ref{eq4.11}), (\ref{eq4.12}) do not matter, and can be neglected from
the start.

\smallskip

Using such an additional circular (and center-of-mass) reduction, we get a
much simplified result for the electric-quadrupole invariant $J_{2e}$,
Eq.~(\ref{eq4.3}), namely,
\begin{multline}
\label{eq4.14}
J_{2e}^{\rm (circ)} = \frac{6M^2X_2^2}{r_{12}^6}\left[1+ 
\epsilon^2\frac{(X_1-3)M}{r_{12}}\right.\\
\left.
-\epsilon^4 \frac{M^2}{28r_{12}^2} (713 X_1^2-805 X_1-336) \right] \, .
\end{multline}
In a similar manner, one gets much simplified results for the other
(subleading) geometrical invariants of tidal significance, namely the magnetic
quadrupolar term $J_{2m}$, Eq.~(\ref{eq4.15}), the electric octupolar term $J_{3e}$,
Eq.~(\ref{eq4.16}), and also for the time-differentiated electric-quadrupole
coupling, say, 
\begin{equation}
\label{eq4.17}
J_{\dot 2 e} \equiv  \left[ \dot G_{ab}^2 \right]_1 = \left[ (u^{\mu} \, 
\nabla_{\mu} \, G_{\alpha\beta})(u^{\nu} \, \nabla_{\nu} \, G^{\alpha\beta}) \right]_1 \, .
\end{equation}
Among these invariants, the 2PN accurate metric allows one (as for $G_{ab}^2$)
to calculate to 2PN fractional accuracy only the electric-octupole term $J_{3e}$.
The other ones can  be computed only at 1PN fractional accuracy because of
their \lq\lq magnetic," or ``$\partial_0 = c^{-1} \partial_t$'' character. Our
explicit ``circular'' results for $J_{2m}$, $J_{3e}$ and $J_{\dot 2 e}$ are

\begin{equation}
\label{eq4.18}
J_{2m}^{\rm (circ)} = \frac{18X_2^2 M^3}{r_{12}^7}\left[1
+\epsilon^2\frac{M}{3r_{12}}(3X_1^2+X_1-9) \right] \, ,
\end{equation}

\begin{multline}
\label{eq4.19}
J_{3e}^{\rm (circ)} =  \frac{90X_2^2 M^2}{r_{12}^8}\left[1
+\epsilon^2 (6X_1-7)\frac{M}{r_{12}} \right.\\
 \left. -\epsilon^4  \frac{M^2}{3r_{12}^2} (61X_1^2+4X_1-98) \right]\, ,
\end{multline}

\begin{equation}
\label{eq4.20}
J_{\dot 2 e}^{\rm (circ)} =  \frac{18X_2^2 M^3}{r_{12}^9}\left[1
+\epsilon^2 (X_1^2-7)\frac{M}{r_{12}}\right] \, .
\end{equation}

To complete the above results, and allow one to compute the corresponding
associated Lagrangians, let us note that the circular value of the redshift
factor is
\begin{align}
\label{eq4.21}
\frac{d\tau_1}{dt}&\equiv\frac{1}{\Gamma_1}=1
-\frac{M(X_1-1)(X_1-3)}{2r_{12}}\epsilon^2 \nonumber\\
&+
\frac{M^2(X_1-1)}{8 r_{12}^2}(3X_1^3-9X_1^2+13X_1-3)\epsilon^4 \, .
\end{align}
Let us also quote the value of the inverse redshift factor, $\Gamma_1$ (analog
to a Lorentz $\gamma$-factor $\gamma = 1/\sqrt{1-{\bm v}^2 / c^2}$), namely
\begin{align}
\label{eq4.22}
\Gamma_1 &\equiv \frac{dt}{d\tau_1}=1+\frac{M(X_1-1)(X_1-3)}{2 r_{12}}
\epsilon^2\nonumber\\
&-\frac{M^2}{8r_{12}^2}(X_1-1)(X_1^3+5X_1^2-17X_1+15)\epsilon^4 \, .
\end{align}

\section{EOB description of the tidal action}\label{sec5}
\setcounter{equation}{0}

We have computed above the effective actions associated with the tidal
parameters $\mu_1^{(2)} , \mu_1^{(3)} , \sigma_1^{(2)}$ and $\mu'^{(2)}_1$.
Before the restriction to circular motions (in the center-of-mass frame) they
have the general form
\begin{equation}
\label{eq5.1}
\mu_1 \int dt \, L_{\mu_1} ({\bm y}_{12} , {\bm v}_1 , {\bm v}_2) \, ,
\end{equation}
where $\mu_1$ denotes a generic tidal parameter, and ${\bm y}_{12} = {\bm y}_1
- {\bm y}_2$. In this section we discuss how one can describe the actions of Eq. 
(\ref{eq5.1}) within the EOB formalism. Let us recall that the EOB formalism
\cite{Buonanno:1998gg, Buonanno:2000ef, Damour:2000we, Damour:2001tu} replaces
the (possibly higher-order) Lagrangian dynamics of two particles by the
Hamiltonian dynamics of an ``effective particle'' embedded within some
``effective external potentials.'' For non-spinning~\footnote{What is important
  for the current discussion (i.e. the application to tidal interactions) is
  the absence of preferred directions in the {\it orbital} dynamics ${\bm y},
  {\bm v}$ or ${\bm y}, {\bm p}_y$. This is the case for the {\it reduced}
  tidal actions (generated by orbital-induced tidal moments).} bodies, the
original (velocity-dependent) two-body interactions become reformulated (and
simplified by means of a suitable contact transformation in phase space) in
terms of three ``EOB potentials'': $A(r_{\rm eff})$, $\bar B (r_{\rm eff})$
and $Q(r_{\rm eff} , p^{\rm eff})$. The first two potentials, $A(r_{\rm eff})$
and $\bar B (r_{\rm eff})$, parametrize an ``effective metric''
\begin{eqnarray}
\label{eq5.2}
ds^2_{\rm eff} &=& g_{\mu\nu} (x_{\rm eff}) \, dx^{\mu}_{\rm eff} \,
dx^{\nu}_{\rm eff} \nonumber\\
&=& -A (r_{\rm eff}) \, c^2 \, dt^2_{\rm eff} + \bar B (r_{\rm eff}) \,
dr^2_{\rm eff} \nonumber\\
&&+ r^2_{\rm eff} (d\theta^2_{\rm eff} + \sin^2 \theta_{\rm eff} \,
d\varphi^2_{\rm eff}) \, ,
\end{eqnarray}
and its associated Hamilton-Jacobi equation, while the third potential $Q
(r_{\rm eff} , p^{\rm eff})$ (which necessarily appears at 3PN
\cite{Damour:2000we}), describes additional contributions to the
(Hamilton-Jacobi) mass-shell condition,
\begin{equation}
\label{eq5.3}
0 = \mu^2 + g_{\rm eff}^{\mu\nu} (x_{\rm eff}) \, p_{\mu}^{\rm eff} \,
p_{\nu}^{\rm eff} + Q (r_{\rm eff} , p^{\rm eff})
\end{equation} 
[where $\mu \equiv m_1 \, m_2 / M \equiv \nu M$ is the reduced mass of the
binary system], that are higher than quadratic in the effective momentum
$p^{\rm eff}$. Following the EOB-simplifying philosophy of
Ref.~\cite{Damour:2000we}, we shall assume that the third potential has been
reduced (by a suitable canonical transformation) to a form where it vanishes
with the radial momentum $p_r^{\rm eff}$.

\smallskip

In addition, EOB theory introduces a {\it dictionary} between the original
dynamical variables (positions, momenta, angular momentum, energy) and the
effective ones. A crucial entry of this dictionary is a non-trivial
transformation between the original ``real'' energy, i.e., the value of the
original (total) Hamiltonian $H$, and the ``effective'' energy $-p_0^{\rm eff}
\equiv H_{\rm eff}$ entering the mass-shell condition of Eq. (\ref{eq5.3}). Because
of this transformation, the final EOB-form of the (original, real) Hamiltonian
reads (here we set $c=1$ for simplicity)
\begin{equation}
\label{eq5.4}
H^{\rm EOB} ({\bm x}_{\rm eff} , {\bm p}_{\rm eff}) = M \, \sqrt{1+2 \, \nu
  \left( \frac{H_{\rm eff}}{\mu} - 1 \right)} \, ,
\end{equation}
where $H_{\rm eff} =H_{\rm eff} ({\bm x}_{\rm eff} , {\bm p}_{\rm eff})$ is given by
\begin{equation}
\label{eq5.5}
H_{\rm eff} = \sqrt{A(r_{\rm eff}) \left( \mu^2 
+ \frac{{\bm J}^2_{\rm eff}}{r^2_{\rm eff}} 
+ \frac{(p_r^{\rm eff})^2}{\bar B (r_{\rm eff})} 
+ Q (r_{\rm eff} , p^{\rm eff}) \right)} \, .
\end{equation}
Here ${\bm J}_{\rm eff} \equiv {\bm x}_{\rm eff} \times {\bm p}_{\rm eff}$
denotes the effective orbital angular momentum, which, by the EOB dictionary,
is actually identified with the original, total (center of mass) orbital
angular momentum ${\bm J}$ of the binary system: ${\bm J}_{\rm eff} \equiv
{\bm J}$.

\subsection{EOB reformulation of tidal actions: general orbits}

Let us now discuss what  the various possible methods are for reformulating an
original action of the type $L_0 ({\bm y}_{12} , {\bm v}_1 , {\bm v}_2 ,
\ldots) + \mu_1 \, L_{\mu_1} ({\bm y}_{12} , {\bm v}_1 , {\bm v}_2)$ [where
$\mu_1$ stands for a sum over a collection of tidal parameters $\mu_1^{(2)} ,
\mu_2^{(2)} , \mu_1^{(3)} , \mu_2^{(3)} , \ldots$] into corresponding
$\mu_1$-deformations of EOB potentials: $A_0 (r_{\rm eff}) + \mu_1 \,
A_{\mu_1} (r_{\rm eff})$, $\bar B_0 (r_{\rm eff}) + \mu_1 \, \bar B_{\mu_1}
(r_{\rm eff})$, $Q_0 (r_{\rm eff} , p^{\rm eff}) + \mu_1 \, Q_{\mu_1} (r_{\rm
  eff} , p^{\rm eff})$. The main difficulty in finding the perturbed EOB
potentials $A_{\mu_1}$, $\bar B_{\mu_1}$, and $Q_{\mu_1}$ that encode the
dynamics of $L_{\mu_1}$ is that such a dynamical equivalence is obtained only
after some a priori unknown phase-space contact transformation between the EOB
phase-space coordinates, say $\xi_{\rm eff} = ({\bm x}_{\rm eff} , {\bm
  p}_{\rm eff})$, and the original (harmonic-coordinate-related) ones, say
$\xi_h = ({\bm y}_{12} , {\bm v}_{12})$. For simplicity, we assume that we
have already performed the reduction of the original harmonic-coordinate
dynamics to its center-of-mass version, in which one can express ${\bm v}_1$
and ${\bm v}_2$ in terms of the relative velocity ${\bm v}_{12} \equiv {\bm
  v}_1 - {\bm v}_2$ and of ${\bm y}_{12} \equiv {\bm y}_1 - {\bm y}_2$. On the
other hand, we do not immediately assume that the original Lagrangian dynamics
is expressed in Hamiltonian form. (Let us recall that, as was found long ago
\cite{DD81,Damour:1982wm}, starting at the 2PN level, the harmonic-coordinate
dynamics {\it does not} admit an ordinary Lagrangian, $L(y,\dot y)$, but only a
higher-order one, $L (y,\dot y , \ddot y)$. In order to express the 2PN
dynamics in Hamiltonian form, one already needs some (higher-order) contact
transformation. However, this transformation is well-known, e.g., Ref. 
\cite{DamourSchaefer85}, and we do not need to complicate our discussion by
explicitly mentionning it. Nonetheless, it will be taken into account in our
calculations below.)

\smallskip

The transformation $T$ between $\xi_{\rm eff}$ and $\xi_h$ will have the
general structure
\begin{equation}
\label{eq5.6}
\xi_h = T_0 (\xi_{\rm eff}) + \mu_1 \, T_{\mu_1} (\xi_{\rm eff}) \, .
\end{equation}

The unperturbed part $T_0 (\xi_{\rm eff})$ is known
from the previous EOB work \cite{Buonanno:1998gg, Damour:2000we}, but the
$O(\mu_1)$ perturbed part $T_{\mu_1} (\xi_{\rm eff})$ is unknown, and,
actually, is part of the problem which must be solved for reformulating the
(perturbed) harmonic-coordinate dynamics in EOB form. This means, in
particular, that it would not be correct to try to compute $A_{\mu_1}$, $\bar
B_{\mu_1}$ and $Q_{\mu_1}$, simply by  replacing in the tidal action in Eq. 
(\ref{eq5.1}) the harmonic variables $\xi_h$ by their unperturbed expression
$T_0 (\xi_{\rm eff})$ in terms of the effective variables $\xi_{\rm eff}$.

\smallskip

For the general case of non-circular orbits, a universal, correct method for
transforming the original Lagrangian $L (\xi_h) = L_0 (\xi_h) + \mu_1 \,
L_{\mu_1} (\xi_h)$ in EOB form consists (as explained in Ref. 
\cite{Buonanno:1998gg}) of the following steps: (i) to transform the original
Lagrangian $L(\xi_h)$ in Hamiltonian form $H(\xi_H) = H_0 (\xi_H) + \mu_1 \,
H_{\mu_1} (\xi_H)$, where $\xi_H = (q,p)$ are canonical coordinates; (ii) to
extract the gauge-invariant content of $H(\xi_H)$ by expressing it in terms of
{\it action variables} $I_a = \frac{1}{2\pi} \oint p_a \, dq_a$, which yields
the {\it Delaunay} Hamiltonian $H({\bm I}) = H_0 ({\bm I}) + \mu_1 \,
H_{\mu_1} ({\bm I})$; (iii) to do the same thing for the EOB Hamiltonian, i.e.
to compute, as a functional of the unknown EOB potentials, its Delaunay form
$H_{\rm EOB} ({\bm I}) = H_0^{\rm EOB} ({\bm I}) + \mu_1 \, H_{\mu_1}^{\rm
  EOB} ({\bm I})$; and finally (iv) to identify the known $H({\bm I})$ to
$H_{\rm EOB} ({\bm I})$, which depends on the unknown functions $A_{\mu_1}$,
$\bar B_{\mu_1}$, $Q_{\mu_1}$. This last step yields (functional) equations
for $A_{\mu_1}$, $\bar B_{\mu_1}$, $Q_{\mu_1}$ and thereby allows one to
determine them. [In practice, the functional dependence on $A,\bar B,Q$ is
replaced by a much simpler parameter-dependence by using the method of
undetermined coefficients for parametrizing general forms of $A,\bar B , Q$.]
An alternative (and equally universal) method for transforming $L(\xi_h)$ in
EOB form (as used in Ref. \cite{Damour:2000we}) is to add the transformation $\xi_h
= T(\xi_{\rm eff})$ to the list of unknowns (using the method of undetermined
coefficients), and to directly solve the set of constraints for $T,A,\bar B$
and $Q$ coming from the requirement that $H_{\rm EOB} (\xi_{\rm eff} , A, \bar
B , Q) = H(T(\xi_{\rm eff}))$. [One must then take into account that $T$ is
constrained to be a {\it canonical} transformation.]

\subsection{EOB reformulation of tidal actions: circular orbits}

The 2PN-accurate results, given for several tidal interactions in the case of
general orbits, in the previous section, can, in principle, be transformed
within the EOB format by using any of the two methods we just explained.
However, from the point of view of current astrophysical applications, one is
mainly interested in knowing the EOB description of (quasi)-{\it circular}
motions. In this case, we  know {\it a priori} that it is only the $A$ radial
potential which matters. Knowing this, the question arises how to compute the
tidal perturbation $A_{\mu_1}$ of the EOB $A$ potential in the most efficient
manner, possibly without having to go through the rather involved, general
universal methods recalled above. Fortunately, it is possible to do so by
using the following facts.

\smallskip

The first useful fact concerns the relation between the tidal perturbation (in
harmonic coordinates) of the Lagrangian of the binary system, say
\begin{equation}
\label{eq5.7}
\delta L^h (y_h , v_h) = \mu_1 \, L_{\mu_1} ({\bm y}_{12} , {\bm v}_1 , {\bm
  v}_2) \, ,
\end{equation}
and the corresponding tidal perturbation (in harmonic-related phase-space
coordinates) of the Hamiltonian, say
\begin{equation}
\label{eq5.8}
\delta H^h (y_h , p_h) \equiv H_{\rm full}^h (y_h , p_h) 
- H_{\rm tidal\mbox{-}free}^h (y_h , p_h) \, .
\end{equation}
[Strictly speaking, as we recalled above, the harmonic-related $(q,p) = (y_h ,
p_h)$ phase-space coordinates involve a supplementary $O(1/c^4)$ gauge
transformation linked to the order reduction of $L_{\rm 2PN} (y,\dot y , \ddot
y)$ into $L_{\rm 2PN}^{\rm red} (y' , \dot y')$.] Note that here and in the
following the notation $\delta Q(\xi)$ will always refer to the tidal
contribution to some function of specified variables, i.e. $\delta Q(\xi)
\equiv Q_{\rm full} (\xi) - Q_{\rm tidal\mbox{-}free} (\xi)$. One has to be
careful about which variables are fixed as, for instance, the transformation
between Lagrangian $(q,\dot q)$ and Hamiltonian $(q,p)$ coordinates does
contain a tidal contribution [because $\delta^{\rm tidal} L(y_h , \dot y_h)$
does depend on velocities]. This being made clear, we have the well-known
universal result about first-order deformations of Lagrangians by small
parameters, $L(q,\dot q) = L_0 (q,\dot q) + \mu_1 \, L_{\mu_1} (q,\dot q)$,
namely
\begin{equation}
\label{eq5.9}
\delta H^h (y_h , p_h) = - \, \delta L^h (y_h , v_h)
\end{equation}
which follows from the properties of the Legendre transform.

\smallskip

Let us now apply the second method recalled above for transforming the
``harmonic'' Hamiltonian $H_{\rm full}^h (y_h , p_h) = H_0^h (y_h , p_h) +
\delta H^h (y_h , p_h)$ (where the index $0$ refers to the {\it unperturbed},
tidal-free dynamics) into its corresponding EOB form $H_{\rm full}^{\rm EOB}
(x_{\rm EOB} , p_{\rm EOB})$, defined in Eqs.~(\ref{eq5.4}), (\ref{eq5.5})
above. [For clarity, we denote here the effective-one-body phase-space
coordinates by $x_{\rm EOB} , p_{\rm EOB}$, instead of $x_{\rm
  eff} , p_{\rm eff}$ as above.] The crucial point is that the EOB
potentials entering the definition of $H_{\rm full}^{\rm EOB}$ must be the
{\it full}, tidally-completed values of $A, \bar B$ and $Q$, e.g.
\begin{eqnarray}
\label{eq5.10}
A_{\rm full} (r_{\rm EOB}) &=& A_0 (r_{\rm EOB}) + \mu_1 A_{\mu_1} (r_{\rm EOB}) \nonumber\\
&\equiv& A_0 (r_{\rm EOB}) + \delta A (r_{\rm EOB}) \, .
\end{eqnarray}
In other words $\delta H^{\rm EOB} (x_{\rm EOB} , p_{\rm EOB}) \equiv H_{\rm
  full}^{\rm EOB} (x_{\rm EOB} , p_{\rm EOB}) - H_0^{\rm EOB} (x_{\rm EOB} ,
p_{\rm EOB})$ is obtained by varying the functions $A,\bar B$ and $Q$ (i.e. $A
(r_{\rm EOB}) = A_0 (r_{\rm EOB}) + \delta A(r_{\rm EOB})$, etc.) in the
definition in Eqs. (\ref{eq5.4}), (\ref{eq5.5}) of $H_{\rm full}^{\rm EOB} [x_{\rm
  EOB} , p_{\rm EOB} ; A(r_{\rm EOB}) , \bar B (r_{\rm EOB}), Q(r_{\rm EOB} ,
p^{\rm EOB})]$.

\smallskip

This second method for mapping $H_{\rm full}^h (\xi_h)$ into $H_{\rm full}^{\rm
  EOB} (\xi_{\rm EOB})$ [where $\xi_h \equiv (y_h , p_h)$, $\xi_{\rm EOB}
\equiv (x_{\rm EOB} , p_{\rm EOB})$] consists of looking for a full, i.e., 
perturbed, (time-independent) contact transformation $\xi_h = T_{\rm full}
(\xi_{\rm EOB}) = T_0 (\xi_{\rm EOB}) + \mu_1 T_{\mu_1} (\xi_{\rm EOB})$ that
transforms $H_{\rm full}^h (\xi_h)$ into $H_{\rm full}^{\rm EOB} (\xi_{\rm
  EOB})$, i.e., such that
\begin{equation}
\label{eq5.11}
H_{\rm full}^h [T_{\rm full} (\xi_{\rm EOB})] = H^{\rm EOB}_{\rm full} (\xi_{\rm EOB}) \, .
\end{equation}
Rewriting the full transformation $T_{\rm full}$ as the composition $T' \circ
T_0$ of the known unperturbed (tidal-free) contact transformation $\xi_h^0 = T_0
(\xi_{\rm EOB})$ mapping $H_0^h (\xi_h^0)$ into $H_0^{\rm EOB} (\xi_{\rm EOB})$
with an {\it unknown} near-identity additional transformation, $\xi_h = T'
(\xi_h^0) = \xi_h^0 + \mu_1 \{ G_{\mu_1} (\xi_h^0) , \xi_h^0 \}$ [where $\{ f,g \}$
denotes a Poisson bracket and where $\mu_1 G_{\mu_1} (\xi_h^0)$ is the
first-order {\it generating} function associated with the {\it canonical}
transformation $T'$], and expanding all functions in Eq.~(\ref{eq5.11}) into
unperturbed plus tidal contributions ($H^h = H_0^h + \delta H^h$, $T =
(1+\delta T') \circ T_0$, $H^{\rm EOB} = H_0^{\rm EOB} + \delta H^{\rm EOB}$),
leads to the condition
\begin{multline}
\label{eq5.12}
\left[ \delta H^h (\xi_h^0) + \{ \delta G(\xi_h^0) , 
H^h (\xi_h^0)\} \right]_{\xi_h^0 = T_0 (\xi_{\rm EOB})} \\
= \delta H^{\rm EOB} (\xi_{\rm EOB}) \, ,
\end{multline}
where $\delta G(\xi_h^0) = \mu_1 G_{\mu_1} (\xi_h^0)$.

\smallskip

In general, $\delta G(\xi_h^0)$ is part of the unknown functions that must be
looked for when writing the condition in Eq. (\ref{eq5.12}). However, another
simplifying fact occurs in the case where one focusses on {\it circular}
motions: The supplementary term $\{ \delta G , H^h \}$
happens to {\it vanish}. Indeed, $\delta G(\xi_h^0)$ is a scalar function and
the Poisson bracket $\{ \delta G , H^h \}$ is equal to the time derivative of
$\delta G(\xi_h^0)$ along the $H^h$-dynamical flow, which clearly vanishes along
circular motions. This allows one to conclude that, along circular motions, we
have the simple condition
\begin{equation}
\label{eq5.13}
\left[ \delta H^h (\xi_h^0) \right]_{\xi_h^0 = T_0 (\xi_{\rm EOB})}^{\rm circ} 
= \left[ \delta H^{\rm EOB} (\xi_{\rm EOB}) \right]^{\rm circ} \, ,
\end{equation}
where the left-hand side is, in principle, fully known.\\

\subsection{Link between the circular tidal action and the tidal contribution to the EOB $A$ potential}

Let us now evaluate the right-hand side of Eq.~(\ref{eq5.13}). When
restricting the definition of Eqs. (\ref{eq5.4}), (\ref{eq5.5}) of the EOB Hamiltonian
to circular motions, the terms $(p_r^{\rm EOB})^2 / \bar B$ and $Q(r_{\rm EOB}
, p^{\rm EOB})$ disappear (because one works with a gauge-reduced $Q$ which
vanishes with $p_r^{\rm EOB}$). As a consequence, $H_{\rm EOB}^{\rm circ}
(r_{\rm EOB} , J)$ only depends on the $A$ potential. The difference, $\delta
H_{\rm EOB}^{\rm circ} \equiv H_{\rm EOB}^{\rm circ} [r_{\rm EOB} , J , A_{\rm
  full}] - H_{\rm EOB}^{\rm circ} [r_{\rm EOB} , J , A_0]$, can then be simply
computed by varying $A$ ($A_{\rm full} = A_0 + \delta A$) within $H_{\rm
  EOB}^{\rm circ} [A]$. To write explicitly the result of this variation, it is
convenient to work with dimensionless variables. We can replace the two
phase-space variables $r_{\rm EOB}$, $p_{\varphi}^{\rm EOB} \equiv J$ that
enter $H_{\rm EOB}^{\rm circ}$ by their dimensionless counterparts
\begin{equation}
\label{eq5.14}
u \equiv \frac{GM}{c^2 \, r_{\rm EOB}} \equiv 
\frac{G(m_1+m_2)}{c^2 \, r_{\rm EOB}} \, ,
\end{equation}
and 
\begin{equation}
\label{eq5.15}
j \equiv \frac{c \, J}{GM \, \mu} \equiv \frac{c \, J}{G \, m_1 \, m_2} \, .
\end{equation}

In terms of these variables, the explicit expression of 
$\left[H_{\rm full}^{\rm EOB}\right]^{\rm circ}$ reads
\begin{multline}
\label{eq5.16}
\left[H_{\rm full}^{\rm EOB}(u,j)\right]^{\rm circ}  \\= M \, c^2 \sqrt{1+2 \, \nu 
\left( -1 + \sqrt{A(u)(1+j^2 \, u^2)} \right)} \, .
\end{multline}
Varying $A(u)$ in Eq. (\ref{eq5.16}) then yields the following explicit 
expression for the right-hand side of Eq. (\ref{eq5.13}):
\begin{widetext}
\begin{eqnarray}
\label{eq5.17}
&&\left[ \delta H^{\rm EOB} (u,j) \right]^{\rm circ}= \frac12 \, M \, \nu \,
c^2 \, \sqrt{\frac{1+j^2 \, u^2}{A(u) \left[ 1+2 \, \nu 
\left( - \, 1+\sqrt{A(u)(1+j^2 \, u^2)} \right) \right]}} \ \delta A(u) \, .
\end{eqnarray}
\end{widetext}
In addition, one must take into account the constraint coming from the
reduction to circular motions, namely, from $\dot p_r^{\rm EOB} = - \partial
H^{\rm EOB} / \partial \, r_{\rm EOB}$, the fact that $\partial_u [A(u)(1+j^2
\, u^2)]=0$, i.e. the fact that $j^2$ is the following function of $u$ (using
a prime to denote the $u$-derivative):
\begin{equation}
\label{eq5.18}
j^2 = j_{\rm circ}^2 (u) \equiv - \, \frac{A'(u)}{(u^2 \, A(u))'} \, .
\end{equation}
Note that this relation depends on the value of the radial potential $A(u)$.
If one is considering the full, tidally-perturbed circular motions one must
use $A_{\rm full} (u) = A_0 + \delta A$ in Eq.~(\ref{eq5.18}). On the other
hand, as we are now interested in considering the (first-order) tidal
perturbations $\delta H^h$ and $\delta H^{\rm EOB}$, and their link in
Eq.~(\ref{eq5.13}), we can evaluate $\delta H^{\rm EOB}_{\rm circ}$ with
sufficient accuracy by replacing in the coefficient of $\delta A (u)$, on the
right-hand side of Eq.~(\ref{eq5.17}), $A(u)$ and $j^2$ by their unperturbed,
tidal-free expressions $A_0 (u)$ and $j_{A_0}^2 (u)$ (obtained by replacing $A \to
A_0$ on the right-hand side of Eq. (\ref{eq5.18}). [This remark applies to several other results below; notably Eqs. (\ref{eq5.20}) and (\ref{eq5.22})].

\smallskip

Combining our results of Eqs. (\ref{eq5.9}), (\ref{eq5.13}) and (\ref{eq5.17}), we
finally get a very simple link between the tidal variation of the
harmonic-coordinate Lagrangian $\delta L (y_h , v_h)$ and the corresponding
tidal variation $\delta A (u)$ of the EOB $A$ potential, namely,
\begin{equation}
\label{eq5.19}
\delta A(u) = - \, \frac{2}{M \, \nu \, c^2} \, \sqrt{F(u)} \, \left[ \delta L
  (y_h , v_h) \right]_{r_h = T_0 (u)}^{\rm circ} \, ,
\end{equation}
where
\begin{widetext}
\begin{equation}
\label{eq5.20}
F(u) \equiv \left[ \frac{A(u)}{1+j^2 \, u^2} \left( 1 + 2 \, \nu \left(-1 
+ \sqrt{A(u)(1+j^2 \, u^2)} \right)\right) \right]_{A=A_0}^{\rm circ} \, .
\end{equation}
\end{widetext}
Here, the superscript ``circ'' means that $j^2$ must be replaced by $j_{\rm
  circ}^2 (u)$, Eq.~(\ref{eq5.18}). (Note that the replacement $A\to A_0$ indicated as a subscript must be done
both in the explicit occurrence of $A$ in Eq. (\ref{eq5.20}) and in the definition in Eq. 
(\ref{eq5.18}) of $j_{\rm circ}^2 (u)$). Finally, if we introduce the
short-hand notation
\begin{equation}
\label{eq5.21}
\tilde A (u) \equiv A(u) + \frac12 \, u \, A'(u) \, ,
\end{equation}
$F(u)$, Eq.~(\ref{eq5.20}), can be rewritten in the explicit form
\begin{equation}
\label{eq5.22}
F(u) = \tilde A (u) \left[ 1+2 \, \nu \left( -1 + 
\frac{A(u)}{\sqrt{\tilde A (u)}} \right) \right] \, ,
\end{equation}
which is valid along circular orbits, and applies for any relevant (exact or approximate) 
value of the $A$ potential.
On the other hand,
as we computed $\delta L$ only to the 2PN fractional accuracy, it
is sufficient to use a value of $F(u)$ which is also only fractionally
2PN-accurate. 
One might think {\it a priori}  that this would mean using for $A(u)$ in Eq. (\ref{eq5.22}) 
the tidal-free approximation $A_0(u)$
truncated at the 2PN order, namely 
$A_0^{\rm 2PN} (u) = 1-2 \, u + 2 \, \nu \, u^3$.
However, the contribution $2 \, \nu \, u^3 = 2 \, \nu (GM/(c^2 \, r_{\rm
  EOB}))^3$ is $O(1/c^6)$ compared to one, which is the leading-order value of
$F(u)$, which starts as $F(u) = 1 + O(u) = 1 + O(1/c^2)$. The same
consideration applies to $\tilde A (u)$. [The situation would have been
different if $F(u)$ had been, say, $\propto A'(u)$.] This means that, at the
2PN fractional accuracy, we can use the value of $F(u)$ obtained from the
leading-order, ``Schwarzschild-like'' value of $A_0 (u)$, namely $A_0^{\rm
  1PN} (u) = 1-2 \, u$. The corresponding $\tilde A$ function is then: $\tilde
A_0^{\rm 1PN} (u) = 1-3 \, u$, so that
\begin{equation}
\label{eq5.23}
F^{\rm 2PN} (u) = (1-3 \, u) \left[ 1+2 \, \nu \left( - \, 1 
+ \frac{1-2 \, u}{\sqrt{1-3 \, u}} \right)\right] \, .
\end{equation}
Consistently with the fractional 2PN accuracy, and remembering, that $u =
O(1/c^2)$, we could as well use the 2PN-accurate series expansion of Eq. 
(\ref{eq5.23}), say $F^{\rm 2PN} (u) = 1+f_1(\nu) \, u + f_2(\nu) \, u^2 +
O(u^3)$. However, it is better to retain the information contained in Eq. 
(\ref{eq5.23}) that, in the test-mass limit $\nu \to 0$ (where $A_0 (u) \to
1-2 \, u$), the exact value of $F(u)$ becomes $1-3 \, u$ (see later).

\smallskip

There remains only one missing piece of information to be able to use our
result in Eq. (\ref{eq5.19}) for computing the various tidal contributions to $A(u)$.
We need to work out the explicit form of the unperturbed transformation $T_0$
between $r_{\rm EOB}$ and $r_h$.

\smallskip

A first method for getting the transformation $T_0$ (at 2PN) is to compose the
transformation $\xi_h^0 \to \xi_{\rm ADM}$ (obtained at 2PN in Ref. 
\cite{DamourSchaefer85}, and at 3PN in Ref. \cite{Damour:2000ni}) with the
transformation $\xi_{\rm ADM} \to \xi_{\rm EOB}$ (obtained at 2PN in Ref. 
\cite{Buonanno:1998gg}, and at 3PN in Ref. \cite{Damour:2000we}). For our present
purpose, it is enough to restrict these transformations to the circular case,
i.e. to transformations $r_h \to r_{\rm ADM}$ and $r_{\rm ADM} \to r_{\rm
  EOB}$.

\smallskip

The transformation $h \to {\rm ADM}$ starts at 2PN, i.e., ${\bm y}_A^h = {\bm
  x}_A^{\rm ADM} + c^{-4} \, Y_A^{\rm 2PN} ({\bm x}^{\rm ADM} , {\bm p}^{\rm
  ADM})$, with $Y_A^{\rm 2PN} ({\bm x}^{\rm ADM} , {\bm p}^{\rm ADM})$ given,
e.g., in Eq.~(4.5) of Ref. \cite{Damour:2000ni}. Its circular, and center-of-mass,
reduction (with ${\bm n}_{12} \cdot {\bm p}_A = 0$, ${\bm p}_1 = - {\bm p}_2
\equiv {\bm p}$, and $({\bm p} / \mu)^2 = GM / r_{12} + O(1/c^2)$) yields at 2PN
\begin{equation}
\label{eq5.24}
r_{12}^h = r_{12}^{\rm ADM} \left[ 1 + \left( \frac14 
+ \frac{29}{8} \, \nu \right) 
\left( \frac{GM}{c^2 \, r_{12}} \right)^2 \right] \, .
\end{equation}
On the other hand the transformation ADM $\to$ EOB starts at 1PN. To determine
the corresponding radial transformation $r_{12}^{\rm ADM} \to r^{\rm EOB}$, one
could think of using Eq.~(6.22) of Ref. \cite{Buonanno:1998gg}. However, this
equation needs to be completed by the knowledge of the circularity condition
relating $({\bm p}^{\rm ADM} / \mu)^2$ to $GM / r_{12}^{\rm ADM}$ at the 1PN
level included. This 1PN-accurate circularity condition can, e.g., be obtained
from combining the 1PN-accurate $r^{\rm ADM} = r^{\rm ADM} (j)$ relation given
in Ref. \cite{Damour:1999cr} (see below), with the fact that (setting $u_{\rm ADM}
\equiv GM / (c^2 \, r_{\rm ADM})$) $({\bm p}_{\rm ADM} / (\mu c))^2 = j^2 \, u_{\rm
  ADM}^2$. This yields $({\bm p}_{\rm ADM} / (\mu c))^2 = u_{\rm ADM} + 4 \,
u_{\rm ADM}^2$, and therefrom the relation between $r_{\rm ADM}$ and $r_{\rm
  EOB}$. 

Another method (which we have checked to give the same result) for determining
the $r_{12}^{\rm ADM} \to r_{\rm EOB}$ transformation does not need to use
Eq.~(6.22) of Ref. \cite{Buonanno:1998gg}. It consists of directly eliminating the
dimensionless angular momentum $j$ between the two relations $r^{\rm ADM} =
r^{\rm ADM} (j)$ and $r^{\rm EOB} = r^{\rm EOB} (j)$. The former relation was
derived at 3PN in Ref.~\cite{Damour:1999cr} and reads, at 2PN,
\begin{equation}
\label{eq5.25}
r_{12}^{\rm ADM} = \frac{GM}{c^2} \, j^2 \left[ 1 - \frac{4}{j^2} 
- \frac18 \, (74-43 \, \nu) \, \frac{1}{j^4} \right] \, ,
\end{equation}
while the latter one is obtained by inverting the 2PN-accurate version of
Eq.~(\ref{eq5.18}), namely, using $A_{\rm 2PN} (u) = 1-2 \, u + 2 \, \nu \,
u^3$:
\begin{equation}
\label{eq5.26}
\frac{1}{j^2} = \frac{u(1-3 \, u + 5 \, \nu \, u^3)}{1-3 \, \nu \, u^2} \, .
\end{equation}
Inserting Eq. (\ref{eq5.26}) into Eq. (\ref{eq5.25}) yields (at 2PN)
\begin{equation}
\label{eq5.27}
\frac{GM}{c^2 \, r_{12}^{\rm ADM}} = u \left[ 1+u+\left( \frac54 
- \frac{19}{8} \, \nu \right) u^2 \right] \, .
\end{equation}
Then, combining Eq.~(\ref{eq5.27}) and Eq.~(\ref{eq5.24}) yields the looked
for transformation $r^{\rm EOB} \to r_{12}^h$, at 2PN accuracy,
\begin{equation}
\label{eq5.28}
r_{12}^h + \frac{GM}{c^2} = r^{\rm EOB} \left( 1
+6 \, \nu \left( \frac{GM}{c^2 \, r^{\rm EOB}} \right)^2 \right) \, ,
\end{equation}
or, setting $u_h \equiv GM / (c^2 \, r_{12}^h)$ by analogy with 
$u \equiv GM / (c^2 \, r_{\rm EOB})$,
\begin{equation}
\label{eq5.29}
u_h = \frac{u}{1-u} \, (1-6 \, \nu \, u^2) \, .
\end{equation}
We have written the transformation of Eqs. (\ref{eq5.28}), (\ref{eq5.29}) so as to
exhibit the exact form of the transformation $r_h \to r_{\rm EOB}$ in the
extreme mass ratio limit $\nu \to 0$, namely $r_h = r_{\rm EOB} - GM / c^2 +
O(\nu)$.

\bigskip

Summarizing: The (first-order) tidal contribution $\delta A(u) = \mu_1
A_{\mu_1} (u)$ to the main EOB radial potential, associated with any tidal
parameter $\mu_1$ ($=\mu_1^{(2)} , \mu_2^{(2)} , \mu_1^{(3)} , \ldots$), is
given in terms of the corresponding harmonic-coordinate tidal contribution to
the action $\delta L (y_h , v_h) = \mu_1 L_{\mu_1} (y_h , v_h)$, for circular
motion, by Eq.~(\ref{eq5.19}), where $F(u)$ is given (at 2PN) by
Eq.~(\ref{eq5.23}), and where the transformation between the harmonic radial
separation $r_{12}^h$ and the EOB radial coordinate $r_{\rm EOB} \equiv GM /
(c^2 \, u)$ is given by Eqs.~(\ref{eq5.28}) or (\ref{eq5.29}).

\section{EOB description of tidal actions}\label{sec6}
\setcounter{equation}{0}

\subsection{Tidal actions for comparable-mass systems}

We have explained in the previous section how to convert each contribution
$\sim \mu_1 \, L_{\mu_1} (y_h , v_h)$ to the (reduced) tidal action into a
corresponding additional contribution $\mu_1 A_{\mu_1} (u)$ to the main EOB
radial potential $A(u)$. For instance, if we consider the dominant tidal
parameter, i.e. the electric quadrupolar one, $\mu_1^{(\ell = 2)}$ (or
$\mu_2^{(\ell=2)}$, after exchanging $1 \leftrightarrow 2$), the combination
of the result of Eq. (\ref{eq4.2}) for the associated Lagrangian, with
Eq.~(\ref{eq5.19}) yields
\begin{equation}
\label{eq6.1}
\mu_1^{(2)} A_{\mu_1^{(2)}} (u) = - \, \frac{1}{2 \, c^2} \, 
\frac{\mu_1^{(2)}}{M \nu} \, \sqrt{F(u)} \ \frac{d\tau_1}{dt} \, 
[G_{\alpha\beta} \, G^{\alpha\beta}]_1 \, .
\end{equation}
In other words, apart from a (negative) numerical coefficient, and the
rescaled tidal parameter $\mu_1^{(2)} / (M\nu)$ (where $M\nu = \mu = m_1 \, m_2
/ (m_1 + m_2)$ is the reduced mass of the system), the corresponding tidal
contribution to $A(u)$ is the product of {\it three factors}: $\sqrt{F(u)}$,
$d\tau_1 / dt$ and the geometrical invariant associated with the considered
tidal parameter, e.g., $[G_{\alpha\beta} \, G^{\alpha\beta}]_1$ for the
electric quadrupole along the first worldline. In addition, two of these
factors, $d\tau_1 / dt$ and the geometrical invariant, must be reexpressed as
functions of the EOB coordinates by using Eq.~(\ref{eq5.28}).

\smallskip

Let us start by applying this procedure to the dominant tidal action term: the
electric-quadrupole one in Eq. (\ref{eq6.1}). We have given above, in
Eq.~(\ref{eq4.14}), the 2PN-accurate value of $J_{2e} \equiv [G_{\alpha\beta} \,
G^{\alpha\beta}]_1$ in harmonic coordinates. Using the transformation of Eq. 
(\ref{eq5.29}) to replace $1/r_{12}^h$ in terms of $1/r_{\rm EOB}$ leads to
\begin{eqnarray}
\label{eq6.2}
J_{2e}^{\rm (circ)} &=& \frac{6M^2X_2^2}{r_{\rm EOB}^6}\left[1
+ \epsilon^2\frac{(X_1+3)M}{r_{\rm EOB}}\right.\nonumber\\
&& \left.+\epsilon^4 \frac{M^2}{28r_{\rm EOB}^2}(295 X_1^2-7X_1+336) \right] \, .
\end{eqnarray}
In addition the reexpression of the time-dilation factor $d\tau_1 / dt$,
Eq.~(\ref{eq4.21}), in terms of $1/r_{\rm EOB}$ yields
\begin{align}
\label{eq6.3}
\frac{d\tau_1}{dt}&=\frac{1}{\Gamma_1}=1
-\frac12(X_1-1)(X_1-3)u\epsilon^2\nonumber\\
&+\frac38 u^2 (X_1-1)(X_1^3-3X_1^2+3X_1+3)\epsilon^4\,.
\end{align}
Their product yields the electric-quadrupole tidal Lagrangian (stripped of its
prefactor $\frac14 \, \mu_1^{(2)}$) in EOB coordinates, at the 2PN accuracy,
namely
\begin{equation}
\label{eq6.4}
G_{ab}^2\frac{d\tau_1}{dt}= \frac{J_{2e}^{\rm (circ)}}{\Gamma_1} 
= \frac{6(X_1-1)^2u^6}{M^4}\hat {\mathcal L}_{2e} \, ,
\end{equation}
where
\begin{eqnarray}
\label{eq6.5}
\hat {\mathcal L}_{2e} &=&  1-\frac12 u(X_1^2-6X_1-3)\epsilon^2\\
&+& \frac{u^2}{56}(21 X_1^4-112X_1^3+744X_1^2+238X_1+357)\epsilon^4  
\nonumber \, .
\end{eqnarray}
Adding the further factor $\sqrt{F(u)}$, as well as the prefactor, leads to
the corresponding contribution to the EOB $A$ potential, namely
\begin{equation}
\label{eq6.6}
\mu_1^{(2)} A_{\mu_1^{(2)}} (u) = A_{\rm 1 \, electric}^{(2){\rm LO}} 
(r_{\rm EOB}) \, \hat A_{\rm 1 \, electric}^{(2)} (u) \, ,
\end{equation}
where
\begin{equation}
\label{eq6.7}
A_{\rm 1 \, electric}^{(2) {\rm LO}} (r_{\rm EOB}) = - \, \frac{3 \, G^2}{c^2}
 \, \frac{\mu_1^{(2)} M}{\nu} \, \frac{X_2^2}{r_{\rm EOB}^6} \, ,
\end{equation}
and
\begin{equation}
\label{eq6.8}
\hat A_{\rm 1 \, electric}^{(2)} (u) = \sqrt{F(u)} \, \hat{\mathcal L}_{2e} 
= 1+ \alpha_1^{2e} \, u + \alpha_2^{2e} \, u^2 + O(u^3) \, ,
\end{equation}
with
\begin{eqnarray}
\label{eq6.9}
\alpha^{2e}_1&=& \frac{5}{2}X_1 \, ,\\
\label{eq6.10}
\alpha^{2e}_2&=& \frac{337}{28}X_1^2+\frac{1}{8}X_1+3 \, .
\end{eqnarray}
The leading-order (i.e., Newtonian-level) $A$ potential of Eq. (\ref{eq6.7}) is
equivalent to Eqs.~(\ref{eq1.6}) and (\ref{eq1.7}) above (i.e., Eqs.~(23), (25) of Ref. 
\cite{Damour:2009wj}), using the link
\begin{equation}
\label{eq6.11}
G \mu_A^{(\ell)} = \frac{1}{(2\ell - 1)!!} \, 2 \, k_A^{(\ell)} \ R_A^{2\ell + 1} \, .
\end{equation}
The term of order $u$ (i.e., 1PN) in the relativistic amplification factor
$\hat A_{\rm 1 \, electric}^{(2)} (u)$, Eq.~(\ref{eq6.8}), coincides with the
result computed some time ago (see Eq.~(38) in Ref. \cite{Damour:2009wj}). By
contrast, the (2PN) term of order $u^2$ in $\hat A_{\rm 1 \, electric}^{(2)}
(u)$ is the main new result of our present work. Let us discuss its
properties.

\smallskip

Similarly to the 1PN coefficient $\alpha_1^{2e} = \frac52 \, X_1$, which was
positive, and monotonically increasing (from $0$ to $5/2$) as $X_1 \equiv m_1
/ M$ varies between $0$ and $1$, the 2PN coefficient $\alpha_2^{2e}$ is also
positive, and increases as $X_1$ varies between $0$ and $1$. When $X_1 = 0$
(i.e. in the limit $m_1 \ll m_2$), $\alpha_2^{2e}$ takes the value $+ \, 3$, while
when $X_1 = 1$ (i.e., in the limit $m_1 \gg m_2$), it takes the value
\begin{equation}
\label{eq6.12}
\alpha_2^{2e} (X_1 = 1) = \frac{849}{56} = 15.16071429 \, .
\end{equation}
Note that this is about 5 times larger than its value when $X_1 = 0$. Of
most interest (as neutron stars are expected to have rather similar masses
$\sim 1.4 \, M_{\odot}$) is the equal-mass value of $\alpha_2^{2e}$, which is
\begin{equation}
\label{eq6.13}
\alpha_2^{2e} \left( X_1 = \frac12 \right) = \frac{85}{14} = 6.071428571 \, .
\end{equation}

In other words, the distance-dependent amplification factor of the electric
quadrupole reads, in the equal-mass case
\begin{eqnarray}
\label{eq6.13bis}
\left[ \hat A_{\rm 1 \, electric}^{(2)} (u) \right]^{\rm equal\mbox{-}mass} 
&= &1 + \frac54 \, u + \frac{85}{14} \, u^2 + O(u^3) \nonumber \\
&= &1 + 1.25 \, u + 6.071429 \, u^2 \nonumber\\
&& + O(u^3) \, .
\end{eqnarray}
We will comment further  on these results for $\hat A_{\rm 1 \,
  electric}^{(2)} (u)$ and on the recent comparisons
between numerical simulations and the EOB description of tidal interactions below.
For the time being, let us give the corresponding results of our analysis for
some of the sub-leading tidal interactions.

\smallskip

The EOB-coordinate value of the electric octupole invariant, $J_{3e}^{({\rm circ})}$,
Eq.~(\ref{eq4.19}), reads
\begin{eqnarray}
\label{eq6.14}
J_{3e}^{\rm (circ)} &=& \frac{90X_2^2 M^2}{r_{\rm EOB}^8}\left[1+\epsilon^2 
(6X_1+1)\frac{M}{r_{\rm EOB}}\right.\nonumber\\
&& \left. +\epsilon^4  \frac{M^2}{3r_{\rm EOB}^2} (83 X_1^2+14 X_1+17)
\right] \, .
\end{eqnarray}
Its corresponding action (stripped of its prefactor) is
\begin{equation}
\label{eq6.15}
G_{abc}^2\frac{d\tau_1}{dt} = \frac{J_{3e}^{\rm (circ)}}{\Gamma_1} 
= \frac{90X_2^2 u^8}{M^6}\hat {\mathcal L}_{3e}
\end{equation}
with
\begin{eqnarray}
\label{eq6.16}
\hat {\mathcal L}_{3e} &=& 1- \frac{1}{2}(X_1^2-16X_1+1) u\epsilon^2\\
&+&  \frac{1}{24}(9X_1^4-108X_1^3+994X_1^2-56X_1+73) u^2 \epsilon^4  \nonumber
\end{eqnarray}
while the corresponding contribution to the EOB $A$ potential reads
\begin{equation}
\label{eq6.17}
\mu_1^{(3)} A_{\mu_1^{(3)}} (u) = A_{\rm 1 \, electric}^{(3){\rm LO}} 
(r_{\rm  EOB}) \, \hat A_{\rm 1 \, electric}^{(3)} (u) \, ,
\end{equation}
where
\begin{equation}
\label{eq6.18}
A_{\rm 1 \, electric}^{(3){\rm LO}} (r_{\rm EOB}) = - \, \frac{15 \, G^2}{c^2}
 \, \frac{\mu_1^{(3)} \, M}{\nu} \, \frac{X_2^2}{r_{\rm EOB}^8} \, ,
\end{equation}
and
\begin{equation}
\label{eq6.19}
\hat A_{\rm 1 \, electric}^{(3)} (u) = \sqrt{F(u)} 
\, \hat {\mathcal L}_{3e} = 1 + \alpha_1^{3e} \, u + \alpha_2^{3e} \, u^2 + O(u^3) \, ,
\end{equation}
with
\begin{eqnarray}
\label{eq6.20}
\alpha^{3e}_1&=& \frac{15}{2}X_1-2 \, , \\
\label{eq6.21}
\alpha^{3e}_2&=& \frac{110}{3}X_1^2-\frac{311}{24}X_1+\frac{8}{3} \, .
\end{eqnarray}
Here, both results in Eqs. (\ref{eq6.20}) and (\ref{eq6.21}) are new. Note that,
contrary to the quadrupolar case where $\alpha_1$ and $\alpha_2$ were always
both positive (so that $\hat A_{\rm 1 \, electric}^{(2)} (u)$ was always an
{\it amplification} 
factor) the electric-octupole factor $\hat A_{\rm 1 \, electric}^{(3)} (u)$ is
smaller than $1$ (for large separations) when $X_1 < \frac{4}{15} \simeq
0.2667$. Moreover, while the $X_1$-variation of $\alpha_1^{3e}$ is monotonic
(going from $- \, 2$ to $\frac{11}{2}$ as $X_1$ increases from $0$ to $1$),
$\alpha_2^{3e} (X_1)$ first decreases from $\alpha_2^{3e} (0) = \frac83 = 2.666667$
to $\alpha_2^{3e} (X_1^{\rm min}) = 42853/28160 = 1.521768$ as $X_1$ increases
from $0$ to $X_1^{\rm min} = 311/1760 = 0.1767046$, before increasing as $X_1$
goes from $X_1^{\rm min}$ to $1$, to reach the final value $\alpha_2^{3e} (1) =
211/8 = 26.375$ for $X_1 = 1$. Note, however, that when (as expected) the two
masses are nearly equal the factor $\hat A_{\rm 1 \, electric}^{(3)} (u)$ is
an amplification factor. In particular, its equal-mass value is
\begin{eqnarray}
\label{eq6.22}
\left[\hat A_{\rm 1 \, electric}^{(3)} (u) \right]^{\rm equal\mbox{-}mass} &=
&1 + \frac74 \, u + \frac{257}{48} \, u^2 + O(u^3) \nonumber \\
&= &1 + 1.75 \, u + 5.354167 \, u^2 \nonumber\\
&& + O(u^3)
\end{eqnarray}
which is similar to its corresponding quadrupolar counterpart,
Eq.~(\ref{eq6.13bis}).

\smallskip

Let us finally give the corresponding results for the magnetic quadrupole and
time-differentiated electric quadrupole. For the magnetic quadrupole (at the
1PN fractional accuracy), we found
\begin{eqnarray}
\label{eq6.23}
\frac14 H_{ab}^2 &\equiv& J_{2m}^{\rm (circ)} \\
&=& \frac{18X_2^2 M^3}{r_{\rm EOB}^7}\left[1
+\epsilon^2\frac{M}{3r_{\rm EOB}}(3X_1^2+X_1+12) \right] \, , \nonumber
\end{eqnarray}
\begin{equation}
\label{eq6.24}
\frac14 H_{ab}^2 \, \frac{d\tau_1}{dt} \equiv 
\frac{18 \, X_2^2}{M^4} \, u^7 \hat{\mathcal L}_{2m} \, ,
\end{equation}
\begin{equation}
\label{eq6.27}
\hat {\mathcal L}_{2m} = 1+ \frac16 (X_1+3)(3X_1+5) u\epsilon^2 \, , 
\end{equation}
\begin{equation}
\label{eq6.28}
\hat A_{\rm 1 \, magnetic}^{(2)} (u) = \sqrt{F(u)} \, \hat {\mathcal L}_{2m} 
= 1+\alpha_1^{2m} \, u + O(u^2) \, ,
\end{equation}
with
\begin{equation}
\label{eq6.29}
\alpha^{2m}_1 = X_1^2+\frac{11}{6}X_1+1 \, .
\end{equation}
Here $\alpha_1^{2m} (X_1)$ is always positive, and monotonically increases from
$\alpha_1^{2m} (0) = 1$ to $\alpha_1^{2m} (1) = \frac{23}{6} = 3.833333$, its
equal-mass value being $\alpha_1^{2m} \left( \frac12 \right) = \frac{13}{6} =
2.166667$.

\smallskip

Finally, for the time-differentiated electric quadrupole, we got
\begin{equation}
\label{eq6.30}
\dot G_{ab}^2 = J_{\dot 2 e}^{\rm (circ)} = \frac{18X_2^2 M^3}{r_{\rm EOB}^9}
\left[1+\epsilon^2 (X_1^2+2)\frac{M}{r_{\rm EOB}}\right] \, ,
\end{equation}
\begin{equation}
\label{eq6.31}
\dot G_{ab}^2 \, \frac{d\tau_1}{dt} = \frac{18 \, X_2^2}{M^6} \, 
u^9 \, \hat{\mathcal L}_{\dot 2 e} \, ,
\end{equation}
\begin{equation}
\label{eq6.32}
\hat {\mathcal L}_{\dot 2 e} = 1 +\frac12 u\epsilon^2  (X_1^2+4X_1+1) \, ,
\end{equation}
\begin{equation}
\label{eq6.33}
\hat A_{1\dot G}^{(2)} (u) = \sqrt{F(u)} \, \hat{\mathcal L}_{\dot 2 e} = 1 
+ \alpha_1^{\dot 2 e} \, u + O(u^2) \, ,
\end{equation}
with
\begin{equation}
\label{eq6.34}
\alpha^{\dot 2 e}_1 = \frac12 (X_1+2)(2X_1-1) \,.
\end{equation}

\subsection{Tidal actions of a tidally-deformable test mass}

One of the characteristic features of the EOB formalism for point-mass systems
is the natural incorporation of the exact test-mass limit $\nu \to 0$. Indeed,
in this limit the effective metric in Eq. (\ref{eq5.2}) describing the relative
dynamics reduces to the Schwarzschild metric: $\lim_{\nu \to 0} A(u) = 1-2\, u
= \left( \lim_{\nu \to 0} \bar B (u) \right)^{-1}$. Let us study the test-mass
limit of tidal effects, with the aim of incorporating it similarly in their
EOB description. When considering the nonminimal worldline action of particle
$1$, the simplest test-mass limit to study is the limit $m_1 / m_2 \to 0$.
[When considering tidal effects within body 2, the permutation $1
\leftrightarrow 2$ of our results below allow them to describe the limit $m_2
/ m_1 \to 0$. We leave to future work a study of the limit $m_2 / m_1 \to 0$,
when considering tidal effects taking place within body 1.] In the limit
investigated here, one
is considering a tidally deformable test-mass $(m_1 , \mu_1^{(\ell)} ,
\ldots)$ moving around a large mass $m_2 \gg m_1$. The effective action of
body 1 is then exactly obtained by evaluating the $A=1$ contribution of the
general (two-body) effective action of Eq. (\ref{eq2.12}) within the background
metric generated by the (non-tidally deformable) large mass $m_2$, at rest,
i.e. within a Schwarzschild metric of mass $m_2$. The latter reads
\begin{eqnarray}
\label{eq6.35}
ds^2 (m_2) &=& - \left( 1-2 \, \frac{G \, m_2}{c^2 \, r_s} \right) c^2 \, dt^2 
+ \frac{dr_s^2}{1-2 \, \frac{G \, m_2}{c^2 \, r_s}} \nonumber\\
&&+ r_s^2 (d\theta^2 + \sin^2 \theta \, d\varphi^2)
\end{eqnarray}
in ``Schwarzschild'', or areal, coordinates, and
\begin{align}
\label{eq6.36}
ds^2 (m_2) &= - \, \frac{1- \frac{G \, m_2}{c^2 \, r_h}}
{1+\frac{G \, m_2}{c^2 r_h}} \, c^2 \, dt^2 
+  \frac{1+ \frac{G \, m_2}{c^2 \, r_h}}{1- \frac{G \, m_2}{c^2 r_h}} 
\, dr_h^2 \nonumber\\
& + \left( r_h + \frac{G \, m_2}{c^2} \right)^2 
(d\theta^2 + \sin^2 \theta \, d\varphi^2)
\end{align}
in harmonic coordinates: $r_h = r_s - G \, m_2 / c^2$. As a check on the
results below (and on our codes), we have computed them both in Schwarzschild
coordinates and in harmonic ones.

\smallskip

The geometrical invariants $J_{2e} = G_{ab}^2$, etc., take the following values in
this Schwarzschild limit, and when considering as above circular motions (we
again set $G$ and $c$ to one for simplicity):
\begin{eqnarray}
\label{eq6.37}
G_{ab}^{\rm (S)}{}^2=\bar J_{2e}^{\rm (S)}&=&
\frac{6m_2^2(m_2^2+r^2_h-m_2r_h)}{(r_h-2m_2)^2(r_h+m_2)^6} \nonumber \\
&\sim &\frac{6m_2^2}{r_h^6}\left[1
-\frac{3m_2}{r_h}+\frac{12m_2^2}{r_h^2}+\ldots  \right]\nonumber\\
&=& \frac{6u_S^6}{m_2^4}\frac{1}{(1-3u_S)}
\left[ 1+\frac{3u_S^2}{(1-3u_S)} \right]\nonumber\\
&=& \frac{6u_S^6}{m_2^4}\left[ 1+3u_S\frac{(1-2u_S)}{(1-3u_S)^2} \right] \, ,\\
\label{eq6.38}
\frac14 H_{ab}^{\rm (S)}{}^2=\bar J_{2m}^{\rm (S)}&
=&\frac{18m_2^3(r_h-m_2)}{(r_h-2m_2)^2(r_h+m_2)^6}\nonumber\\
&\sim& \frac{18m_2^3}{r_h^7}\left[1-\frac{3m_2}{r_h}
+\frac{11m_2^2}{r_h^2}+\ldots  \right]\nonumber\\
&=& \frac{18u_S^7}{m_2^4}\left[1+\frac{u_S(4-9u_S)}{(1-3u_S)^2}\right] \, ,
\end{eqnarray}
\begin{align}
\label{eq6.39}
G_{abc}^{\rm (S)}{}^2=\bar J_{3e}^{\rm (S)} &= 
\frac{30 m_2^2(r_h-m_2)(2m_2^2+3r_h^2-3m_2r_h)}
{(r_h-2m_2)^2(r_h+m_2)^9}\nonumber\\
&\sim  \frac{90 m_2^2}{r_h^8}\left[1-7 \frac{m_2}{r_h}
+\frac{98}{3} \frac{m_2^2}{r_h^2}+\ldots \right]\,\nonumber\\
&= \frac{90u_S^8}{m_2^6}\frac{(1-2u_S)}{(1-3u_S)}
\left[1+\frac{8u_S^2}{3(1-3u_S)}  \right] \, ,\\
\label{eq6.40}
\dot G_{ab}^{\rm (S)}{}^2=\bar J_{\dot 2 e}^{\rm (S)}&= 
\frac{18m_2^3(r_h-m_2)^2}{(r_h-2m_2)^2(r_h+m_2)^9}\nonumber\\
&= \frac{18m_2^3}{r_h^9}\left[1-7\frac{m_2}{r_h}
+32\frac{m_2^2}{r_h^2}+\ldots\right]\nonumber\\
&= \frac{18u_S^9}{m_2^6}\frac{(1-2u_S)^2}{(1-3u_S)^2} \, ,
\end{align}
where $u_S \equiv G \, m_2 / (c^2 \, r_s)$. 
We have indicated above the expansions in powers of the inverse
harmonic radius $r_h$ as checks of our 2PN-accurate results, written in
harmonic coordinates; see Eqs.~(\ref{eq4.14}), (\ref{eq4.18})--(\ref{eq4.20}).

In the following, we shall focus on the transformation of the exact test-mass
geometrical invariants above into corresponding contributions to the EOB
$A$ potential. As explained previously, Eqs.~(\ref{eq5.19}), (\ref{eq6.1}), apart
from the universal prefactor $- \, 2 / (M \, \nu \, c^2)$ and the specific
original tidal coefficient multiplying the considered geometrical invariant
(such as $\frac14 \, \mu_1^{(2)}$ for the electric quadrupole), the
contribution to $A(u)$ associated with some given invariant is obtained by
multiplying it by two extra factors: (i) the time-dilation factor $d\tau_1 /
dt$ and (ii) the EOB-rooted factor $\sqrt{F(u)}$. Let us discuss their values
in the test-mass limit $m_1 \ll m_2$ that we are now considering.

\smallskip

The first factor is the square-root of
\begin{equation}
\label{eq6.41}
\left( \frac{d\tau_1}{dt} \right)^2 = 1 - \frac{2 \, G \, m_2}{c^2 \, r_s} 
- \frac{1}{c^2} \, r_s^2 \left( \frac{d\varphi}{dt} \right)^2 \, .
\end{equation}
Denoting, as above, $u_S \equiv G \, m_2 / (c^2 \ r_s)$, and using the
well-known Kepler law for circular orbits in Schwarzschild coordinates,
$\Omega^2 = G \, m_2 / r_s^3$, simply yields
\begin{equation}
\label{eq6.42}
\left( \frac{d\tau_1}{dt} \right)_{\rm circ}^{\rm test\mbox{-}mass} 
= \sqrt{1-3 \, u_S} \, .
\end{equation}

The exact test-mass limit of the second factor is obtained by taking the limit
$\nu \to 0$ in the exact expression of Eq. (\ref{eq5.22}). In this limit, $A(u) \to
1-2 \, u$, so that $\tilde A (u) \to 1-3 \, u$, and
\begin{equation}
\label{eq6.43}
\left( \sqrt{F(u)} \right)_{\rm circ}^{\rm test\mbox{-}mass} = \sqrt{1-3 \, u} \, .
\end{equation}

In addition, as the EOB coordinates reduce to Schwarzschild coordinates in the
test-mass limit $\nu \to 0$, and $M = m_1 + m_2 \to m_2$, we have simply
\begin{equation}
\label{eq6.44}
u_S \equiv \frac{G \, m_2}{c^2 \, r_s} \to u \equiv 
\frac{GM}{c^2 \, r_{\rm EOB}} \, .
\end{equation}
In other words, the two extra factors in Eqs. (\ref{eq6.42}), (\ref{eq6.43}) become
both equal to $\sqrt{1-3 \, u}$. As a consequence the $A$ contribution
corresponding to the various geometrical invariants of Eqs. 
(\ref{eq6.37})--(\ref{eq6.40}) is obtained (apart from a constant prefactor)
by multiplying these invariants by $\left( \sqrt{1-3 \, u} \right)^2 = 1-3 \,
u = 1-3 \, u_S$. Including the universal factor $-2 / (M \, \nu \, c^2)$ and the
various tidal coefficients $\frac12 \, \frac{1}{\ell!} \, \mu_1^{(\ell)}$,
$\frac12 \, \frac{\ell}{\ell +1} \,\frac{1}{\ell!} \,
\frac{\sigma_1^{(\ell)}}{c^2} , \ldots$ (as well as the factor 4 in $H_{ab}^2
= 4 J_{2m}$) yields the following exact, test-mass contributions
\begin{equation}
\label{eq6.45}
\mu_1^{(2)} \, A_{\mu_1^{(2)}}^{\rm test\mbox{-} mass} (u) 
= - \, 3 \, \frac{G^2}{c^2} \, \frac{\mu_1^{(2)}}{m_1} 
\, \frac{(m_2)^2}{r_{\rm EOB}^6} 
\left( 1 + \frac{3 \, u^2}{1-3 \, u} \right) \, ,
\end{equation}
\begin{eqnarray}
\label{eq6.46}
\mu_1^{(3)} \, A_{\mu_1^{(3)}}^{\rm test\mbox{-} mass} (u) &=& 
- \, 15 \, \frac{G^2}{c^2} \, \frac{\mu_1^{(3)}}{m_1} 
\, \frac{(m_2)^2}{r_{\rm EOB}^8} \, (1-2 \, u) \times \nonumber\\
&& \quad
\times \left( 1 + \frac83 \, \frac{u^2}{1-3 \, u} \right) \, ,
\end{eqnarray}
\begin{equation}
\label{eq6.47}
\sigma_1^{(2)} \, A_{\sigma_1^{(2)}}^{\rm test\mbox{-}mass} (u) 
= - \, 24 \, \frac{G^3}{c^4} \, \frac{\sigma^{(2)}_1}{m_1} \, 
\frac{(m_2)^3}{r_{\rm EOB}^7} \, \frac{1-2 \, u}{1-3 \, u} \, ,
\end{equation}
\begin{equation}
\label{eq6.48}
\mu'^{(2)}_1 \, A_{\mu'^{(2)}_1}^{\rm test\mbox{-}mass} (u) 
= - \, 9 \, \frac{G^3}{c^4} \, \frac{\mu'^{(2)}_1}{m_1} 
\, \frac{(m_2)^3}{r_{\rm EOB}^9} \, \frac{(1-2 \, u)^2}{1-3 \, u} \, .
\end{equation}

One easily sees that the various exact, test-mass amplification factors $\hat
A (u)$ exhibited here are compatible with the $X_1 \to 0$ limit of the
2PN-expanded ones $\sim 1 + \alpha_1 u + \alpha_2 \, u^2 + O(u^3)$ derived
above.

\subsection{Light-ring behavior of test-mass tidal actions}

A striking feature of all the amplification factors present in
Eqs.~(\ref{eq6.45})--(\ref{eq6.48}), such as
\begin{equation}
\label{eq6.49}
\hat A_{\rm 1 \, electric}^{\rm (2) \, test\mbox{-}mass} (u) = 1+3 \, \frac{u^2}{1-3 \, u} \, ,
\end{equation}
is that they all formally exhibit a pole $ \propto 1/(1-3 \, u)$
mathematically located at $3 \, u = 1$, i.e. corresponding to formally letting
particle 1 tend to the last unstable circular orbit, located at $3 \, G \, m_2
/ c^2$ (``light-ring'' orbit). This behavior has a simple origin.

\smallskip

The invariant that is simplest to consider in order to see this is $J_{2e} = G_{ab}^2$.
From Eq.~(\ref{eq4.3}) its covariant expression reads
\begin{equation}
\label{eq6.50}
G_{ab}^2 = R_{\alpha\mu\beta\nu} \, R_{ \, \bullet \, \kappa \, \bullet \,
  \lambda}^{\alpha \ \, \beta} \ u^{\mu} \, u^{\nu} \, u^{\kappa} \,
u^{\lambda} \, .
\end{equation}
Let us study its mathematical behavior in the formal limit where particle 1
tends to the light-ring orbit. Using the language of Special Relativity, we
consider the Schwarzschild coordinates as defining a ``lab-frame.'' With
respect to this lab-frame, particle $1$ becomes ultra-relativistic as it
approaches the light ring. More precisely, near the light ring the lab-frame
components of the 4-velocity $u^{\mu} = (dt / d\tau_1) (c,v^i)$ tend
towards infinity proportionally to $dt/d\tau_1 = \Gamma_1 = 1/\sqrt{1-3 \,
  u}$, while the lab-frame components of $R_{\alpha\mu\beta\nu}$ (and of the
metric) stay finite. As $G_{ab}^2$ is quartic in the lab-frame components of
$u^{\mu}$, it will tend towards infinity like $\Gamma_1^4 = (dt/d\tau_1)^4 =
(1-3 \, u)^{-2}$. The corresponding contribution to $A(u)$ is obtained by
multiplying $G_{ab}^2$ by the factor $(d\tau_1 / dt)^2 = \Gamma_1^{-2} = (1-3
\, u)^{+1}$, which reduces the blow-up of $G_{ab}^2$ to the milder $(1-3 \,
u)^{-2+1} = (1-3 \, u)^{-1}$ behavior that is apparent in Eqs.~(\ref{eq6.45})
or (\ref{eq6.49}).

\smallskip

A different way of phrasing this result uses the law of transformation of the
electric and magnetic components of the Weyl tensor, $G_{ab}$ and $H_{ab}$,
under a boost. Using, for instance, the fact that, under a boost with velocity
$\beta = \tanh \varphi$ in the $x$ direction, the complex tensor $F_{ab} =
G_{ab} + i \, H_{ab}$ undergoes a complex rotation of angle $\psi = i \,
\varphi$ in the $yz$ plane \cite{LandauLifshitz}, one easily finds that the
transverse traceless components of $F_{ab}$ (in the $yz$ plane) acquire, under
such a boost, a factor of order $\cos^2 \psi = \cosh^2 \phi
=(1-\beta^2)^{-1} \equiv \Gamma_1^2$. Because of the special structure of the
tensor $F_{ab} \propto {\rm diag} \, (-1,-1,2)$, with the third axis $z$
labelling the radial direction, this reasoning shows that boosts in the radial
$(z)$ direction leave $F_{ab}$ invariant. However, we are mainly interested
here in boosts in a \lq\lq tangential" direction, say $x$, associated with the
fast motion of a circular orbit, and therefore orthogonal to the radial
direction, which do introduce a factor $\Gamma_1^2$ in some of the boosted
components of $F_{ab}$. For completeness, let us indicate that because of
this special structure of $F_{ab}$, the invariant $J_{2e} = G_{ab}^2$ for
general, {\it non-circular} orbits is equal to
\begin{equation}
\label{eq6.50new}
J_{2e} = G_{ab}^2 = \frac{6m_2^2}{r_s^6} \left( 1 + 3 {\bf u}_{\rm tg}^2 
+ 3 {\bf u}_{\rm tg}^4  \right) \, ,
\end{equation}
where ${\bf u}_{\rm tg}^2 \equiv r_s^2 ( (u^\theta)^2 + \sin^2 \theta
(u^\phi)^2 )$ is the square of the part of the 4-velocity $u^\mu$ that is
tangent to the sphere. [The radial component of the 4-velocity brings no
contribution to $J_{2e}$.] \smallskip

The behavior near the light ring of the magnetic-quadrupole invariant $J_{2m} =
\frac14 H_{ab}^2$ is understood in the same way as that of $J_{2e} = G_{ab}^2$.
Concerning the other invariants, one can note that $J_{3e} = G_{abc}^2$ can be
written as the sum
\begin{equation}
\label{eq6.51}
J_{3e} = G_{abc}^2 = C_{\alpha\beta\gamma} \, C^{\alpha\beta\gamma} + \frac{1}{3 \, c^2} \, J_{\dot 2 e}
\end{equation}
where
\begin{equation}
\label{eq6.52}
C_{\alpha\beta\gamma} = {\rm Sym}_{\alpha\beta\gamma} \, \nabla_{\alpha} \, 
R_{\beta\mu\gamma\nu} \, u^{\mu} \, u^{\nu} \, ,
\end{equation}
and
\begin{equation}
\label{eq6.53}
J_{\dot 2 e} = \dot G_{ab}^2 = \dot G_{\alpha\beta} \, \dot G^{\alpha\beta} 
\end{equation}
with 
\begin{equation}
\label{eq6.54}
\dot G_{\alpha\beta} = u^{\lambda} \, \nabla_{\lambda} \,
R_{\alpha\mu\beta\nu} \, u^{\mu} \, u^{\nu} \, .
\end{equation}
Similarly to $G_{ab}^2$, Eq.~(\ref{eq6.50}), the term
$C_{\alpha\beta\gamma}^2$ in Eq.~(\ref{eq6.51}) is quartic in $u^{\mu}$ and is
therefore expected to blow up like $\Gamma_1^4$. On the other hand, though
$\dot G_{\alpha\beta}$, Eq.~(\ref{eq6.54}), is cubic in $u^{\mu}$, it only
blows up like $\Gamma_1^2$ (so that $J_{\dot 2 e} \sim \Gamma_1^4$ and $J_{3e} \sim C^2 +
J_{\dot 2 e} \sim \Gamma_1^4$) because of the special geodetic-precession properties of
the proper-time derivative operator $\nabla / d\tau = u^{\lambda} \,
\nabla_{\lambda}$ (see, e.g., Sec.~3.6 of Ref. \cite{Straumann}).

\subsection{A suggested \lq\lq resummed" version of comparable-mass tidal actions}

Having understood that the formal pole-like behavior, $\sim (1-3 \, u)^{-1}$,
in the test-mass limit of the electric-quadrupole $A$ potential is linked to
simple boost properties of $G_{ab}$ near the light-ring orbit, and knowing
that the EOB formalism predicts the existence of a formal analog of the usual
Schwarzschild light ring at the EOB dimensionless radius $\hat r_{\rm LR}
\equiv 1/u_{\rm LR}$, defined as the solution of
\begin{equation}
\label{eq6.55}
\tilde A (u_{\rm LR}) = 0 \, ,
\end{equation}
with $\tilde A (u)$ defined in Eq.~(\ref{eq5.21}), it is natural to expect
the (unknown) exact two-body version of the electric-quadrupole
$A$ potential to mathematically exhibit an analogous pole-like behavior of
the form $\sim (1-\hat r_{\rm LR} \, u)^{-1}$. As we shall discuss elsewhere,
such a mathematical behavior, linked to considering (within the
EOB-simplifying approach advocated in Ref. \cite{Damour:2000we}) what would happen
if one formally considered (unstable) circular orbits with $u \to u_{\rm
  LR}$, does not mean that there is a real physical singularity in the EOB
dynamics near $u = u_{\rm LR}$, but it indicates that higher-than-2PN
contributions to the electric-quadrupole amplification factor $\hat A_{\rm 1
  \, electric}^{(2)} (u) = 1 + \alpha_1^{2e} \, u + \alpha_2^{2e} \,u^2 + \alpha_3^{2e}
\,u^3 + \cdots$ will probably be slowly convergent, and will tend to 
amplify further the corresponding tidal interaction. Such an extra amplification
might, for instance, be physically important in the last orbits of
comparable-mass neutron-star binaries (which will reach contact for values of
$u$ smaller than $ u_{\rm LR}$).

This leads us to suggest that a more accurate value (for $u < u_{\rm LR}$) of
the electric-quadrupole amplification factor is the following ``resummed''
version of Eq.~(\ref{eq6.8}):
\begin{equation}
\label{eq6.56}
\hat A_{\rm 1 \, electric}^{(2)} (u) 
= 1 + \alpha_1^{2e} \, u + \alpha_2^{2e} \, \frac{u^2}{1-\hat r_{\rm LR} \, u} \, ,
\end{equation}
where $ \alpha_1^{2e}$ and $\alpha_2^{2e}$ are given by Eqs.~(\ref{eq6.9}) and
(\ref{eq6.10}), and where $\hat r_{\rm LR} \equiv 1/u_{\rm LR}$ is the
solution of Eq.~(\ref{eq6.55}). Similar resummed versions of the other
amplification factors can be defined by incorporating in their PN-expanded
versions the formal light-ring behaviors exhibited by the exact test-mass
results of Eqs. (\ref{eq6.45})--(\ref{eq6.48}).

\smallskip

Let us finally discuss several possible approximate values for $\hat r_{\rm
  LR}$ in the proposed Eq.~(\ref{eq6.56}). The simplest approximation consists
of using the ``Schwarzschild'' value $\hat r_{\rm LR}^S = 3$. However, a
better value might be obtained by taking a solution of Eq.~(\ref{eq6.55}) that
incorporates more physical effects. This might require solving
Eq.~(\ref{eq6.55}) numerically, with $A(u)$ being the full $A$ potential
(containing both Pad\'e-resummed two-point-mass effects and the various
tidal contributions). In order to have a feeling for the modification of $\hat
r_{\rm LR}$ brought by incorporating these changes, let us consider
solving Eq.~(\ref{eq6.55}) when using the following approximation to the full
$A$ potential:
\begin{equation}
\label{eq6.57}
A_{\rm approx} (u) = 1-2 \, u + 2 \, \nu \, u^3 - \kappa \, u^6 \, 
\end{equation}
where
\begin{eqnarray}
\label{eq6.58}
\kappa &=& \kappa_1^{(2)} + \kappa_2^{(2)} 
= 2 \, k_1^{(2)} \, \frac{m_2}{m_1} 
\left( \frac{R_1 \, c^2}{G(m_1+m_2)} \right)^5 \nonumber\\
&&+ 2 \, k_2^{(2)} \, \frac{m_1}{m_2} 
\left( \frac{R_2 \, c^2}{G(m_1 + m_2)} \right)^5 \, .
\end{eqnarray}
Here, the term $+ \, 2 \, \nu \, u^3$ is the 2PN-accurate point-mass
modification of $A(u)$, while the term $- \kappa \, u^6$ is the leading-order
tidal modification. Note that they have opposite signs. The corresponding
expression of $\tilde A (u)$ reads
\begin{equation}
\label{eq6.59}
\tilde A_{\rm approx} (u) = 1 - 3 \, u + 5 \, \nu \, u^3 - 4 \, \kappa \, u^6 \, .
\end{equation}
The corresponding value of $u_{\rm LR} \equiv 1/\hat r_{\rm LR}$ is the
solution close to $1/3$ of the equation
\begin{equation}
\label{eq6.60}
u_{\rm LR} = \frac13 \left[ 1+5 \, \nu \, u_{\rm LR}^3 
- 4 \, \kappa \, u_{\rm LR}^6 \right] \, .
\end{equation}
If we could treat both $\nu$ and $\kappa$ as small deformation parameters,
this would imply that, to first order in these two deformation parameters, the
value of $u_{\rm LR} (\nu , \kappa)$ would be obtained by inserting the
leading-order value $u_{\rm LR} \simeq 1/3$ in the right-hand side of
Eq.~(\ref{eq6.60}). This would yield
\begin{equation}
\label{eq6.61}
u_{\rm LR} (\nu , \kappa) = \frac13 \left[ 1 + \frac{5}{3^3} \, \nu 
- \frac{4}{3^6} \, \kappa + O(\nu^2 , \nu\kappa , \kappa^2) \right] \, ,
\end{equation}
and
\begin{equation}
\label{eq6.62}
\hat r_{\rm LR} (\nu,\kappa) = 3 \left[ 1-\frac{5}{3^3} \, \nu 
+ \frac{4}{3^6} \, \kappa + O(\nu^2 , \nu\kappa , \kappa^2) \right] \, .
\end{equation}
Note that while comparable-mass corrections $(\propto \nu)$ have the effect of
decreasing $\hat r_{\rm LR}$, tidal ones $(\propto \kappa)$ have the opposite
effect of increasing $\hat r_{\rm LR}$. Let us focus on the tidal effects, and
consider the equal-mass case with $R_1 = R_2$ and $k_1^{(2)}= k_2^{(2)}$. One
has a first order increase of $\hat r_{\rm LR}$ equal to
\begin{equation}
\label{eq6.63}
\delta^{\rm tidal} \, \hat r_{\rm LR} \simeq 16 \, k_1^{(2)} 
\left( \frac{R_1 \, c^2}{6 \, Gm_1} \right)^5 = 16 \, k_1^{(2)} 
\, \frac{1}{(6 \, {\mathcal C}_1)^5} \, ,
\end{equation}
where ${\mathcal C}_1 \equiv Gm_1 / (c^2 \, R_1)$ denotes the common compactness
of the two neutron stars. This simple approximate analytical formula shows
that $\delta^{\rm tidal} \, \hat r_{\rm LR}$ is very sensitive to the value of
the compactness of the neutron star. If ${\mathcal C}_1 = 1/6 = 0.166667$,
i.e., $R_1 = 6 \, Gm_1 / c^2$ (roughly corresponding to a radius of 12~km for a
$1.4 \, M_\odot$ neutron star), then $\delta^{\rm tidal} \, \hat r_{\rm LR} = 1.44
\, (k_1^{(2)} / 0.09)$ will be of order $1$ [the value $k_1^{(2)} = 0.09$
being typical for ${\mathcal C}_1 = 1/6$; see, e.g., Table~II in
Ref. \cite{Damour:2009wj}]. On the other hand, if $\hat R_1 \equiv R_1 \, c^2 /
(Gm_1)$ is slightly smaller than $6$, $\delta^{\rm tidal} \, \hat r_{\rm LR}$
will quickly become much smaller than $1$, while if $\hat R_1$ is slightly
larger than $6$, $\delta^{\rm tidal} \, \hat r_{\rm LR}$ will quickly become
formally large (thereby invalidating the first-order analytical estimate of Eq. 
(\ref{eq6.63}), which assumed $\delta \, \hat r_{\rm LR} \ll 3$). These rough
estimates indicate that in many cases, tidal effects on $\hat
r_{\rm LR}$ will be quite important and will significantly increase the
numerical value of $\hat r_{\rm LR}$. Note that an increased value of $\hat
r_{\rm LR}$ will, in turn, {\it increase} the effect of the conjectured
resummed 2PN contribution $\alpha_2^{2e} \, u^2 / (1-\hat r_{\rm LR} \, u)$ to
$\hat A_{\rm 1 \, electric}^{(2)} (u)$.

\section{Summary and conclusions}\label{sec7}
\setcounter{equation}{0}

Using an effective action technique, we have shown how to compute the
additional terms in the {\it reduced} (Fokker) two-body Lagrangian $L({\bm
  y}_1 , {\bm y}_2 , \dot{\bm y}_1 , \dot{\bm y}_2)$ that are linked to tidal
interactions. Thanks to a general property of perturbed Fokker actions
[explained at the end of Sec.~II, see Eq.~(\ref{eq2.20})], the additional
tidal terms are correctly obtained (to first order in the tidal perturbations)
by replacing in the complete, unreduced action $S [g_{\mu\nu} ; y_1 , y_2]$
the gravitational field $g_{\mu\nu}$ by the solution of Einstein's equations
generated by two structureless point masses $m_1 , {\bm y}_1 ; m_2 , {\bm
  y}_2$. This allowed us to compute in a rather straightforward manner the
reduced tidal action at the  2PN fractional accuracy
by using the known, explicit form of the 2PN-accurate two-point-mass metric
\cite{OOKH73, DamourMG3, Schaefer:1986rd, Blanchet:1998vx}. The main technical
subtlety in this calculation is the regularization of the self-field effects
associated with the computation of the various nonminimal tidal-action terms
$\sim \int d\tau (R_{\alpha\mu\beta\nu} \, u^{\mu} \, u^{\nu})^2 + \ldots$,
where, e.g., $R_{\alpha\mu\beta\nu} (x;y_1,y_2)$ is to be evaluated on one of
the worldlines that generate the metric $g_{\mu\nu}$ (so that
$R_{\alpha\mu\beta\nu} (y_1 ; y_1 , y_2)$ is formally infinite). We explained
in detail (in Sec.~III) one (algorithmic) way to perform this regularization,
using Hadamard regularization (which is equivalent to dimensional
regularization at the 2PN level). We then computed the regular parts of the
brick potentials that parametrize the 2PN metric, from which we derived the
regularized values of several of the geometrical invariants
entering the nonminimal worldline tidal action terms. [See
Eqs.~(\ref{eq4.4})--(\ref{eq4.10}) for the 2PN-accurate Lagrangians (for
general orbits) of the three leading tidal terms (electric quadrupole,
electric octupole and magnetic quadrupole)]. We then focussed on the most physically
 useful information contained in these actions, namely the corresponding
contributions to the  EOB main radial potential, $A(u)$,
with $u = G(m_1 + m_2) / (c^2 \, r_{\rm EOB})$. Our Eqs.~(\ref{eq5.19}),
(\ref{eq5.20}), (\ref{eq5.28}) gave the explicit transformation between the
previously derived harmonic-coordinates tidal Lagrangians and their
corresponding contributions to the EOB $A$ potential. Using this
transformation, we could finally explicitly compute the most important tidal
contributions to the EOB $A$ potential to a higher accuracy than had been
known before: namely, we computed the quadrupolar $(\ell = 2)$ and octupolar
$(\ell = 3)$ gravito-electric tidal contributions to 2PN fractional accuracy,
i.e., with the inclusion of a relativistic distance-dependent factor of the
type $u^{2\ell + 2} (1+\alpha_1 \, u + \alpha_2 \, u^2)$ [see
Eqs.~(\ref{eq6.6})--(\ref{eq6.10}) and (\ref{eq6.17})--(\ref{eq6.21})]. We also
computed the quadrupolar gravito-magnetic tidal contribution, as well as a
newly introduced time-differentiated electric quadrupolar tidal term, to 1PN
fractional accuracy [see Eqs.~(\ref{eq6.24})--(\ref{eq6.29}),
(\ref{eq6.31})--(\ref{eq6.34})]. Of most interest among these results is the
obtention of the 2PN coefficient $\alpha_2^{2e}$ entering the distance-dependence
of the electric quadrupolar term. We found that this coefficient,
Eq.~(\ref{eq6.10}), is always positive and varies between $+ \, 3$ and $+ \,
15.16071$ as the mass fraction $X_1 = m_1 / (m_1 + m_2)$ of the considered
tidally deformed body varies between $0$ and $1$. In the equal-mass case, $m_1
= m_2$, i.e. $X_1 = \frac12$, we found that $\alpha_2^{2e} = 6.07143$. This value
shows that, when the neutron stars near their contact, 2PN effects are
comparable to 1PN ones. Indeed, contact occurs when the separation $r \simeq
R_1 + R_2 = Gm_1 / (c^2 \, {\mathcal C}_1) + Gm_2 / (c^2 \, {\mathcal C}_2)$
(where ${\mathcal C}_A \equiv Gm_A / (c^2 \, R_A)$, $A=1,2$, are the two
compactnesses). In the equal-mass case (with ${\mathcal C}_1 = {\mathcal
  C}_2$), this shows that, at contact, $u = G(m_1+m_2)/(c^2 \, r)$ is
approximately equal to $u_{\rm contact} \simeq {\mathcal C}_1$. If we consider
as typical neutron star a star of mass $1.4 \, M_\odot$ and radius 12~km, we
expect ${\mathcal C}_1 \sim 1/6$, i.e. $u_{\rm contact} \sim 1/6$. The
successive PN contributions to the distance-dependent amplification factor
$\hat A_{\rm 1\, electric}^{(2){\rm 2PN}} (u) = 1+\alpha_1^{2e} \, u + \alpha_2^{2e} \,
u^2$ of the electric quadrupolar tidal interaction for the first body then becomes, at contact,
\begin{align}
\label{eq7.1}
\hat A_{\rm 1\, electric}^{\rm (2) 2PN} (u_{\rm contact}) &\simeq 1 
+ \alpha_1^{2e} \, {\mathcal C}_1 + \alpha_2^{2e} \, {\mathcal C}_1^2 \nonumber \\ 
&\sim 1 + \frac{1.25}{6} + \frac{6.07143}{6^2} \, ,
\end{align}
where one sees that the 2PN $(O(u^2))$ contribution is numerically comparable
to the 1PN one. This suggests that the PN-expanded form of the tidal
amplification factor $\hat A_{\rm 1\, electric}^{(2)} (u)$ is slowly converging
and could get comparable or even larger contributions from higher powers of
$u$ (i.e., 3PN and higher terms). In order to get a feeling about the possible
origin of this slow convergence of the PN expansion, we followed the approach
of Ref. \cite{Damour:1997ub}, i.e., we looked for the existence of a nearby pole (in
the complex $u$ plane) within the formal analytic continuation of the considered
function $\hat A_{\rm 1\, electric}^{(2)} (u)$. [Ref.~\cite{Damour:1997ub}
considered the energy flux $F$ as a function of $x = (GM \, \Omega /
c^3)^{2/3}$; it pointed out that $F(x)$ had (in the test-mass limit) a pole at
the light-ring value $x = 1/3$ and recommended  improving the PN expansion of
$F(x)$ (for $x < 1/3$) by a Pad\'e-type resummation incorporating the
existence of this pole in $F(x)$.] By computing the exact test-mass limit of
the function $\hat A_{\rm 1\, electric}^{(2)} (u)$, we found that it formally
exhibits a pole located at the light-ring value $u_{\rm LR}^{\rm test \, mass}
= 1/3$ [see Eq.~(\ref{eq6.45})]. Such a pole is also present in other
amplification factors [see Eqs.~(\ref{eq6.46})--(\ref{eq6.48})], and we
discussed its origin. [Note that two equal-mass neutron stars will get in
contact before reaching this pole. However the idea here is that the
hidden presence of this pole in the analytical continuation of the
function $\hat A_{\rm 1\, electric}^{(2)} (u)$ is behind the bad convergence of
the Taylor expansion of this function in powers of $u$.] This led us to
suggest that one might get an improved value of the tidal amplification
factor $\hat A_{\rm 1\, electric}^{(2)} (u)$ by formally incorporating the
presence of this pole in the following Pad\'e-resummed manner:
\begin{equation}
\label{eq7.2}
\hat A_{\rm 1\, electric}^{(2)} (u) = 1 + \alpha_1^{2e} \, u + \alpha_2^{2e} 
\, \frac{u^2}{1-\hat r_{\rm LR} \, u} \, ,
\end{equation}
where $\hat r_{\rm LR} \equiv 1 / u_{\rm LR}$ is the (EOB-defined)
dimensionless light-ring radius, i.e., the solution of Eq.~(\ref{eq6.55}), with
$\tilde A (u)$ defined by Eq.~(\ref{eq5.21}). Let us point out that
Eq.~(\ref{eq7.2}) is equivalent to saying that the 2PN coefficient
$\alpha_2^{2e}$ becomes replaced by the effective distance-dependent coefficient
$\alpha_2^{\rm eff} (u) \equiv \alpha_2^{2e} / (1-\hat r_{\rm LR} \, u)$. Note
that $\alpha_2^{\rm eff} (u) > \alpha_2^{2e}$. In particular, for the ``typical''
compactness ${\mathcal C}_1 = {\mathcal C}_2 \sim 1/6$ considered above, and
when using the unperturbed value of $\hat r_{\rm LR}$, i.e. $\hat r_{\rm
  LR}^{(0)} = 3$, the effective value $\alpha_2^{\rm eff} (u)$ will, at
contact (i.e. when $u = u_{\rm contact} \simeq {\mathcal C}_1 \sim 1/6$), be
equal to $\alpha_2^{\rm eff} (u_{\rm contact}) \simeq \alpha_2^{2e} / (1-3 \,
{\mathcal C}_1) \sim \alpha_2^{2e} / (1-3/6) \sim 2 \, \alpha_2^{2e} \sim 12$. We
recalled in the Introduction that several comparisons between the analytical
(EOB) description of tidal effects and numerical simulations of tidally
interacting binary neutron stars \cite{Damour:2009wj, Baiotti:2010xh,
  Baiotti:2011am} have suggested the need for significant amplification
factors $\hat A_{\rm 1\, electric}^{(2)} (u)$ parametrized by rather large
values of $\alpha_2^{2e}$. However, up to now, the numerical results that have
been used have been affected by numerical errors that have not been fully
controlled. In particular, in the recent comparisons \cite{Baiotti:2010xh,
  Baiotti:2011am}, one did not have in hand sufficiently many simulations with
different resolutions for being able to compute and subtract the
finite-resolution error. We hope that  a more complete analysis will be
performed soon (see, in this respect, Refs. \cite{Bernuzzi:2011aq, Bernuzzi2012}). 
We recommend comparing resolution-extrapolated numerical data
to the pole-improved amplification factor of Eq. (\ref{eq7.2}). As discussed in
Sec.~\ref{sec6}, it might be necessary to use as value of $\hat r_{\rm LR}$
the improved estimate obtained from the full (tidally modified) value of the
$A$ potential. This suggests (especially for compactnesses ${\mathcal C}_1
\lesssim 1/6$) as discussed above that $\hat r_{\rm LR}$ might be
significantly larger than 3, thereby further amplifying the effective value of
$\alpha_2^{2e}$ during the last stages of the inspiral.

\smallskip

The present study has focused on the 2PN tidal effects in the interaction
Hamiltonian. There is also a 2PN tidal effect in the radiation reaction, which
has contributions from various tidally modified multipolar waveforms. The
tidal contribution to each (circular) multipolar gravitational waveform can be
parametrized (following Refs.~\cite{Damour:2009wj, Baiotti:2011am}) as an
additional term of the form
\begin{equation}
\label{eq7.3}
h_{\ell m}^{\rm tidal} (x) = \sum_J h_{\ell m}^{(J) \, {\rm LO}} (x) 
\, \hat h_{\ell m}^{(J) \, {\rm tail}} (x) \, \hat h_{\ell m}^{(J) \, {\rm PN}} (x) \, ,
\end{equation}
where $x \equiv (G(m_1 + m_2) \,\Omega / c^3)^{2/3}$; $J$ labels the various
tidal geometrical invariants, such as $J_{2e} \equiv G_{\alpha\beta} \,
G^{\alpha\beta}$; $h_{\ell m}^{(J) \, {\rm LO}} (x)$ denotes the leading-order
(i.e., Newtonian-order) tidal waveform; $\hat h_{\ell m}^{(J) \, {\rm tail}}
(x)$ the effect of tails \cite{Blanchet:1992br, Blanchet:1995fr} and their
resummed EOB form \cite{Damour:2007yf}; while
\begin{equation}
\label{eq7.4}
\hat h_{\ell m}^{(J) \, {\rm PN}} (x) = 1 + \beta_1^{(J \ell m)} x 
+ \beta_2^{(J \ell m)} x^2 + \ldots
\end{equation}
denotes the effect of higher PN contributions. The 1PN coefficient
$\beta_1^{(J_{2e} 2 2)}$ is known \cite{Vines:2011ud, DamourNagarVillain2012}. The
other 1PN coefficients needed for deriving a 2PN-accurate flux can be obtained
from applying the simple 1PN-accurate formalism of Eq. \cite{Blanchet:1989ki}. It
is more challenging to compute the 2PN coefficient $\beta_2^{(J_{2e} 22)}$.
Indeed, this requires applying the 2PN-accurate version \cite{Blanchet:1995fr}
of the Blanchet-Damour-Iyer wave-generation formalism
\cite{Blanchet:1985sp, Blanchet:1989ki, Damour:1990gj, Damour:1990ji} to the
tidal-modified Einstein equations (\ref{eq2.13}). Let us, however, note that 
although from a PN point of view, the 2PN coefficient $\beta_2^{(J_{2e} 22)}$
contributes to the phasing of coalescing binaries at the same formal level as
the dynamical 2PN coefficient $\alpha_2^{2e}$ determined above, it has been found
in Refs. \cite{Baiotti:2010xh, DamourNagarVillain2012} that (if $\beta_2^{(J_{2e}22)}
\sim \alpha_2^{2e}$) it has a significantly smaller observable effect.

\smallskip

Let us finally point out that our general result in Eq. (\ref{eq2.20}) also opens  the
possibility of computing the 3PN coefficient $\alpha_3^{2e}$ in the PN-expanded
amplification factor of the electric quadrupolar tidal interaction $\hat
A_{\rm 1\, electric}^{(2)} (u) = 1 + \alpha_1^{2e} \, u + \alpha_2^{2e} \, u^2 +
\alpha_3^{2e} \, u^3 + O(u^4)$. This computation would, however, be much more
involved than the calculation of $\alpha_2^{2e}$ because of the technical
subtleties in the regularization of self-field effects at the 3PN level
\cite{Jaranowski:1997ky, Blanchet:2000ub, Blanchet:2000nu, Blanchet:2000cw}
that necessitate   using dimensional regularization
\cite{Damour:2001bu, Blanchet:2003gy} instead of Hadamard regularization.

\vglue 1cm

\begin{acknowledgments}
  T.D. thanks Gilles Esposito-Far\`ese for useful discussions at an early
  stage of this work. D.B. thanks ICRANet for support, and IHES for
  hospitality during the start of this project.
\end{acknowledgments}

\vglue 1cm

\appendix

\section{Explicit forms of the (time-symmetric) 2PN-accurate brick potentials}

The explicit forms of the (time-symmetric) 2PN-accurate brick potentials $V$, $V_i$, etc. are
\cite{Blanchet:1998vx}
\begin{widetext}
\begin{eqnarray}
\label{eq3.7}
V &=& \frac{G m_1}{r_1} +\frac{G m_1}{c^2} \left(
-\frac{(n_1v_1)^2}{2r_1} +\frac{2v_1^2}{r_1} +
Gm_2 \left(-\frac{r_1}{4 r_{12}^3}-\frac{5}{4r_1r_{12}} +
\frac{r_2^2}{4r_1r_{12}^3}\right) \right) \nonumber \\
&+&
\frac{G m_1}{c^4r_1} \left( \frac{3 (n_1v_1)^4}{8}-
\frac{3(n_1v_1)^2 v_1^2}{2} + 2 v_1^4 \right) \nonumber \\
&+& \frac{G^2 m_1 m_2}{c^4} \left\{ v_1^2 \left( \frac{3 r_1^3}{16 r_{12}^5} - 
\frac{37 r_1}{16 r_{12}^3} - \frac{1}{r_1 r_{12}} - 
\frac{3 r_1 r_2^2}{16 r_{12}^5} + \frac{r_2^2}{r_1 r_{12}^3} \right) 
\right. \nonumber \\ & & \qquad \qquad +
v_2^2 \left( \frac{3 r_1^3}{16 r_{12}^5} + \frac{3 r_1}{16 r_{12}^3} + 
\frac{3}{2 r_1 r_{12}} - 
  \frac{3 r_1 r_2^2}{16 r_{12}^5} + \frac{r_2^2}{2 r_1 r_{12}^3} \right)
\nonumber \\ & & \qquad \qquad + 
(v_1v_2) \left(-\frac{3 r_1^3}{8 r_{12}^5} + \frac{13 r_1}{8 r_{12}^3} - 
\frac{3}{r_1 r_{12}} + 
  \frac{3 r_1 r_2^2}{8 r_{12}^5} - \frac{r_2^2}{r_1 r_{12}^3} \right)
\nonumber \\ & & \qquad \qquad
+(n_{12}v_1)^2 \left(-\frac{15 r_1^3}{16 r_{12}^5}+\frac{57 r_1}{16 r_{12}^3}+ 
  \frac{15 r_1 r_2^2}{16 r_{12}^5} \right)
\nonumber \\ & & \qquad \qquad
+(n_{12}v_2)^2 \left( -\frac{15 r_1^3}{16 r_{12}^5} - 
\frac{33 r_1}{16 r_{12}^3} + \frac{7}{8 r_1 r_{12}} + 
  \frac{15 r_1 r_2^2}{16 r_{12}^5} - \frac{3 r_2^2}{8 r_1 r_{12}^3} \right)
\nonumber \\ & & \qquad \qquad
+(n_{12}v_1)(n_{12}v_2) \left( \frac{15 r_1^3}{8 r_{12}^5} - 
\frac{9 r_1}{8 r_{12}^3} - \frac{15 r_1 r_2^2}{8 r_{12}^5} \right)
\nonumber \\ & & \qquad \qquad+
(n_1v_1)(n_{12}v_1) \left( -\frac{3 r_1^2}{2 r_{12}^4} + 
\frac{3}{4 r_{12}^2} + \frac{3 r_2^2}{4 r_{12}^4} \right)+
(n_{1}v_2)(n_{12}v_1) \left( \frac{3 r_1^2}{4 r_{12}^4}+
\frac{2}{r_{12}^2} \right) \nonumber \\ & & \qquad \qquad
+(n_1v_1)(n_{12}v_2) \left( \frac{3 r_1^2}{2 r_{12}^4} + 
\frac{13}{4 r_{12}^2} - \frac{3 r_2^2}{4 r_{12}^4} \right)+
(n_{1}v_2)(n_{12}v_2) \left( -\frac{3 r_1^2}{4 r_{12}^4} - 
\frac{3}{2 r_{12}^2} \right)
\nonumber \\ & & \qquad \qquad \left. +
(n_1v_1)^2 \left(-\frac{r_1}{8 r_{12}^3} +\frac{7}{8 r_1 r_{12}}
 - \frac{3 r_2^2}{8 r_1 r_{12}^3} \right)+
\frac{(n_1v_1) (n_1v_2) r_1}{2 r_{12}^3} \right\}
\nonumber \\ &+&
\frac{G^3 m_1^2 m_2}{c^4} \left( -\frac{r_1^3}{8 r_{12}^6} +
 \frac{5 r_1}{8 r_{12}^4} + \frac{3}{4 r_1 r_{12}^2} + 
  \frac{r_1 r_2^2}{8 r_{12}^6} - \frac{5 r_2^2}{4 r_1 r_{12}^4} \right)
\nonumber \\ &+&
\frac{G^3 m_1 m_2^2}{c^4} \left(-\frac{r_1^3}{32 r_{12}^6} + 
\frac{43 r_1}{16 r_{12}^4} + \frac{91}{32 r_1 r_{12}^2} - 
  \frac{r_1 r_2^2}{16 r_{12}^6} - \frac{23 r_2^2}{16 r_1 r_{12}^4} + 
  \frac{3 r_2^4}{32 r_1 r_{12}^6} \right)+O(6)+1\leftrightarrow 2  \ ,  
\end{eqnarray}

\begin{eqnarray}
\label{eq3.8}
V_i &=&  \frac{G m_1 v_1^i}{r_1}+
       n^i_{12} \frac{G^2 m_1 m_2}{c^2 r_{12}^2} \left((n_1v_1)+ 
\frac{3 (n_{12}v_{12}) r_1}{2 r_{12}}\right)
\nonumber \\ &+& \frac{v_1^i}{c^2} \left\{
\frac{G m_1}{ r_1} \left(-\frac{(n_1v_1)^2}{2} + v_1^2 \right)
+ G^2 m_1 m_2 \left( -\frac{3 r_1}{4 r_{12}^3} + 
\frac{r_2^2}{4 r_1 r_{12}^3} - \frac{5}{4 r_1 r_{12}} \right) \right\}\nonumber\\
&
+& v_2^i \frac{G^2 m_1 m_2 r_1}{2 c^2 r_{12}^3} +
O(4)+1\leftrightarrow 2 \ , 
\end{eqnarray}

\begin{eqnarray}
\label{eq3.9}
{\hat W}_{ij} &=&
\delta^{ij} \left(  -\frac{G m_1 v_1^2}{r_1}-\frac{G^2 m_1^2}{4 r_1^2}+
\frac{G^2 m_1 m_2}{r_{12} S} \right)
+\frac{G m_1 v_1^i v_1^j}{r_1} +\frac{G^2 m_1^2 n_1^i n_1^j}{4 r_1^2}
\nonumber \\ &+&
G^2 m_1 m_2 \left\{ \frac{1}{S^2} \left(n_1^{(i} n_2^{j)} 
+ 2n_1^{(i} n_{12}^{j)} \right)-
n_{12}^i n_{12}^j \left(\frac{1}{S^2}+\frac{1}{r_{12} S} \right) \right\}
+ O(2)+1\leftrightarrow 2 \ , 
\end{eqnarray}
\end{widetext}

\begin{widetext}
\begin{eqnarray}
\label{eq3.10}
{\hat R}_i &=&
G^2 m_1 m_2 n_{12}^i \left\{ -\frac{(n_{12}v_1)}{2S} \left(\frac{1}{S}+
\frac{1}{r_{12}} \right)
-\frac{2 (n_2v_1)}{S^2}+\frac{3(n_2v_2)}{2S^2} \right\}
\nonumber \\ &+&
n_1^i \left\{ \frac{G^2 m_1^2 (n_1v_1)}{8 r_1^2}
+\frac{G^2 m_1 m_2}{S^2} \left( 2 (n_{12}v_1) - \frac{3(n_{12}v_2)}{2}
+2 (n_2v_1) - \frac{3(n_2v_2)}{2} \right) \right\}
\nonumber \\ &+& 
v_1^i \left\{ -\frac{G^2 m_1^2}{8r_1^2}+
G^2 m_1 m_2 \left( \frac{1}{r_1 r_{12}} + \frac{1}{2 r_{12}S} \right) \right\}
-v_2^i\frac{G^2 m_1 m_2}{r_1 r_{12}} 
+ O(2)+1\leftrightarrow 2 \ , 
\end{eqnarray}

\begin{eqnarray}
\label{eq3.11}
{\hat X} &=&
\frac{G^2 m_1^2}{8r_1^2} \left((n_1v_1)^2-v_1^2\right)
+G^2 m_1 m_2 v_1^2 \left(\frac{1}{r_1 r_{12}} 
+ \frac{1}{r_1 S} + \frac{1}{r_{12} S} \right)
\nonumber \\ &+&
G^2 m_1 m_2 \left\{v_2^2 \left(-\frac{1}{r_1 r_{12}} + \frac{1}{r_1 S} 
+ \frac{1}{r_{12} S} \right)
-\frac{(v_1v_2)}{S} \left(\frac{2}{r_1}+
\frac{3}{2 r_{12}} \right)
-\frac{(n_{12}v_1)^2}{S} \left(\frac{1}{S}+\frac{1}{r_{12}}\right) \right.
\nonumber \\ & & \qquad \qquad
-\frac{(n_{12}v_2)^2}{S} \left(\frac{1}{S} + \frac{1}{r_{12}} \right)
+\frac{3(n_{12}v_1)(n_{12}v_2)}{2S} \left(\frac{1}{S} + 
\frac{1}{r_{12}}\right)+\frac{2(n_{12}v_1)(n_1v_1)}{S^2}
\nonumber \\ & & \qquad \qquad
-\frac{5(n_{12}v_2)(n_1v_1)}{S^2}
-\frac{(n_1v_1)^2}{S} \left(\frac{1}{S} + \frac{1}{r_1} \right)
+\frac{2(n_{12}v_2)(n_1v_2)}{S^2} \nonumber \\ & & \qquad \qquad
+\frac{2(n_1v_1)(n_1v_2)}{S} \left(\frac{1}{S} + \frac{1}{r_1} \right)
-\frac{(n_1v_2)^2}{S} \left(\frac{1}{S} + \frac{1}{r_1}\right)
-\frac{2(n_{12}v_2)(n_2v_1)}{S^2} \nonumber \\ 
& & \left. \qquad \qquad +\frac{2(n_1v_2)(n_2v_1)}{S^2}
-\frac{3(n_1v_1)(n_2v_2)}{2 S^2} \right\}+\frac{G^3 m_1^3}{12 r_1^3}
\nonumber \\ 
&+& G^3 m_1^2 m_2 \left(\frac{1}{2 r_1^3} + \frac{1}{16 r_2^3} + 
\frac{1}{16 r_1^2 r_2} - \frac{r_2^2}{2 r_1^2 r_{12}^3} + 
\frac{r_2^3}{2 r_1^3 r_{12}^3} - \frac{r_1^2}{32 r_2^3 r_{12}^2} 
- \frac{3}{16 r_2 r_{12}^2} + 
\frac{15 r_2}{32 r_1^2 r_{12}^2} \right. \nonumber \\ 
& & \left. \qquad \qquad - \frac{r_2^2}{2 r_1^3 r_{12}^2} 
- \frac{r_2}{2 r_1^3 r_{12}} - \frac{r_{12}^2}{32 r_1^2 r_2^3} \right)
+G^3 m_1 m_2^2 \left( -\frac{1}{2 r_{12}^3} + \frac{r_2}{2 r_1 r_{12}^3} 
-\frac{1}{2 r_1 r_{12}^2} \right) \nonumber \\
& & \qquad \qquad + \, O(2)+1\leftrightarrow 2 \ . 
\end{eqnarray}
\end{widetext}

Here ${\bm r}_1 \equiv {\bm x} - {\bm y}_1$, $r_1 \equiv \vert {\bm r}_1
\vert$, ${\bm n}_1 \equiv {\bm r}_1 / r_1$, ${\bm r}_2 \equiv {\bm x} - {\bm
  y}_2$, etc., ${\bm y}_{12} \equiv {\bm y}_1 - {\bm y}_2$, $r_{12} \equiv
\vert {\bm y}_{12} \vert$, ${\bm n}_{12} \equiv {\bm y}_{12} / r_{12}$, ${\bm
  v}_{12} \equiv {\bm v}_1 - {\bm v}_2$, $(n_{12} \, v_1) \equiv {\bm n}_{12}
\cdot {\bm v}_1$. In addition, the notation $1\leftrightarrow 2$ means adding the terms obtained 
by exchanging the  particle labels 1 and 2, while the quantity $S$ denotes the perimeter of the triangle
defined by ${\bm x}$, ${\bm y}_1$ and ${\bm y}_2$, viz.
\begin{equation}
\label{eq3.12}
S \equiv r_1 + r_2 + r_{12} \, .
\end{equation}


\end{document}